\documentclass[
  aps,
  prx,          
  reprint,
  superscriptaddress,
  floatfix
]{revtex4-2}

\usepackage{graphicx}
\usepackage{adjustbox}
\usepackage{amssymb,amsmath}
\usepackage{xcolor}
\usepackage{braket}
\usepackage{booktabs}
\usepackage{placeins}
\usepackage{microtype}
\usepackage{titlesec}

\newsavebox{\hepbox}
\newlength{\aceneH}
\newlength{\acnW}
\newlength{\tldW}
\newlength{\ovlp}
\newlength{\tldUp}
\newlength{\tldRight}

\titleformat{\paragraph}[block]
  {\normalfont\normalsize\itshape}
  {}
  {0pt}
  {}

\titlespacing*{\paragraph}
  {0pt}
  {3.25ex plus 1ex minus .2ex}
  {1ex}
  
\usepackage[
  colorlinks=true,
  linkcolor=blue,
  citecolor=blue,
  urlcolor=blue
]{hyperref}

\usepackage{quantikz}

\newcommand{\ii}{\mathrm{i}}
\newcommand{\CX}{\mathrm{CX}}
\newcommand{\tUPS}{\ensuremath{\mathrm{tUPS}}}
\newcommand{\ctUPS}{\ensuremath{\mathrm{c\text{-}tUPS}}}

\usepackage{algorithm}
\usepackage{algpseudocode}

\begin{document}

\title{Hardware-Efficient Ground-State Preparation using\\
Variational Imaginary-Time Majorana Evolution}

\author{Federico Santona}
\email{federicosntn@gmail.com}
\affiliation{IQM Quantum Computers GmbH, Georg-Brauchle-Ring 23--25,
80992 Munich, Germany}
\affiliation{RPTU Kaiserslautern-Landau, 67663 Kaiserslautern, Germany}

\author{Manuel G. Algaba}
\affiliation{IQM Quantum Computers GmbH, Georg-Brauchle-Ring 23--25,
80992 Munich, Germany}
\affiliation{PhD Programme in Condensed Matter Physics, Nanoscience and Biophysics,
Doctoral School, Universidad Aut\'onoma de Madrid}

\author{Aeishah Ameera Anuar}
\affiliation{IQM Quantum Computers GmbH, Georg-Brauchle-Ring 23--25,
80992 Munich, Germany}
\affiliation{Technical University of Munich, CIT, Department of Computer Science, Boltzmannstra{\ss}e 3, 85748 Garching, Germany}

\author{Anna M. Wernbacher}
\affiliation{IQM Quantum Computers GmbH, Friedrichstraße 68,  10117 Berlin, Germany}

\author{Prachi Sharma}
\affiliation{IQM Quantum Computers GmbH, Georg-Brauchle-Ring 23--25,
80992 Munich, Germany}

\author{Fedor Šimkovic IV}
\email{fedor.simkovic@iqm.tech}
\affiliation{IQM Quantum Computers GmbH, Georg-Brauchle-Ring 23--25,
80992 Munich, Germany}

\begin{abstract}
Compact ground-state preparation circuits are essential for early fault-tolerant quantum chemistry. We introduce Variational Imaginary-time Majorana Evolution (VIME), a classical pre-training algorithm that extends operator-projected variational quantum imaginary-time evolution (OVQITE) to molecular electronic structure using a single compiled Majorana-propagation (MP) surrogate graph.  We pair VIME with a compressed tiled Unitary Product State (c-tUPS) ansatz that we develop to reduce both classical simulation and quantum state-preparation costs. In the strongly correlated ruthenium complex TLD-1411, active-space calculations requiring up to 52 qubits achieve chemical precision against DMRG reference energies, with energy errors approximately $400\times$ smaller than those from ADAPT-VQE-based variational Majorana-propagation (ADAPT-VMPE). Across the strongly multireference acene series up to heptacene (60 qubits), VIME matches or improves on the MP-reported energy accuracy of the ADAPT-VMPE comparison
points with up to $26\times$ fewer CNOTs and $167\times$ lower two-qubit depth. These results establish imaginary-time classical pre-training of compressed variational ansätze as a promising route to resource-efficient molecular ground-state preparation.
\end{abstract}

\maketitle

\section{Introduction}

Accurate descriptions of strongly correlated electronic systems underpin
challenging problems in molecular catalysis \cite{reiher2017elucidating},
energy-storage materials \cite{delgado2022batteries}, and quantum materials
\cite{ma2020nearterm}. Their multi-configurational character can render
single-reference approximations unreliable \cite{lyakh2012multireference},
while exact treatments are limited by the combinatorial growth of the
electronic Hilbert space \cite{bauer2020quantum,cao2019quantum}.
Quantum computers can represent this space with a linear number of qubits
\cite{feynman1982simulating,lloyd1996universal}, and fault-tolerant
algorithms such as quantum phase estimation (QPE) can systematically
resolve molecular energies \cite{abrams1999quantum,aspuru2005simulated}.
Their resource requirements, however, depend critically on preparing
an initial state with sufficient ground-state overlap
\cite{lee2021evenmore,lee2023evaluating}.
Compact circuits for approximate ground-state preparation are therefore
an important component of electronic-structure algorithms, especially in early fault-tolerant regimes.

Variational algorithms construct approximate ground states through
parameterised circuits and classical optimization
\cite{mcclean2016theory,cerezo2021vqa,endo2021hybrid}.
The Variational Quantum Eigensolver (VQE) is the canonical example
\cite{peruzzo2014vqe,tilly2022vqe}, but repeated measurements and
difficult optimization landscapes, including barren plateaus and
noise-induced gradient suppression, limit its practical performance
\cite{mcclean2018barren,cerezo2021cost,wang2021noiseinduced}.
Classical pre-training offers a way to perform the parameter search
entirely on a classical computer, using approximate expectation values
to guide the optimization.
The trained circuit can then be executed on a quantum processor as a
state-preparation subroutine for downstream quantum algorithms~\cite{lee2021evenmore,
lee2023evaluating,
kosugi2022imaginary,
silva2023fragmented,
gluza2026double,
liu2026quantumchannelpolynomialprocessing,
benavidesriveros2026augmenting}.
The challenge is to keep these classical evaluations accurate and affordable, while ensuring the corresponding circuits remain compact enough for quantum execution.

Majorana Propagation (MP) provides an attractive route to fully classical pre-training of fermionic quantum circuits. MP propagates observables in a Majorana-monomial basis and controls the resulting expansion through truncation, with polynomial cost in the system size at fixed truncation cutoff. In particular, monomial \emph{length truncation} is independent of the variational parameters, allowing the propagation to be compiled into a reusable surrogate graph for efficient repeated energy evaluations \cite{miller2025majorana,chakraborty2026adaptvmpe}. This capability underpins the Variational Majorana Propagation Eigensolver (VMPE), which uses MP as a classical surrogate for VQE optimization, replacing repeated quantum measurements with classical energy evaluation \cite{chakraborty2026adaptvmpe}. The performance of VMPE, however, depends on both the choice of variational ansatz and the accuracy of the truncated MP surrogate.

ADAPT-VMPE addresses both aspects by combining adaptive ansatz construction with MP. It constructs the circuit incrementally, building on the ability of ADAPT-VQE to mitigate some of the limitations of fixed ansätze \cite{grimsley2019adaptvqe,chakraborty2026adaptvmpe}, while its relatively unstructured circuits lie in the regime where MP has average-case accuracy guarantees~\cite{miller2025majorana}. However, the molecular circuits produced by ADAPT-VMPE can require on the order of $10^4$ CNOTs at system sizes of scientific interest, even with all-to-all connectivity and using circuit optimization techniques~\cite{chakraborty2026adaptvmpe}. This raises the question of whether the advantages of MP can be preserved in substantially shallower circuits.

Imaginary-time evolution (ITE) provides an alternative route to ground-state preparation without directly minimizing the energy. Normalized evolution under $e^{-\tau \hat H}$ drives a state with nonzero ground-state overlap toward the ground-state subspace \cite{motta2020determining}. Because this evolution is non-unitary, it cannot be implemented directly as a quantum circuit. Variational quantum imaginary-time evolution (VQITE) instead approximates the imaginary-time dynamics within a parameterized circuit manifold \cite{mcardle2019vqite,yuan2019variational}. Conventional VQITE formulations typically involve overlaps between derivatives of the variational state, while MP provides direct access to observable expectation values. Operator-projected VQITE (OVQITE) instead formulates the imaginary-time equations in terms of expectation values of a polynomially sized operator pool and their parameter derivatives \cite{anuar2024operator}, making it particularly well suited for integration with MP.

In this paper, we therefore introduce Variational Imaginary-time Majorana Evolution (VIME), a classical pre-training algorithm that combines OVQITE with MP, using a single compiled surrogate graph to evaluate the expectation values and parameter derivatives required for OVQITE updates. Here, we apply VIME to molecular electronic structure and combine it with c-tUPS (cf. Sec.~\ref{subsec: ctUPS}), a compressed variant of the tiled Unitary Product State (tUPS) ansatz \cite{burton2024tups} developed in this work. The tUPS ansatz applies a fixed tile structure across groups of orbitals with three variational parameters per tile. c-tUPS retains this structure while only using half as many length-four Majorana rotations. This compression reduces the cost of truncated MP while yielding lower variational energies compared to tUPS. By jointly synthesizing the operations within each tile, c-tUPS can be implemented using only $10$ CNOTs per tile under all-to-all connectivity, compared to previously reported 20–21 CNOTs for tUPS~\cite{burton2024tups,shaik2026_12cnot}. 

We benchmark VIME and c-tUPS on hydrogen chains, TLD-1411,
and the acene series. For hydrogen chains, we find that MP truncation errors remain manageable for the structured c-tUPS ansatz, while the classical computational cost permits simulations at system sizes of potential scientific interest. On the same ansatz, VIME produces both lower energies and larger ground-state overlaps than VMPE. For the strongly correlated ruthenium complex TLD-1411~\cite{chakraborty2026adaptvmpe}, we consider active spaces up to $(26e,26o)$ ($52$ qubits) and obtain energy estimates within chemical precision of density matrix renormalization group (DMRG) references. Direct VMPE energy minimization on the same c-tUPS circuit architectures instead drives the estimated energies substantially below said references.

We further benchmark the energy accuracy and circuit resources of VIME with c-tUPS against ADAPT-VMPE. For TLD-1411, VIME with c-tUPS achieves closer MP-reported agreement with DMRG while using more than an order of magnitude fewer CNOTs and over two orders of magnitude lower two-qubit depth. This circuit-resource advantage extends to the strongly
multireference acene series up to heptacene, within a
$(30e,30o)$ active space ($60$ qubits)~\cite{pelzer2011strong}:
VIME with c-tUPS matches or improves on the MP-reported
energy accuracy of the selected ADAPT-VMPE comparison
points, while reducing CNOT counts by factors of
$3.3$--$26$ and two-qubit depth by $36$--$167$.
Across both molecular families, these circuit-resource
reductions are obtained with allocated CPU-hours of the
same order of magnitude as ADAPT-VMPE.

The paper is organized as follows. In Sec.~\ref{sec:methods}, we review MP and OVQITE, formulate VIME, and introduce the c-tUPS ansatz. In Sec.~\ref{sec:results}, we validate the quantities entering the VIME parameter update against exact statevector calculations and assess MP truncation errors across structured circuit architectures. We then compare VIME with VMPE and ADAPT-VMPE in terms of molecular energy accuracy and quantum circuit resources. Finally, in Sec.~\ref{sec:conclusions}, we discuss the implications for classical pre-training of molecular state-preparation circuits and outline directions for future research.

\section{Methods}
\label{sec:methods}

\subsection{Majorana Propagation (MP)}
\label{subsec: mp}

Majorana Propagation (MP) \cite{miller2025majorana} is an entirely classical
framework for simulating fermionic circuits, which evaluates expectation values
in the Heisenberg picture at a cost that, for a fixed monomial-length cutoff,
is linear in circuit depth and polynomial in system size. For a system of $N$
fermionic modes, let $\hat{a}_j^\dagger$ and $\hat a_j$ denote the fermionic creation and
annihilation operators associated with mode $j$. We introduce the corresponding
$2N$ Majorana operators
$\hat m_{2j-1}\equiv \hat a_j^\dagger+\hat a_j$ and
$\hat m_{2j}\equiv i(\hat a_j^\dagger-\hat a_j)$, which are Hermitian, unitary, and satisfy
$\{\hat m_i,\hat m_k\}=2\delta_{ik}$. Throughout this work, $[A,B]=AB-BA$ and $\{A,B\}=AB+BA$ denote the commutator and anticommutator, respectively.  Ordered products of distinct Majorana operators
define the Majorana monomials
$\hat M_{\vec b}=i^{r_{\vec b}}\hat m_1^{b_1}\cdots \hat m_{2N}^{b_{2N}}$, with
$\vec b\in\{0,1\}^{2N}$ and $r_{\vec b}$ chosen such that $\hat M_{\vec b}$ is
Hermitian. The $4^N$ monomials form an orthogonal basis of the operator algebra,
and we call $w=\|\vec b\|_1$ the length of $\hat M_{\vec b}$. Fermionic Hamiltonians of chemical interest are spanned by low-length monomials, in particular those of length two
and four, so their representation in this basis is compact.

More concretely, MP evaluates

\begin{equation}
\label{eq: MP expectation value}
\begin{aligned}
f(\theta)
&= \langle\hat H\rangle_\theta
 = \operatorname{Tr}\!\left[
 \hat U(\theta)\hat \rho \hat U^\dagger(\theta) \hat H
 \right],\\
\hat U(\theta)
&= \prod_{j=1}^{L} e^{-i\theta_j \hat M_{\vec b_j}/2},
\end{aligned}
\end{equation}

where $\rho$ is the reference-state density operator and
$\hat H=\sum_{\vec b}\alpha_{\vec b}\hat M_{\vec b}$
is the observable of interest. The expansion includes
the length-zero monomial $\hat M_{\mathbf0}=\mathbf1$,
whose coefficient $\alpha_{\mathbf0}$ contains the full,
constant identity contribution. In this work, we take the reference state to be a Fock state,
$\rho = |\mathbf{n}\rangle\langle\mathbf{n}|$, with occupation string
$\mathbf{n}\in\{0,1\}^{N}$.
MP evaluates Eq.~\eqref{eq: MP expectation value} by back-propagating each
Majorana monomial of $\hat{H}$ through the gates composing $\hat U(\theta)$. The action
of a single gate $\hat U_j=e^{-i\theta_j\hat M_{\vec b_j}/2}$ is exact and takes one of
two forms,

\begin{equation}
\label{eq:majorana-propagation}
\hat M_{\vec b}\ \xrightarrow{\,\hat{U}_j\,}\
\begin{cases}
\hat M_{\vec b},
& [\hat M_{\vec b},\hat M_{\vec b_j}]=0,\\[2pt]
c_j \hat M_{\vec b}+i s_j \hat M_{\vec b_j}\hat M_{\vec b},
& \{\hat M_{\vec b},\hat M_{\vec b_j}\}=0,
\end{cases}
\end{equation}

with $c_j=\cos\theta_j$ and $s_j=\sin\theta_j$. After all gates have been
applied, the expectation value is evaluated using
$f(\theta)=\sum_{\vec b}c_{\vec b}(\theta)
\operatorname{Tr}[\hat \rho \hat M_{\vec b}]$.
For a Fock state, this trace vanishes unless $\hat{M}_{\vec b}$ contains an even
number of Majorana operators on every fermionic mode. We refer to the
surviving terms as paired monomials.\\

Repeated application of the branching rule in
Eq.~\eqref{eq:majorana-propagation} can cause the number of propagated
monomials to grow exponentially with the number of branching events.
The standard MP procedure controls this growth by truncating the
expansion after each Majorana rotation. Here, we use Majorana-length
truncation supplemented by \emph{paired acceptance}, following
Ref.~\cite{chakraborty2026adaptvmpe}.
Let
\begin{equation}
\mathcal P
=
\left\{
\vec b:
b_{2j-1}=b_{2j}\ \text{for all }j
\right\}
\end{equation}
denote the set of paired monomials. For a chosen Majorana-length cutoff $w^\star$, we retain $\hat M_{\vec b}$ at each truncation point whenever
\begin{equation}
\|\vec b\|_1\leq w^\star
\qquad\text{or}\qquad
\vec b\in\mathcal P .
\label{eq:paired-acceptance}
\end{equation}
Thus, the length cutoff is applied strictly only to unpaired monomials;
paired monomials are retained even when their length exceeds
\(w^\star\). This exception is important because paired monomials are
precisely those with potentially nonzero expectation value on the Fock
reference. A strict length cutoff would otherwise discard contributing
terms solely because of their length, which can substantially degrade
the accuracy of the resulting expectation values. We denote the
corresponding approximation by \(f^{(w^\star)}(\theta)\).

The effect of a gate on the monomial length follows directly from the Majorana
algebra. Consider a propagated monomial $\hat M_{\vec b}$ of length
$w=\|\vec b\|_1$ and a gate generated by $\hat M_{\vec b_j}$ of length
$w_j=\|\vec b_j\|_1$. We define
$s =\vec b_j\cdot\vec b$ as the number of Majorana operators shared by the
generator and the propagated monomial. Branching occurs if and only if
$s$ is odd, in which case the branching term
$\hat M_{\vec b_j}\hat M_{\vec b}$ has length

\begin{equation}
\label{eq:length-update}
w' = w + w_j - 2s .
\end{equation}

Even-length generators therefore preserve the parity of the monomial length.
In particular, for one-body generators ($w_j=2$), branching requires
$s=1$ and therefore $w'=w$, so the branching term remains the same length.
For two-body generators ($w_j=4$), branching can instead change the monomial
length by $\pm2$, allowing monomials to cross the truncation threshold.\\

For monomial-length truncation, Ref.~\cite{miller2025majorana} establishes
average-case accuracy guarantees for an ensemble in which the circuit
generators are independently and uniformly sampled length-four monomials,
the angles are uniformly distributed over $[0,2\pi]$, and the observable is
a random homogeneous operator. Within this ensemble, the mean-squared
truncation error is bounded by
\begin{equation}
\label{eq: MP error bound}
2^{1-N}
+
\left(\frac{e w^\star}{N}\right)^{w^\star/2},
\end{equation}
with an additional depth-dependent suppression for sufficiently deep
circuits. The bound therefore decreases exponentially with the retained
monomial length. Two properties of the random ensemble underpin Eq.~\eqref{eq: MP error bound}. First,
high-length monomials are unlikely to be paired and hence are unlikely to
contribute to the final trace against a Fock state. Second, propagation back
towards lower monomial lengths is suppressed: for uniformly sampled
generators, large overlaps $s$ in Eq.~\eqref{eq:length-update} are rare, and
the ratio of length-decreasing to length-increasing branching events scales
as $O(1/N^2)$. These properties need not hold for structured circuits, where
successive generators can have correlated support and the circuit parameters
need not follow the distribution assumed by the random ensemble. The
guarantees of Ref.~\cite{miller2025majorana} therefore do not directly extend
to arbitrary structured circuits.\\

Monomial length is, however, not the only criterion that can be used to control the
growth of the propagated operator. We additionally consider two alternative
truncation schemes based, respectively, on the propagation history of a
monomial and on its spatial support.\\

The first is a sine-order truncation. According to
Eq.~\eqref{eq:majorana-propagation}, every branching event that generates a
new monomial contributes a factor $\sin\theta_j$ to the corresponding
propagation path. A path $\gamma$ therefore acquires a coefficient of the form
\begin{equation}
C_\gamma(\theta)
\propto
\prod_{j\in\mathcal J_\gamma^{\cos}}\cos\theta_j
\prod_{j\in\mathcal J_\gamma^{\sin}}\sin\theta_j ,
\end{equation}
where $\mathcal J_\gamma^{\cos}$ and
$\mathcal J_\gamma^{\sin}$ contain the rotation indices
at which the path acquires a cosine or sine factor,
respectively. We define the sine order of a path as
the number of sine factors,
$|\mathcal J_\gamma^{\sin}|$. For sufficiently small circuit angles, contributions from higher sine orders are suppressed, since $\sin\theta_j\simeq\theta_j$. This motivates retaining only paths satisfying $|\mathcal S_\gamma| \leq p^\star $, where $p^\star$ is the prescribed sine-order cutoff.

The second alternative is \emph{orbital truncation}, a truncation scheme we introduce here that, to our knowledge, has not previously been considered for MP. For a Majorana monomial $\hat M_{\vec b}$, we define its orbital support size $o(\vec b)$ as the number of distinct spatial orbitals on which it acts nontrivially. Monomials of equal Majorana length can differ in how their Majorana operators are distributed among spatial orbitals. Orbital truncation retains only monomials satisfying $o(\vec b) \leq o^\star$ , for a chosen support cutoff $o^\star$. This criterion preferentially retains
operators that remain localized on relatively few spatial orbitals and should
not be confused with an active-space approximation: the molecular Hamiltonian
and its orbital space are left unchanged, and only the propagated operator
expansion is truncated.\\

Under such parameter-agnostic truncations, the propagation structure is independent of
the circuit parameters $\theta$. The propagation can therefore be compiled once into a
fixed surrogate graph, whose nodes represent the monomials retained at each
circuit layer. Its edges encode the branching rule of
Eq.~\eqref{eq:majorana-propagation}: a commuting monomial propagates 
unchanged with weight $1$, while an anticommuting monomial contributes one
branch proportional to $\cos\theta_j$ and a second branch proportional to
$\pm\sin\theta_j$. For new values of $\theta$, only these edge weights need to
be updated, so evaluating $f^{(w^\star)}(\theta)$ only requires a traversal of
the graph already compiled. This separates the one-off construction of the
surrogate from its substantially cheaper repeated evaluation
\cite{miller2025majorana,chakraborty2026adaptvmpe}.\\

MP provides expectation values of fermionic observables for
a parameterised circuit, but does not itself prescribe how the circuit
parameters should be updated. The direct approach is to minimize the
energy reported by the surrogate, as in VMPE \cite{chakraborty2026adaptvmpe}. Here, instead, we use the surrogate to determine the evolution of the circuit parameters in imaginary time. Operator-projected variational quantum imaginary-time evolution (OVQITE) \cite{anuar2024operator} provides a suitable parameter-update rule expressed through observable expectation values and their
parameter derivatives, making it directly compatible with our MP
construction. We review its formulation in the following section.

\subsection{Operator-projected Variational Quantum Imaginary-Time Evolution (OVQITE)}
\label{subsec: ovqite}

Imaginary-time evolution (ITE) drives an initial state with nonzero
ground-state overlap toward the ground-state subspace by suppressing
excited-state components under $e^{-\tau\hat H}$ as the imaginary
time $\tau$ increases
\cite{motta2020determining,mcardle2019vqite}. For a normalized state described by the density matrix $\hat\rho_\tau$, the corresponding imaginary-time evolution can be written as~\cite{yuan2019variational}
\begin{equation}
\label{eq: density ITE}
\partial_\tau\hat\rho_\tau
=
-\{\hat H,\hat\rho_\tau\}
+
2\langle\hat H\rangle_{\hat\rho_\tau}\hat\rho_\tau.
\end{equation}
Consequently, the
expectation value of any Hermitian operator $\hat O$ evolves according to
\begin{equation}
\label{eq: Ehrenfest theorem}
\begin{aligned}
\partial_\tau\langle\hat O\rangle_{\hat\rho_\tau}
&=
v(\hat O,\hat\rho_\tau), \\
v(\hat O,\hat\rho_\tau)
&\equiv 
-\langle\{\hat H,\hat O\}\rangle_{\hat\rho_\tau}
+
2\langle\hat H\rangle_{\hat\rho_\tau}
 \langle\hat O\rangle_{\hat\rho_\tau}.
\end{aligned}
\end{equation}
We refer to $v(\hat O,\hat\rho_\tau)$ as the imaginary time (IT) velocity hereafter.
Operator-projected imaginary-time evolution (OITE)
\cite{anuar2024operator} replaces the requirement that the full state obey
Eq.~\eqref{eq: density ITE} by requiring Eq.~\eqref{eq: Ehrenfest theorem} to be satisfied
only for a chosen operator pool
$S_\tau=\{\hat O_k\}_{k=1}^{|S_\tau|}$, whose size $|S_\tau|$ is selected to grow at most
polynomially with the system size. When $S_\tau$ is a complete operator basis, the evolution of its expectation
values uniquely determines the full density-matrix evolution and OITE is
equivalent to exact ITE.

OVQITE restricts the OITE dynamics to a parametrized circuit state
$\ket{\psi_\theta}$, with
$\hat\rho_\theta=\ket{\psi_\theta}\bra{\psi_\theta}$ and
$\theta\in\mathbb{R}^{N_\theta}$. Rather than propagating the state directly,
the circuit parameters are evolved in imaginary time. Their induced change
in each pool-operator expectation value is
\begin{equation}
\label{eq: operator chain rule}
\partial_\tau
\langle\hat O_k\rangle_{\hat\rho_\theta}
=
\sum_{j=1}^{N_\theta}
\frac{\partial\langle\hat O_k\rangle_{\hat\rho_\theta}}
     {\partial\theta_j}
\dot\theta_j.
\end{equation}
The OITE equations, Eq.~\eqref{eq: Ehrenfest theorem}, prescribe a target IT velocity for each
operator in the pool, whereas the variational circuit can only generate
velocities obtainable by changing its parameters. We therefore define the Jacobian $J(\theta)$ of the pool expectation values
with respect to the circuit parameters and the IT velocity vector $v$ by
\begin{equation}
\label{eq: J and v}
J_{kj}(\theta)
=
\frac{\partial\langle\hat O_k\rangle_{\hat\rho_\theta}}
     {\partial\theta_j},
\qquad
v_k(\theta)
=
v(\hat O_k,\hat\rho_\theta).
\end{equation}
Using these definitions, the chain-rule expression of Eq. \eqref{eq: operator chain rule} shows that evolving the circuit parameters with velocity $\dot\theta=d\theta/d\tau$ changes the pool expectation values at rates given by $J\dot\theta$.
Reproducing the OITE dynamics would therefore require
$J\dot\theta=v$. For a restricted variational ansatz, this system
need not have an exact solution. We instead choose $\dot\theta$
to minimize the squared mismatch between the circuit-induced
and target rates,
\begin{equation}
\label{eq: OVQITE McLachlan distance}
L(\dot\theta\mid\theta,S_\tau)
=
\frac{1}{2}\left\|J\dot\theta-v\right\|_2^2.
\end{equation}
This least-squares prescription is the operator-projected
form of McLachlan's variational principle
\cite{mclachlan1964variational,yuan2019variational,anuar2024operator}.
Minimizing $L$ with respect to $\dot\theta$ gives the normal equations
\begin{equation}
\label{eq: OVQITE linear system}
G_{\theta,S_\tau}\dot\theta
=
b_{\theta,S_\tau},
\end{equation}
where $G=J^TJ$ is the Gram matrix of the Jacobian columns and
$b=J^Tv$. Thus, at each imaginary-time step and for the
current parameter point $\theta$, solving Eq.~\eqref{eq: OVQITE linear system}
selects the parameter velocity $\dot\theta$ whose induced change
$J\dot\theta$ in the pool expectation values closely approximates the
target IT velocity $v$.

The Gram matrix $G$ is positive semi-definite and satisfies
$\operatorname{rank}G\leq\min(N_\theta,|S_\tau|)$. For arbitrary  ans\"atze, $G$ can be strongly rank deficient, and we therefore regularize the normal equations
with a Tikhonov shift $\lambda$,
\begin{equation}
\label{eq: regularized OVQITE linear system}
\left(G+\lambda I \right)\dot\theta
=
b.
\end{equation}
After solving the linear system the parameters are evolved in imaginary time as
$\theta_{\tau+\delta}=\theta_\tau+\delta\dot\theta$ for an imaginary time step
$\delta>0$.

We follow Ref. ~\cite{anuar2024operator} and choose $S_\tau=S_H$, where the Hamiltonian pool $S_H$ consists of the operators in the expansion $\hat H=\sum_k c_k\hat O_k$. Here we apply this choice to molecular electronic structure Hamiltonians. In a basis of $N$ spin orbitals, the second-quantized Hamiltonian is
\begin{equation}
\label{eq: electronic Hamiltonian}
\hat H
=
E_{\mathrm{const}}
+\sum_{p,q=1}^{N} h_{pq} \hat a_p^\dagger \hat a_q
+\frac14\sum_{p,q,r,s=1}^{N}
\langle pq\Vert rs\rangle
\hat a_p^\dagger \hat a_q^\dagger \hat  a_s \hat a_r,
\end{equation}
where $h_{pq}$ and $\langle pq\Vert rs\rangle$ are the one-electron
and antisymmetrized two-electron integrals, respectively, and
$E_{\mathrm{const}}$ includes nuclear and frozen-core contributions.
We use its expansion in Hermitian Majorana monomials,
so that $S_H$ contains only
monomials of length two and four.
This choice connects the projected observable dynamics directly
to the energy. Multiplying the OITE equations, Eq.~\eqref{eq: Ehrenfest theorem}, for each
$\hat O_k$ by $c_k$ and summing gives
\begin{equation}
\label{eq: variance descent}
\partial_\tau\langle\hat H\rangle_{\hat\rho_\tau}
=
-2\,\operatorname{Var}_{\hat\rho_\tau}(\hat H)
\leq 0.
\end{equation}
Thus, if Eq.~\eqref{eq: Ehrenfest theorem} is satisfied exactly for every operator in $S_H$, the energy decreases monotonically and is stationary only when the state is supported entirely within a single eigenspace of $\hat H$.

Using this Hamiltonian pool in the variational update, however, requires evaluating expectation values and parameter derivatives for every pool operator at each step. For molecular systems, the number of these quantities creates a substantial computational
bottleneck. OVQITE was originally demonstrated for the transverse-field Ising model, whose Hamiltonian contains $O(N)$ terms \cite{anuar2024operator}. The generic molecular Hamiltonian of Eq.~\eqref{eq: electronic Hamiltonian}, by contrast, gives a
Hamiltonian pool of $O(N^4)$ operators.
Although each observable has low Majorana length, the total number of observables
can be large: every pool operator requires an
anticommutator expectation value for $v$ and $N_\theta$ parameter derivatives for its row of $J$. Compatibility with MP alone consequently does not establish that the full OVQITE update is practical for molecular systems.
Addressing this evaluation cost is central to the VIME construction introduced in Sec. \ref{subsec: vime}.

\subsection{Variational Imaginary-time Majorana Evolution (VIME)}
\label{subsec: vime}

\begin{figure*}[t!]
    \centering
    \includegraphics[width=\textwidth]{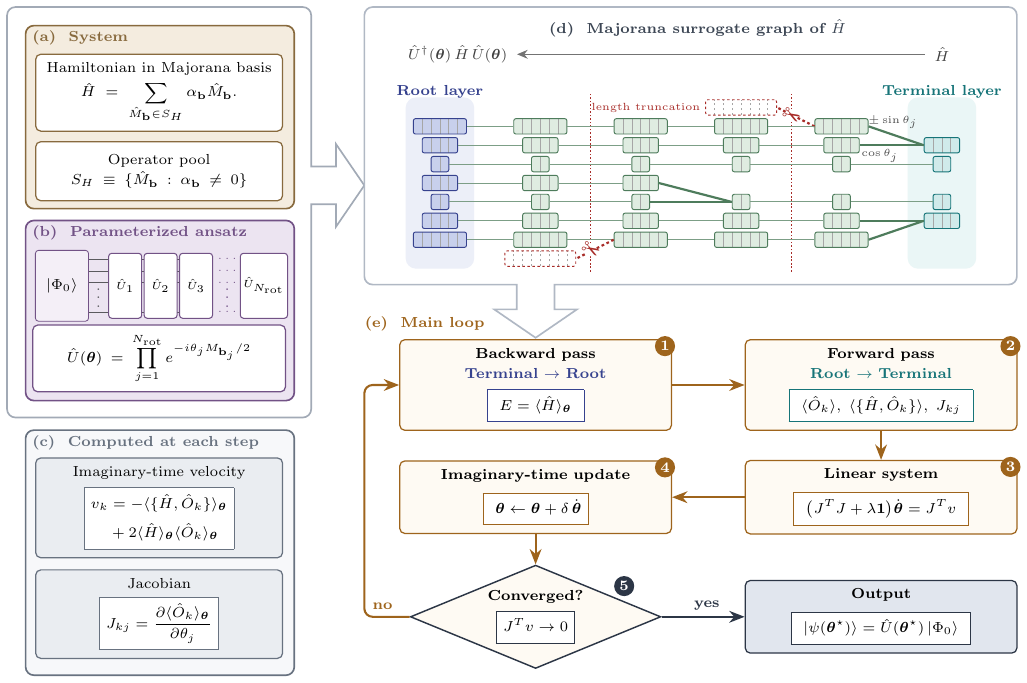}
  \caption{\textbf{Overview of VIME.}
(a) The active-space Hamiltonian is expressed in the Majorana
basis, with its constituent monomials defining the operator
pool $S_\tau=S_H$.
(b) A chosen parameterised circuit and Fock reference define
the variational ansatz.
(c) The IT velocity $v$ and Jacobian $J$ are evaluated at each iteration.
(d) A single Majorana surrogate graph is compiled for
$\hat U^\dagger(\theta)\hat H\hat U(\theta)$, with length
truncation at fermionic-block boundaries.
(e) Each iteration reuses this graph: a backward pass evaluates
the energy (1), and a joint forward pass evaluates the pool
expectation values, anticommutator expectation values, and
Jacobian (2).
These quantities define the regularized linear
system, which determines $\dot\theta$ by fitting $J\dot\theta$
to $v$ (3), followed by an imaginary-time parameter update (4).
Iterations stop when $\|J^T v\|_2 \rightarrow 0$ or the update direction fails the energy-descent check as described in Appendix  \ref{app:stopping}
 (5).
The output is the trained circuit parameters for subsequent
quantum state preparation.}
    \label{fig: VIME schematic}
\end{figure*}

Let us now introduce Variational Imaginary-time Majorana Evolution (VIME),
which combines OVQITE with MP to classically pre-train a
fixed fermionic circuit. Starting from an initial
parameter point $\theta_0$, VIME evaluates the OVQITE equations using a
compiled MP surrogate, solves the resulting linear system for the
parameter velocity $\dot\theta$, and advances the circuit parameters in
imaginary time until convergence. The complete VIME workflow is summarized in
Fig.~\ref{fig: VIME schematic}. The remainder of this subsection describes how the full VIME update is
obtained efficiently from a single compiled Hamiltonian surrogate graph.

A standard MP surrogate graph is compiled for a chosen observable and subsequently
returns its expectation value as a function of the circuit parameters
\cite{miller2025majorana}. A naive application to OVQITE would therefore
require separate surrogate evaluations for the many observables entering
$v$ and $J$. For each $\hat O_k\in S_\tau$, the velocity $v_k$ requires
$o_k (\theta) \equiv \langle\hat O_k\rangle_{\hat\rho_\theta}$ and
$C_k(\theta) \equiv\langle\{\hat H,\hat O_k\}\rangle_{\hat\rho_\theta}$, together with the
energy $E (\theta) =\langle\hat H\rangle_{\hat\rho_\theta}$ shared across the pool,
while the corresponding row of $J$ requires the $N_\theta$ derivatives
$\partial_{\theta_j}o_k(\theta)$. If each derivative were evaluated through two parameter-shifted
expectation values, a single OVQITE update would therefore require
\begin{equation}
\label{eq: Nexp}
N_{\mathrm{exp}}
=
2\,|S_\tau|\,(N_\theta+1)+1
\end{equation}
separate expectation-value evaluations: $2N_\theta$ per pool
operator for the Jacobian, two per pool operator for $o_k$ and
$C_k$, and one shared energy evaluation. For the Hamiltonian pool
$S_\tau=S_H$, the number of terms in a generic electronic-structure
Hamiltonian scales as $|S_H|=O(N^4)$, giving
$N_{\mathrm{exp}}=O(N^4N_\theta)$. Compiling an independent MP surrogate graph for each of these expectation values is therefore impractical: it would replicate essentially the same circuit-dependent propagation structure across an enormous number of closely related graphs.

For this reason VIME evaluates all quantities required for the update using a single Hamiltonian propagation graph. A backward pass provides
the propagated Hamiltonian coefficients and energy, while a
joint forward pass supplies the pool expectation values,
anticommutator expectation values, and parameter derivatives.
We outline the construction here and present the full derivation
in Appendix~\ref{app: single surrogate pass}.
Throughout, the subscript $\mathrm{MP}$ denotes quantities obtained with Majorana Propagation.

For a circuit containing $N_{\mathrm{rot}}$ elementary Majorana
rotations, the compiled graph has layers
$j=0,\ldots,N_{\mathrm{rot}}$. At each layer, let
\begin{equation}
\mathcal M_j=\{\hat M_a^{(j)}\}_{a=1}^{D_j}
\end{equation}
be the ordered set of distinct retained Majorana monomials,
with $D_j=|\mathcal M_j|$ and $a$ indexing monomials within
that layer. The \emph{terminal layer},
$j=N_{\mathrm{rot}}$, contains the Hamiltonian's initial
monomial expansion; the \emph{root layer}, $j=0$, contains
the monomials retained after back-propagation through the
full circuit. We use the superscripts ``$\mathrm{term}$'' and
``$\mathrm{root}$'' for these two endpoint bases, with
$D_{\mathrm{term}}=D_{N_{\mathrm{rot}}}$ and
$D_{\mathrm{root}}=D_0$. For the Hamiltonian pool $S_\tau=S_H$, the terminal basis
also spans every pool operator $\hat O_k$, where
$k=1,\ldots,|S_\tau|$. Thus,
\begin{equation}
\label{eq:terminal-expansions}
\begin{aligned}
\hat H
&=\sum_{a=1}^{D_{\mathrm{term}}}
\alpha_a \hat M_a^{(\mathrm{term})},\\
\hat O_k
&=\sum_{a=1}^{D_{\mathrm{term}}}
(A_k)_a \hat M_a^{(\mathrm{term})},
\end{aligned}
\end{equation}
where $\vec\alpha,\vec A_k\in\mathbb R^{D_{\mathrm{term}}}$
are coefficient vectors independent of the circuit parameters
$\theta$.

The sparse map
$\hat L_j(\theta)\in\mathbb R^{D_j\times D_{j+1}}$
propagates monomial coefficients from layer $j+1$ to layer $j$.
The complete terminal-to-root map is therefore
\begin{equation}
\label{eq: graph operator}
\mathsf{\hat P}(\theta)
=
\hat L_0(\theta)\dots
\hat L_{N_{\mathrm{rot}}-1}(\theta)
\in\mathbb R^{D_{\mathrm{root}}\times D_{\mathrm{term}}}.
\end{equation}
The retained graph structure is fixed at compilation; only
the transition weights depend on $\theta$. We apply these
maps through their sparse transitions without explicitly
forming $\mathsf{\hat P}$.

The fixed reference state $\hat \rho$ defines the
root expectation vector
\begin{equation}
r_a=\operatorname{Tr}[\hat \rho \hat M_a^{(\mathrm{root})}],
\qquad
\vec r\in\mathbb R^{D_{\mathrm{root}}}.
\end{equation}

The backward pass applies Eq.~(\ref{eq: graph operator}) to the Hamiltonian coefficient vector $\vec \alpha$, yielding the back-propagated Hamiltonian coefficient vector
\begin{equation}
\vec h_{\mathrm{root}}
=\mathsf{\hat P}(\theta)\vec\alpha,
\end{equation}
Contracting these coefficients with the root expectation
vector $\vec r$ gives the energy,
\begin{equation}
E_{\mathrm{MP}}
=\vec r^{\,T}\vec h_{\mathrm{root}}.
\end{equation}

To obtain all of the $o_k(\theta)$, we traverse the same graph in the reverse direction, starting from $\vec r$. Specifically, the forward pass recursively computes
\begin{equation}
 \vec u_0=\vec r,
\qquad
\vec u_{j+1}=\hat L_j^T\vec u_j.
\end{equation}
Upon reaching the terminal layer, the resulting vector is
\begin{equation}
 \label{eq: forward pass main}
\vec m_{\mathrm{MP}}^{\mathrm{term}} =
\vec u_{N_{\mathrm{rot}}} =
\mathsf{\hat P}^{T}\vec r,   
\end{equation}
whose components correspond to the MP expectation values of the terminal monomials in the reference state. Every pool-operator expectation value therefore follows from
\begin{equation}
o_{k,\mathrm{MP}}
=
\vec A_k^{\,T}\vec m_{\mathrm{MP}}^{\,\mathrm{term}},
\end{equation}
without requiring a different compiled graph for each operator.

The anticommutator estimates $C_{k,\mathrm{MP}}$
are obtained through the same forward pass, with a different
initial vector. To construct this vector, we define
\begin{equation}
W_{ab}
=
\operatorname{Tr}\!\left[
\hat\rho
\{\hat M_a^{(\mathrm{root})},\hat M_b^{(\mathrm{root})}\}
\right].
\end{equation}
Each entry of $W$ is the expectation value, in the fixed
reference state $\hat\rho$, of the anticommutator of two
root-layer monomials. Using the Hamiltonian coefficient
vector $\vec h_{\mathrm{root}}$ from the backward pass,
the forward recurrence becomes
\begin{equation}
\vec B_0=W\vec h_{\mathrm{root}},
\qquad
\vec B_{j+1}=\hat L_j^T\vec B_j,
\end{equation}
for $j=0,\ldots,N_{\mathrm{rot}}-1$. All anticommutator
estimates then follow from the contraction at the terminal layer:
\begin{equation}
\label{eq: anticommutator main}
C_{k,\mathrm{MP}}
=
\vec A_k^{\,T}\vec B_{N_{\mathrm{rot}}}
=
\vec A_k^{\,T}\mathsf{\hat P}^{T}\vec B_0.
\end{equation}
The vectors $\vec B_j$ and $\vec u_j$ can therefore be
computed together in a single forward traversal. We assess the accuracy of this construction under truncation
in Sec.~\ref{subsec: VIME scaling}. Its derivation is given
in Appendix~\ref{app: A anticommutators}.

The anticommutator construction imposes an additional requirement on the
root layer. In a standard energy-only MP evaluation on a Fock reference,
only paired monomials can have a nonzero expectation value,
$\operatorname{Tr}[\hat \rho \hat M]\neq0$, so all unpaired root monomials may be
discarded after the circuit has been fully propagated. This pruning cannot
be applied to the anticommutator channel: two individually unpaired
monomials can have a paired product and therefore contribute to
$\operatorname{Tr}[\hat{\rho}\{\hat M_i,\hat M_j\}]$. The surrogate must consequently
retain all propagation paths leading to the surviving root monomials,
resulting in a larger graph than for the corresponding energy-only
evaluation.

For a Fock reference, the structure of $W$ allows the product
$\vec B_0=W\vec h_{\mathrm{root}}$ to be evaluated without
forming or storing the full matrix. Computing $\vec B_0$
requires $O(D_{\mathrm{root}})$ operations and
$O(D_{\mathrm{root}})$ storage in addition to the compiled
graph, avoiding the quadratic storage cost of an explicit
representation of $W$. The construction is derived in
Appendix~\ref{app: A W structure}.

The remaining quantity needed to assemble the OVQITE linear
system is the Jacobian $J$. In VIME, its entries are obtained
by differentiating the forward recurrence for $\vec u_j$
with respect to each variational parameter $\theta_i$, with $i=1,\ldots,N_\theta$. Define
$\vec d_j^{\,(i)}=\partial_{\theta_i}\vec u_j$
as the derivative of the expectation vector at layer $j$.
Since the reference vector is independent of $\theta$,
$\vec d_0^{\,(i)}=0$, and the derivative product rule gives
\begin{equation}
\vec d_{j+1}^{\,(i)}
=
\hat L_j^T\vec d_j^{\,(i)}
+
\left(\partial_{\theta_i}\hat L_j^T\right)\vec u_j,
\end{equation}
for $j=0,\ldots,N_{\mathrm{rot}}-1$.
The second term accounts for the parameter dependence
of the current transition and vanishes when that
transition is independent of $\theta_i$. This recurrence
also includes all contributions when several rotations
share the same parameter.

At the terminal layer, contraction with the fixed
pool coefficients gives
\begin{equation}
(J_{\mathrm{MP}})_{ki}
=
\partial_{\theta_i}o_{k,\mathrm{MP}}
=
\vec A_k^{\,T}\vec d_{N_{\mathrm{rot}}}^{\,(i)}.
\end{equation}
Thus, the Jacobian is computed by propagating one
derivative vector per parameter alongside the expectation
vectors in the same forward traversal. The transition
derivatives are detailed in
Appendix~\ref{app: A jacobian}.

Figure~\ref{fig: VIME schematic}(e) summarizes the complete
VIME update. One backward pass provides $\vec h_{\mathrm{root}}$, from which we evaluate $E_{\mathrm{MP}}$ and construct $\vec B_0$. A joint forward pass then propagates $N_\theta+2$ channels: $\vec u_j$ for the pool expectation values, $\vec B_j$ for the anticommutator estimates, and one derivative vector $\vec d_j^{\,(i)}$ per variational parameter for the Jacobian. Terminal contractions yield $o_{k,\mathrm{MP}}$, $C_{k,\mathrm{MP}}$, and $(J_{\mathrm{MP}})_{ki}$ for every pool operator. This joint pass replaces the $O(|S_\tau|N_\theta)$ separate expectation-value evaluations of the naive approach with a number of forward channels that scales linearly with $N_\theta$. This channel count does not imply that the total cost is
independent of the Hamiltonian size: the Hamiltonian expansion
sets the terminal basis and thereby influences the size of
the compiled graph. We examine the resulting scaling in Sec.~\ref{subsec: VIME scaling}.

At each iteration, the quantities returned by the backward and joint forward passes determine the IT velocity
$v_{k,\mathrm{MP}}= -C_{k,\mathrm{MP}}+2E_{\mathrm{MP}}o_{k,\mathrm{MP}}$ and the linear system
in Eq.~\eqref{eq: OVQITE linear system}, whose solution gives the parameter velocity $\dot\theta$. To advance the parameters along this direction, we select the imaginary-time step size using the adaptive line search described in Appendix ~\ref{app:integration}, corresponding to step (4) in Fig.~\ref{fig: VIME schematic}(e). The line-search slope follows directly from the pool Jacobian, $s=\sum_k c_k(J_{\mathrm{MP}}\dot\theta)_k$, and trial energies require only backward evaluations of the compiled graph.
After a step is accepted, a joint forward pass evaluates the pool quantities and Jacobian at the updated parameters.
These quantities serve both to check the stopping criterion defined in Appendix ~\ref{app:stopping}, corresponding to step (5) in Fig.~\ref{fig: VIME schematic}(e), and to construct the next update if convergence has not yet been reached.

\subsection{The compressed tiled Unitary Product State (c-tUPS) ansatz  }
\label{subsec: ctUPS}

VIME requires a circuit ansatz that balances three requirements: sufficient expressivity, a depth shallow enough for execution on quantum hardware, and efficient classical simulability via Majorana Propagation. Among
fixed-structure, symmetry-preserving fermionic ansätze, the tiled Unitary
Product State (tUPS) of Ref.~\cite{burton2024tups} stands out in providing this combination
of properties and therefore forms the starting point for the circuit family
constructed below.

Let $N$ be the number of fermionic modes and $n=N/2$ the number of spatial
orbitals. A tUPS circuit with $L$ layers is
\begin{equation}
\label{eq:tups-ansatz}
\ket{\Psi_{\mathrm{tUPS}}}
=
\prod_{m=1}^{L}
\left(
\prod_{j=1}^{A}
\hat{U}_{2j+1,\,2j}^{(m)}
\right)
\left(
\prod_{k=1}^{B}
\hat{U}_{2k,\,2k-1}^{(m)}
\right)
\ket{\Phi_{0}},
\end{equation}
where $A=\left\lfloor\frac{n-1}{2}\right\rfloor$ and $B=\left\lfloor\frac{n}{2}\right\rfloor.$
The two inner products are half-layers acting on alternating sets of disjoint
nearest-neighbor orbital pairs. Here,
$\ket{\Phi_{0}}$ is the reference occupation-number determinant, while the
parenthesized superscript $(m)$ labels the layer.

Each two-orbital block contains three independent variational parameters and
has the Euler form
\begin{equation}
\label{eq:tups-operator}
\hat{U}_{pq}^{(m)}
=
\exp\!\left(
\theta_{pq,1}^{(m)}\hat{\kappa}_{pq}^{(1)}
\right)
\exp\!\left(
\theta_{pq,2}^{(m)}\hat{\kappa}_{pq}^{(2)}
\right)
\exp\!\left(
\theta_{pq,3}^{(m)}\hat{\kappa}_{pq}^{(1)}
\right),
\end{equation}
where the spin-adapted one-body and paired two-body generators are
\begin{align}
\label{eq:kappa-generators}
\hat{\kappa}_{pq}^{(1)}
&=
\hat{E}_{pq}-\hat{E}_{qp},
&
\hat{\kappa}_{pq}^{(2)}
&=
\hat{E}_{pq}^{\,2}-\hat{E}_{qp}^{\,2},
\end{align}
with $\hat{E}_{pq} = \sum_{\sigma\in\{\alpha,\beta\}} \hat{a}_{p,\sigma}^{\dagger}\hat{a}_{q,\sigma}$. Here, $\hat{E}_{pq}$ is the singlet excitation operator
\cite{helgaker2000molecular}, while
$\hat{a}_{p,\sigma}^{\dagger}$ and $\hat{a}_{q,\sigma}$ create and annihilate,
respectively, an electron of spin $\sigma \in\{\alpha,\beta\}$ in spatial orbitals $\phi_p$ and
$\phi_q$. Both generators conserve particle number, spin projection, and
total spin:
\begin{equation}
\left[\hat{\kappa}_{pq}^{(r)},\hat{\mathcal N}\right]
=
\left[\hat{\kappa}_{pq}^{(r)},\hat{S}_z\right]
=
\left[\hat{\kappa}_{pq}^{(r)},\hat{S}^{2}\right]
=0,
\label{eq:tups-symmetries}
\end{equation}
where $r\in\{1,2\}$. Consequently,  the  tUPS circuit preserves
the corresponding quantum numbers of $\ket{\Phi_0}$.

In the three-dimensional singlet subspace of two electrons in two spatial
orbitals, $\hat{\kappa}_{pq}^{(1)}$ and
$\hat{\kappa}_{pq}^{(2)}$ generate rotations about orthogonal axes.
Consequently, the Euler sequence in Eq.~\eqref{eq:tups-operator} can realize
an arbitrary real rotation within this subspace. Tiling these blocks over nearest-neighbor orbitals extends this local expressivity across the active space. 

We adopt two additional refinements introduced with tUPS in
Ref.~\cite{burton2024tups}. First, $\ket{\Phi_0}$ is chosen to alternate
between doubly occupied and empty spatial orbitals. The blocks in the first
half-layer therefore act on occupied--empty pairs, giving this half-layer the
structure of a perfect-pairing wave function. Second, we include a one-body orbital transformation over the
active space following the tiled circuit, which we refer to as the
\emph{active orbital rotations}. Although represented as a trailing unitary
$\hat{U}_{\mathrm{orb}}$, these rotations need not be executed on the quantum
processor. Their effect is incorporated classically by transforming the
one- and two-electron integrals, or equivalently by replacing
$\hat H$ with
$\hat U_{\mathrm{orb}}^\dagger \hat H \hat U_{\mathrm{orb}}$.
This transformation preserves the one- and two-body form of
the Hamiltonian and hence its $O(N^4)$ term-count scaling.
Previous benchmarks of this orbital-optimized, perfect-pairing
tUPS architecture were limited to relatively small active
spaces~\cite{burton2024tups}. For applications to substantially larger systems, we seek to reduce the resources needed for both the classical MP simulation and the quantum execution of this architecture.

\begin{figure*}[t!]
    \centering
    \includegraphics[width=\textwidth]{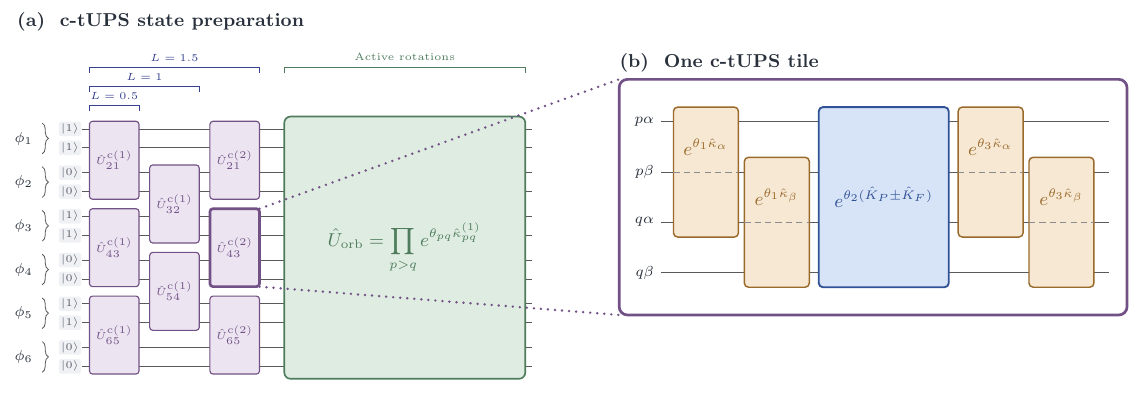}
\caption{\textbf{The c-tUPS ansatz.}
(a) Circuit for six spatial orbitals, with alternating doubly
occupied and empty reference orbitals and staggered half-layers
of nearest-neighbor tiles. Blue brackets indicate circuit
depths $L=0.5$, $1$, and $1.5$.
The green box denotes the active-space orbital rotation
unitary $\hat U_{\mathrm{orb}}$, which can be absorbed into
the Hamiltonian for quantum execution once its parameters
have been classically determined.
(b) Each tile combines outer one-body rotations with a
central factor $e^{\theta_2(\hat K_P\pm\hat K_F)}$.
The $+$ and $-$ signs define cp-tUPS and cm-tUPS, respectively.}
    \label{fig: ctups ansatz}
\end{figure*}

To this end, we introduce the
\emph{compressed tiled Unitary Product State (c-tUPS)} ansatz, illustrated in Fig.~\ref{fig: ctups ansatz}. This family retains the two-half-layer brickwork structure,
perfect-pairing reference, active orbital rotations, and the two outer one-body factors generated by $\hat{\kappa}_{pq}^{(1)}$, while replacing the central
two-body factor generated by $\hat{\kappa}_{pq}^{(2)}$.
In ordinary tUPS, this central factor
directly generates pair transfer between the two spatial orbitals, whereas
the complementary opposite-spin exchange process can arise only indirectly
through the composition of blocks across multiple layers. Because both processes act
on the same four spin orbitals, exchange can be incorporated without requiring
additional connectivity. We do so without introducing another variational
parameter by locking its angle to that of the pair transfer, in the spirit of
overlapping coupled-exchange constructions \cite{ramoa2025ovpceo}.

For the local mode ordering
$(p\alpha,q\alpha,p\beta,q\beta)$, define
\begin{equation}
\label{eq: KP KF}
\begin{aligned}
\hat P
&=
a^{\dagger}_{p\alpha}a^{\dagger}_{p\beta}
a_{q\beta}a_{q\alpha},
&
\hat K_{P}
&=
\hat P-\hat P^{\dagger}
=\tfrac12\hat\kappa^{(2)}_{pq},
\\
\hat F
&=
a^{\dagger}_{p\alpha}a^{\dagger}_{q\beta}
a_{p\beta}a_{q\alpha},
&
\hat K_{F}
&=
\hat F-\hat F^{\dagger}.
\end{aligned}
\end{equation}
Here $\hat K_{P}$ generates the closed-shell transition
$\ket{0101}\leftrightarrow\ket{1010}$, while $\hat K_{F}$ generates the
open-shell transition
$\ket{0110}\leftrightarrow\ket{1001}$. These transitions act on disjoint
subspaces such that
$\hat K_{P}\hat K_{F}=\hat K_{F}\hat K_{P}=0$, and therefore
\begin{equation}
e^{\theta(\hat K_{P}\pm\hat K_{F})}
=
e^{\theta\hat K_{P}}e^{\pm\theta\hat K_{F}}
\end{equation}
holds exactly. Thus, we define compressed tile 
\begin{equation}
\label{eq: compressed block}
\hat U_{pq}^{\mathrm{c},\pm\,(m)}
=
e^{\theta_{pq,1}^{(m)}\hat\kappa_{pq}^{(1)}}
e^{\theta_{pq,2}^{(m)}(\hat K_{P}\pm\hat K_{F})}
e^{\theta_{pq,3}^{(m)}\hat\kappa_{pq}^{(1)}}.
\end{equation}
Replacing the original tUPS tiles with
$\hat U_{pq}^{\mathrm{c},+(m)}$ or
$\hat U_{pq}^{\mathrm{c},-(m)}$ defines the
compressed-plus tUPS (cp-tUPS) and compressed-minus tUPS
(cm-tUPS) families, respectively.
Both variants retain three parameters per tile and the brickwork connectivity of tUPS. They conserve $\hat N_{\alpha}$ and $\hat N_{\beta}$,
and consequently $\hat N$ and $\hat S_z$, while preserving real-valued
amplitudes. However, they do not conserve total spin in general:
the pair-transfer generator satisfies $[\hat K_P,\hat S^2]=0$, whereas
for the exchange generator  $[\hat K_F,\hat S^2]\neq0$, yielding
\begin{equation}
[\hat K_P\pm\hat K_F,\hat S^2]
=
\pm[\hat K_F,\hat S^2]\neq0.
\end{equation}
The cp-tUPS and cm-tUPS variants are related by global
exchange of the $\alpha$ and $\beta$ spin modes.
For the spin-free Hamiltonians and closed-shell references
used here, this symmetry gives identical exact energy
landscapes, as shown in Appendix~\ref{app:ctups-spin-exchange}.
We thus refer to the two variants collectively as c-tUPS.
Locking the exchange and pair-transfer angles changes the variational
family rather than adding an independent degree of freedom. Consequently, $\ctUPS$ does not constitute a variational extension
of tUPS. Both belong to a broader family of ans\"atze that retain the same
tiling, perfect-pairing reference, and orbital dressing, while varying the generators and parameter constraints within each tile.
Additional variants are defined in
Table~\ref{tab: C catalogue} of
Appendix~\ref{app:masks-catalogue}.

In the direct decomposition used for MP, each exponential generated by a single Majorana monomial constitutes a \emph{Majorana rotation}. Each outer one-body factor contains four length-two rotations and is unchanged between tUPS and c-tUPS. The reduction occurs in the central factor: $\hat K_P$ and $\hat K_F$ expand over the same eight length-four monomials, with matching coefficients on four and opposite coefficients on the other four. Their sum
or difference therefore retains only four terms, halving the number of length-four rotations and reducing the total from $4+8+4=16$ for tUPS to $4+4+4=12$ for c-tUPS. Whereas length-two rotations preserve monomial length, length-four rotations can increase it and move propagated terms across the truncation threshold. Their reduction therefore targets a principal source of complexity in truncated MP.

The length-four rotations within each central factor
together represent a single fermionic exponential.
During its propagation, terms can temporarily exceed
$w^\star$ and return below the cutoff in subsequent
rotations. Truncating after each rotation would discard these contributions prematurely, so we propagate the central factor in full before applying the cutoff. Because the outer one-body factors preserve monomial length, this is equivalent to filtering only at the tile boundary, which we call
\emph{blockwise Majorana truncation}. We quantify the additional compilation and evaluation time required by blockwise relative to regular truncation
in Appendix~\ref{app:  C majorana-block-analysis}.

\begin{table}[!t]
\centering
\caption{
Majorana rotations and two-qubit resources per tile.
$n_{\mathrm{rot}}$ counts the elementary Majorana rotations
used for MP, and $G_4$ counts those generated by length-four
Majorana monomials. $N_{\mathrm{CNOT}}$ and $D_{2Q}$ are the CNOT count and two-qubit depth under all-to-all connectivity.
Other connectivities are discussed in
Appendix~\ref{app: CNOT counts}.
}
\label{tab:block-content}
\medskip
\begin{tabular}{lcccc}
\toprule
Tile & $n_{\mathrm{rot}}$ & $G_4$
& $N_{\mathrm{CNOT}}$ & $D_{2Q}$\\
\midrule
tUPS   & $16$ & $8$ & $14$ & $8$\\
c-tUPS & $12$ & $4$ & $10$ & $8$\\
\bottomrule
\end{tabular}
\end{table}

The c-tUPS ansatz also reduces quantum-execution costs.
Each layer of either ansatz contains $n-1$ two-orbital tiles.
Previously reported componentwise decompositions of tUPS
required $21$ CNOTs per tile on all-to-all connectivity
\cite{burton2024tups,anselmetti2021qnp,yordanov2020circuits},
subsequently reduced to $20$ by an improved implementation
of the paired double excitation~\cite{shaik2026_12cnot}.
Corresponding two-qubit depths were not reported.
These constructions synthesize the three factors of
Eq.~\eqref{eq:tups-operator} separately, leaving shared
structure across the factor boundaries unexploited.

To exploit this structure, we develop a hardware-aware
procedure that jointly synthesizes the three factors of
each tUPS or c-tUPS tile. We first express the tile as a
sequence of Pauli rotations,
\begin{equation}
R_P(\varphi)=e^{-i\varphi \hat P/2},
\end{equation}
where $\hat P$ is a tensor product of single-qubit Pauli operators
and identities. The synthesis combines exact algebraic reductions with a bounded circuit search and preserves the full three-parameter tile unitary for arbitrary parameter values. Here we report resources for
all-to-all connectivity. The synthesis procedure, explicit circuits, verification, and results for square-grid and linear nearest-neighbor connectivity are given in Appendix~\ref{app: CNOT counts}.

Under all-to-all connectivity, the resulting c-tUPS circuit uses $10$ CNOTs at a two-qubit depth of $8$, halving the previously reported tUPS cost of $20$ CNOTs per tile. Joint synthesis also reduces the tUPS cost to $14$ CNOTs at depth $8$. Thus, c-tUPS uses four fewer CNOTs per tile than our improved tUPS circuit. The cp-tUPS and cm-tUPS variants have identical resource requirements. Table~\ref{tab:block-content} summarizes the Majorana content and all-to-all circuit resources of both
ansatz families.

\section{Results}
\label{sec:results}


We proceed to assess the numerical accuracy and computational scaling of VIME and the suitability of c-tUPS for ground-state preparation with Majorana Propagation. Thereafter, we will benchmark VIME against VMPE on identical circuits and initial conditions, extend the comparison to larger molecular active spaces, and compare the accuracy and circuit resources with ADAPT-VMPE~\cite{chakraborty2026adaptvmpe}.

All hydrogen-chain benchmarks use the STO-6G basis; H$_n$ denotes a chain of $n$ hydrogen atoms. Throughout this section, we use the Jordan--Wigner transformation to map fermionic operators to qubits. In all VIME calculations below, we use the Hamiltonian operator pool introduced in Sec.~\ref{subsec: ovqite}, taking each observable to be a single Majorana monomial.
Thus, in the terminal-basis expansion of
Eq.~\eqref{eq:terminal-expansions},
$\hat O_k=\hat M_{a(k)}^{(\mathrm{term})}$ and $(A_k)_a=\delta_{a,a(k)}$.
Unless otherwise stated, MP uses blockwise Majorana-length truncation as described in Sec.~\ref{subsec: ctUPS}, with paired monomials retained according to
Eq.~\eqref{eq:paired-acceptance}.
The length cutoffs and any additional orbital-support restrictions are specified for each benchmark.

\subsection{Validation and computational scaling of the VIME update}
\label{subsec: VIME scaling}

\begin{figure}[t!]
    \centering
    \includegraphics[
        width=\columnwidth
    ]{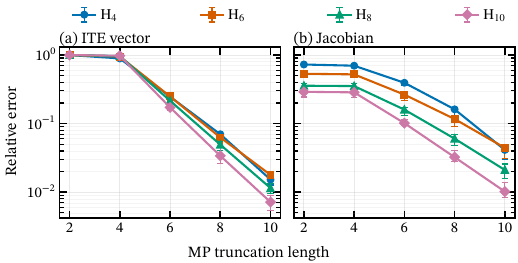}
\caption{\textbf{Convergence of VIME update quantities with MP truncation length.}
Relative errors against statevector benchmarks for
(a) the IT velocity vector $v$ (Euclidean norm) and
(b) the Jacobian $J$ (Frobenius norm), for
H$_4$, H$_6$, H$_8$, and H$_{10}$ hydrogen chains in the STO-6G basis.
Circuits contain 100 random length-four Majorana rotations with
independent angles drawn from $\mathcal{U}[0,2\pi]$ and truncation
after each rotation. Markers show medians over five random seeds,
error bars span the minimum and maximum values, and lines connect
the medians.}
    \label{fig: MP vs statevector truncation error}
\end{figure}

The bound in Eq.~\eqref{eq: MP error bound} controls the truncation error for a single observable evaluated by standard backward propagation. A VIME step instead depends on the velocity vector \(v\) and the Jacobian \(J\), which are evaluated simultaneously using the forward-pass construction of Sec.~\ref{subsec: vime}. Although this construction is algebraically exact for a fixed truncated graph, the single-observable bound does not establish how truncation errors accumulate across the pool expectation values, anticommutators, and parameter derivatives that form \(v\) and \(J\). It therefore remains to be determined whether these two inputs to the VIME update inherit the exponential convergence with \(w^\star\) predicted for a single propagated observable.

Figure~\ref{fig: MP vs statevector truncation error}
compares the relative errors in the IT velocity vector
$v$ and Jacobian $J$ against statevector references for
linear H$_4$, H$_6$, H$_8$, and H$_{10}$ chains, with a uniform interatomic separation of
$0.74\,\text{\AA}$.
For this validation, we use circuits comprising 100
randomly selected length-four Majorana rotations, with
independent parameters drawn from
$\mathcal{U}[0,2 \pi]$ and no active orbital rotations.
We evaluate $v$ and $J$ at these fixed random parameter points, applying truncation after each rotation and without performing imaginary-time evolution. Markers show medians over five random seeds, with error bars spanning the minimum and maximum values.
The statevector Jacobian is computed using central finite differences over all $N_\theta=100$ parameters.

Beyond the initial plateau at small $w^\star$, the errors in both $v$ and $J$
decay approximately exponentially with truncation length. At fixed
$w^\star$, the error also generally decreases with increasing system size,
with H$_{10}$ giving the most accurate results. This behavior is consistent
with Eq.~\eqref{eq: MP error bound}, whose dominant term
$(ew^\star/N)^{w^\star/2}$ decreases with $N$. Reaching the asymptotic regime
$N\gg w^\star$ would further suppress the error but makes full statevector
validation prohibitively expensive. These results establish a convergence trend for the VIME update quantities; whether this translates into reliable ground-state convergence is examined in the following sections.

\begin{figure}[t!]
    \centering
    \includegraphics[
        width=\columnwidth
    ]{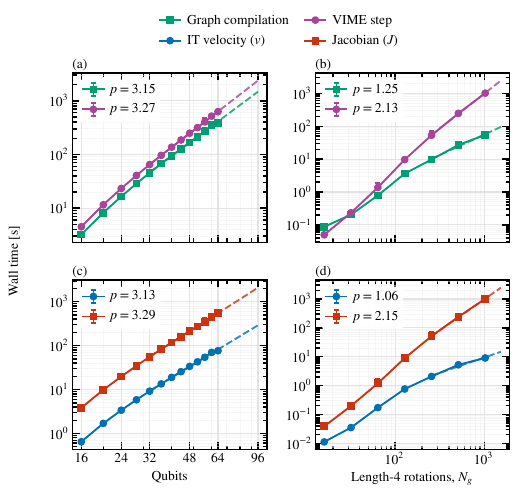}
\caption{\textbf{Wall-time scaling of VIME at $w^\star=6$.}
(a,c) System-size scaling for H$_n$/STO-6G
($n=8,10,\ldots,32$), with 100 length-2 and 100 length-4 Majorana rotations.
(b,d) Circuit-size scaling for H$_8$, with a fixed active-rotation
prefix and $N_g$ independently parameterised length-4 rotations.
Top: graph-compilation and complete VIME-step times.
Bottom: IT velocity ($v$) and Jacobian ($J$)
evaluation times.
Markers and error bars indicate medians and 10th--90th percentiles
across circuit seeds.
Dashed lines extrapolate power-law fits to the large-size regime: 32--64 qubits in (a,c) and
$N_g=256,512,1024$ in (b,d);
legends report the fitted exponents $p$.}
    \label{fig:VIME-scaling}
\end{figure}

We next examine the computational scaling of a VIME imaginary-time update step. For conventional
MP, Ref.~\cite{miller2025majorana} reports polynomial scaling with system size
and linear scaling with circuit depth, with graph compilation typically
dominating the subsequent evaluation of a single observable. VIME instead
constructs an extensive graph that retains all surviving root monomials
and reuses it in a joint forward pass of $N_\theta+2$ channels to obtain
the pool expectation values, anticommutators, and Jacobian required by
the update. This reuse avoids repeated compilation for each required
quantity, but evaluating the $N_\theta+2$ channels introduces an additional
cost that can exceed that of compilation itself.
Figure~\ref{fig:VIME-scaling} compares the time required for graph
compilation and one complete VIME step, showing how these costs scale
with system size and circuit depth.

At fixed truncation length $w^\star=6$, we fit power laws to the median wall times in the large-size regime, using 32--64 qubits
for system-size scaling and $N_g=256,512,1024$ for circuit-size
scaling. For hydrogen chains with 100 length-2 and 100 length-4 rotations, compilation and the complete VIME step have fitted dependences $N^{3.15}$ and $N^{3.27}$, respectively as shown in panel (a). Panel (c) shows that Jacobian and IT velocity evaluation yield similar exponents, $3.29$ and $3.13$. Within the limited system-size range studied, these exponents suggest comparable scaling, with the higher cost of a VIME step relative to graph compilation primarily reflecting the larger prefactor of Jacobian evaluation.

For H$_8$, increasing the number $N_g$ of independently
parameterised length-4 rotations makes this Jacobian overhead grow with circuit length.
Panel~(b) shows that graph compilation scales as $N_g^{1.25}$, whereas the complete VIME step scales as $N_g^{2.13}$.
Panel~(d) shows that this steeper scaling is driven by Jacobian
evaluation, which scales as $N_g^{2.15}$, while IT velocity
evaluation remains cheaper than compilation and scales nearly
linearly, as $N_g^{1.06}$.
Propagating a growing number of derivative channels thus makes
the VIME step more steeply dependent on circuit length.

\subsection{MP energy accuracy across circuit architectures and update rules}
\label{subsec:surrogate-energy accuracy}

Having established the accuracy and computational scaling of a VIME update,
we now assess the energy accuracy of the MP surrogate graph for the structured, tiled
ansätze used in this work. Existing MP truncation guarantees apply to
ensembles of unstructured circuits composed of independently sampled
Majorana rotations. The tiled ansätze considered here instead repeatedly
apply fixed local tiles and couple several rotations through shared
variational parameters, placing them outside the assumptions underlying
those guarantees. The accuracy of MP on such circuits must therefore be assessed
empirically. We examine how it depends on two choices: the two-orbital
tile defining the circuit architecture and the rule used to update
its parameters.

\begin{figure}[t]
    \centering
    \includegraphics[width=\columnwidth]{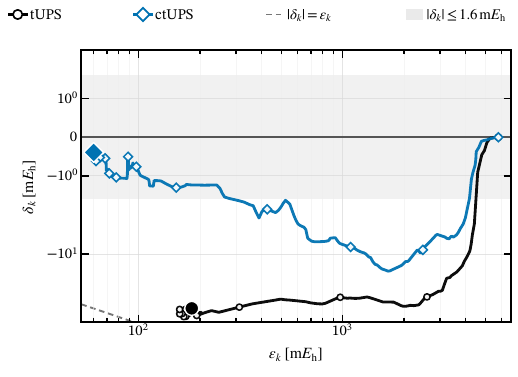}
    \caption{\textbf{MP energy accuracy during VIME convergence.}
VIME convergence curves for one-layer tUPS and c-tUPS circuits
on H$_8$/STO-6G, using $w^\star=6$ and truncation at block
boundaries.
The MP-truncation error
$\delta_k=E_{\mathrm{MP},k}-E_{\mathrm{SV},k}$
is plotted against the statevector energy error
$\varepsilon_k=E_{\mathrm{SV},k}-E_{\mathrm{ref}}$,
where $E_{\mathrm{ref}}$ is the exact active-space ground-state
energy.
Each point gives the median of each error over 20 runs at a
successive optimization step, progressing generally from right
to left; filled symbols mark the last step reached by all runs.
The gray band marks
$|\delta_k|\leq1.6\,\mathrm{m}E_{\mathrm h}$.
The dashed line marks $\delta_k=-\varepsilon_k$, corresponding
to $E_{\mathrm{MP},k}=E_{\mathrm{ref}}$ for an individual run.}
    \label{fig: MP-Statevector error vs family}
\end{figure}

Figure~\ref{fig: MP-Statevector error vs family} compares
the accuracy of MP energies during VIME convergence for
tUPS and c-tUPS on H$_8$.
The figure plots the signed MP-truncation error
$\delta_k=E_{\mathrm{MP},k}-E_{\mathrm{SV},k}$
against the statevector energy error
$\varepsilon_k=E_{\mathrm{SV},k}-E_{\mathrm{ref}}$.
For each circuit configuration, we perform $20$ VIME
runs with different random initial parameter sets.
Each point shows the median of each plotted error
across these runs at a given optimization step. The curves connect successive steps, progressing generally from right to left
as the statevector energy approaches the reference.
Comparing $|\delta_k|$ with $\varepsilon_k$ indicates whether
the discrepancy between MP and statevector energies is small
relative to the remaining energy error.
When the two are comparable, the MP energy cannot reliably
resolve this difference from the ground-state reference.

For both ans\"atze, the median MP-truncation error is initially
close to zero and becomes negative during the first few updates,
indicating that MP underestimates the statevector energy.
For c-tUPS, its magnitude subsequently decreases as the
statevector energy approaches the reference.
The curve ends within the
$|\delta_k|\leq1.6\,\mathrm{m}E_{\mathrm h}$ band, with a median
statevector energy error approximately three times smaller
than that reached by tUPS.
In contrast, tUPS retains a negative MP-truncation error of
several tens of $\mathrm{m}E_{\mathrm h}$ during the later
stages of optimization. As optimization proceeds, this persistent negative bias becomes comparable to the remaining statevector energy error, making the MP energy an unreliable indicator of proximity to the ground-state reference.

A comparison with additional tUPS variants is given in
Appendix~\ref{app:masks-catalogue}.
These results motivate our choice of c-tUPS for the remaining
benchmarks, based on its more accurate MP energies and lower
final statevector energy errors under VIME.

\begin{figure}[t!]
    \centering
    \includegraphics[width=\columnwidth]{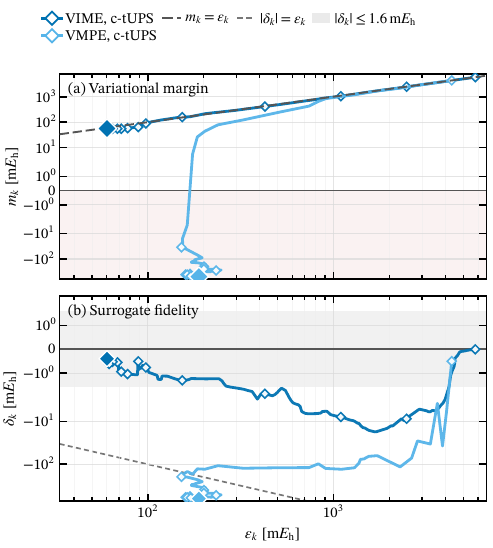}
  \caption{\textbf{MP accuracy under VIME and VMPE.}
Optimization curves for one-layer c-tUPS circuits on H$_8$/STO-6G at $w^\star=6$, progressing generally from right to left in both panels. Each point shows the median of each error over 20 runs; filled symbols mark the last step reached by all runs.
(a) MP-reported energy error $m_k=E_{\mathrm{MP},k}-E_{\mathrm{ref}}$  versus statevector energy error $\varepsilon_k$, both relative to the exact ground-state energy.
The diagonal marks agreement; shading indicates MP energies below the reference.
(b) MP-truncation error $\delta_k$ versus statevector energy error. The dashed line marks $\delta=-\varepsilon$; the gray band denotes $|\delta|\leq1.6\,\mathrm{m}E_{\mathrm h}$.}
    \label{fig: MP-statevector error VIME vs VMPE}
\end{figure}

To isolate the effect of the parameter-update rule, we compare
VIME and VMPE on identical one-layer c-tUPS circuits for H$_8$.
Both methods use $w^\star=6$, truncation at block boundaries,
and the same $20$ random initial parameter sets.
VMPE directly minimizes the MP energy estimate $f^{(w^\star)}$,
whereas VIME follows the imaginary-time parameter velocity
obtained from Eq.~\eqref{eq: OVQITE linear system}.
Figure~\ref{fig: MP-statevector error VIME vs VMPE} shows the
resulting convergence. As in the preceding figure, the curves
connect the medians of each plotted error at successive
optimization steps, progressing generally from right to left
as the statevector energy decreases.

Panel~(a) plots the MP energy error,
$m_k=E_{\mathrm{MP},k}-E_{\mathrm{ref}}$, against the
statevector energy error $\varepsilon_k$.
The VIME curves remain close to the diagonal
$m_k=\varepsilon_k$, showing that MP closely reproduces
the statevector energy. The VMPE curves instead depart
from the diagonal: the median MP energy falls below the
exact ground-state reference while the median statevector
energy remains substantially above it. Its final statevector energy error is
approximately $200\,\mathrm{m}E_{\mathrm h}$, compared with
$60\,\mathrm{m}E_{\mathrm h}$ for VIME. 

Panel~(b) explains this discrepancy through the
MP-truncation error $\delta_k$, plotted against
$\varepsilon_k$. During the later stages of optimization,
VIME moves towards states with smaller MP-truncation
errors, whose medians end within the
$|\delta|\leq1.6\,\mathrm{m}E_{\mathrm h}$ band.
VMPE instead moves towards states with increasingly
negative MP-truncation errors of order $10^2\,\mathrm{m}E_{\mathrm h}$.
This error magnitude eventually exceeds the remaining statevector energy error, causing MP to report energies below the ground-state reference. 

\begin{figure}[t!]
    \centering
    \includegraphics[width=0.9\columnwidth]
    {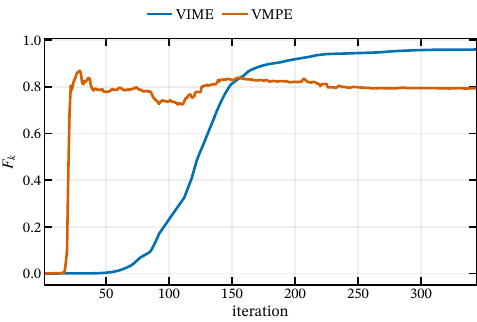}
\caption{\textbf{Ground-state overlap for VIME and VMPE.}
Ground-state overlap
$F_k=|\langle\psi_0|\psi(\theta_k)\rangle|^2$,
evaluated using exact statevectors, versus iteration for one-layer c-tUPS circuits on H$_8$/STO-6G.
Curves show medians over different initializations.}
    \label{fig:ground-state-overlap}
\end{figure}

The difference in the quality of the prepared states is also evident
from the overlap, $F_k=|\langle\psi_0|\psi(\theta_k)\rangle|^2$, with the exact ground state $|\psi_0\rangle$.
Figure~\ref{fig:ground-state-overlap} shows this quantity, evaluated
using exact statevector calculations, for one-layer c-tUPS circuits applied to H$_8$.
At each optimization step, we plot the median overlap over
20 initializations.  VMPE improves the overlap rapidly during the initial updates,
but this improvement is not sustained, and the median overlap
eventually settles near $0.79$.
VIME, on the other hand, improves gradually and reaches a median overlap of
approximately $0.96$. This indicates that, for this benchmark, the lower energy errors of VIME are associated with a larger ground-state component in the prepared states, although extending these statevector-based benchmark conclusions to larger system sizes is computationally prohibitive.

\subsection{Ground-state preparation of large molecular active spaces}
\label{subsec:tld1411-benchmarks}

\begin{figure}[t]
    \centering

    \setlength{\aceneH}{0.180\columnwidth}
    \setlength{\tldW}{0.479\columnwidth}
    \setlength{\tldUp}{12pt}
    \setlength{\tldRight}{0.404\tldUp}

    \sbox{\hepbox}{%
        \adjincludegraphics[
            trim={0.4132\width} {0.0340\height}
                 {0.4138\width} {0.0340\height},
            clip,
            angle=90,
            height=\aceneH
        ]{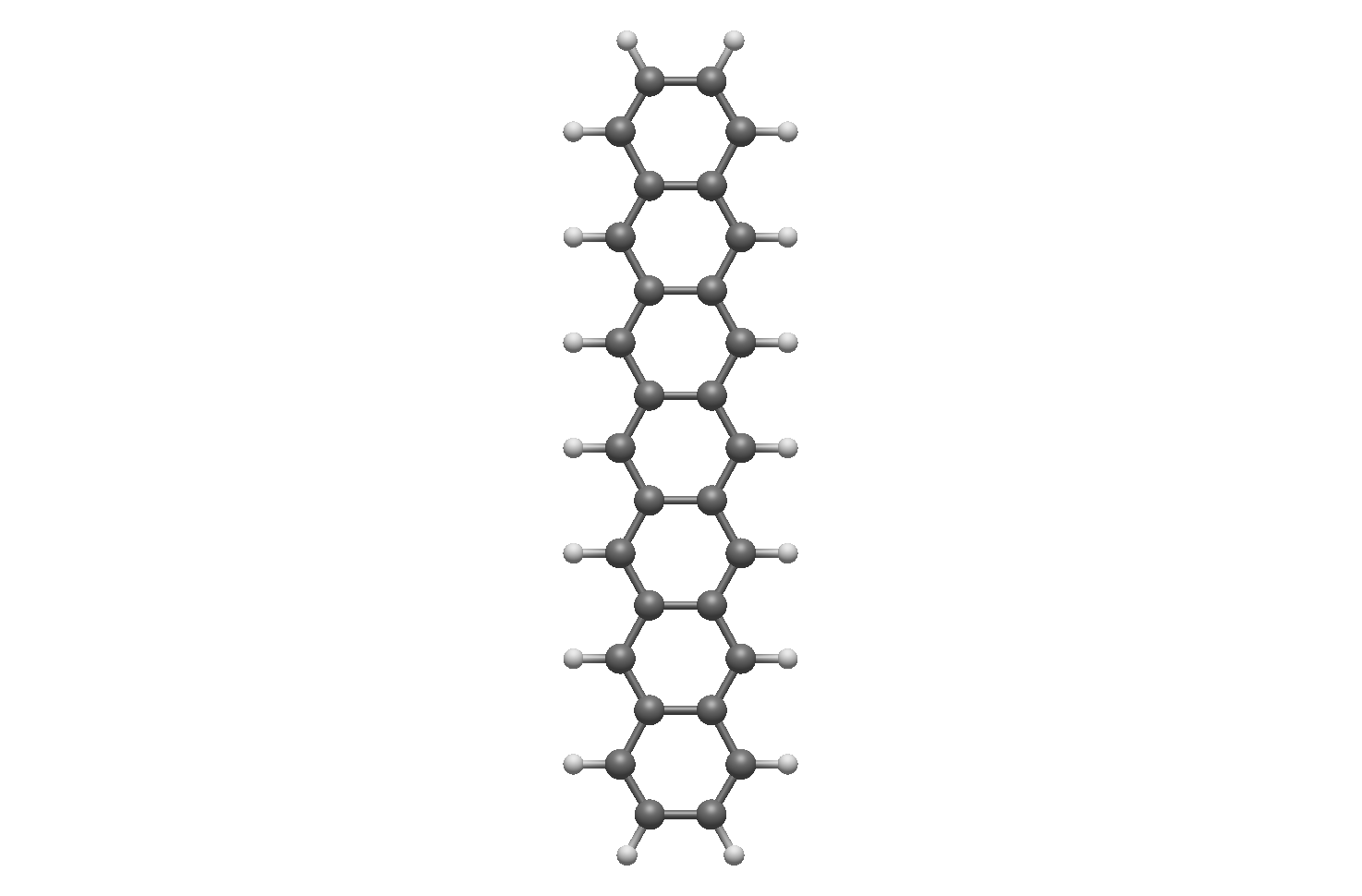}%
    }

    \setlength{\acnW}{\wd\hepbox}
    \setlength{\ovlp}{\tldW}
    \addtolength{\ovlp}{\acnW}
    \addtolength{\ovlp}{-\columnwidth}

    \begin{minipage}[t]{\tldW}
        \centering
        \textbf{(a) TLD-1411}\par
        \smallskip
        \vspace*{-\tldUp}%
        \makebox[\linewidth][l]{%
            \hspace*{\tldRight}%
            \adjustbox{
                trim={0.2975\width} {0.0392\height}
                     {0.2249\width} {0.1083\height},
                clip,
                width=\linewidth
            }{%
                \includegraphics[angle=-60]
                    {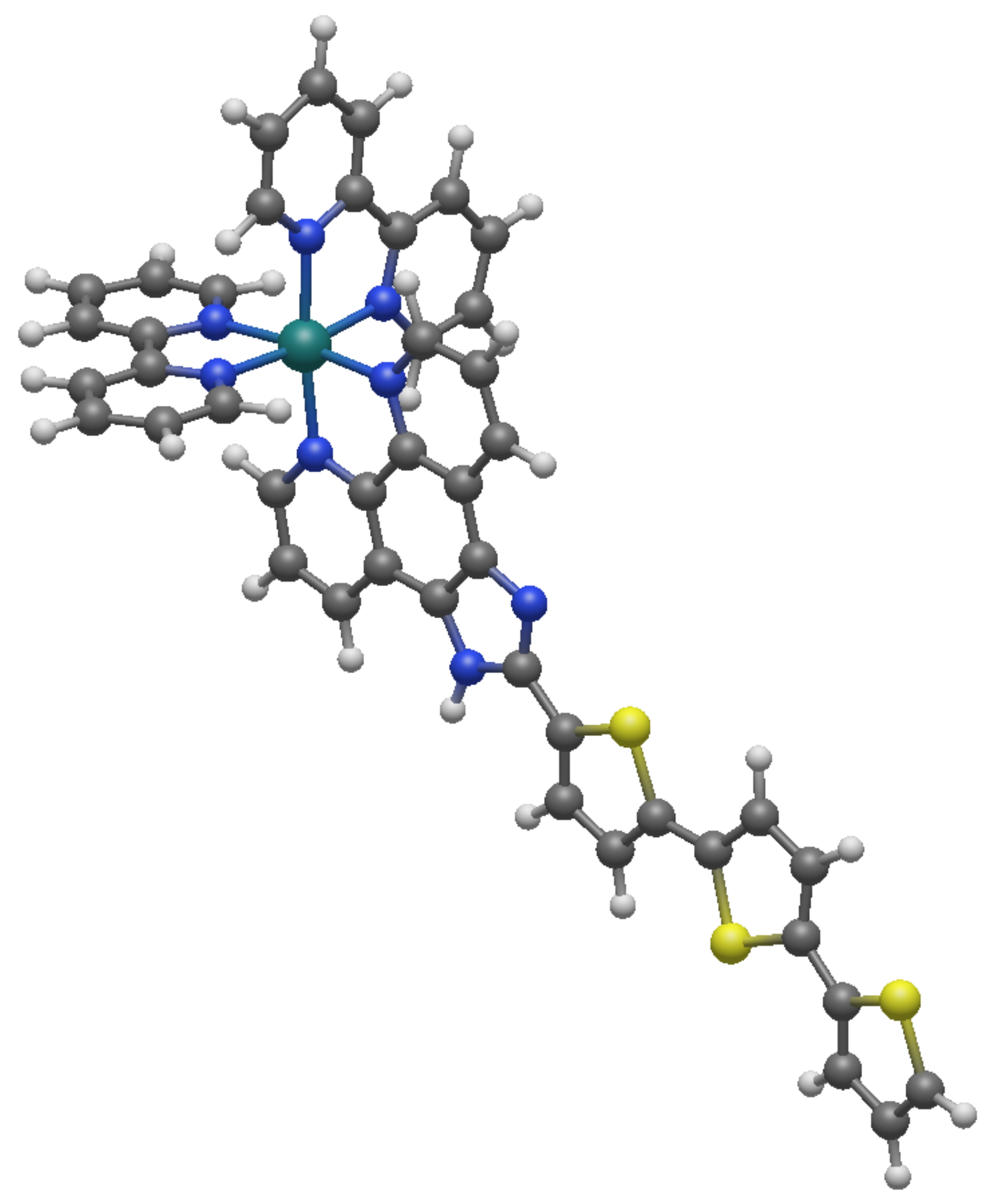}%
            }%
        }
    \end{minipage}%
    \hspace*{-\ovlp}%
    \begin{minipage}[t]{\acnW}
        \raggedleft
        \textbf{(b) Acenes}\par
        \vspace{3pt}
        \smallskip

        {\small Anthracene}\par
        \vspace{3pt}
        \adjincludegraphics[
            trim={0.4138\width} {0.2680\height}
                 {0.4145\width} {0.2680\height},
            clip,
            angle=90,
            height=\aceneH
        ]{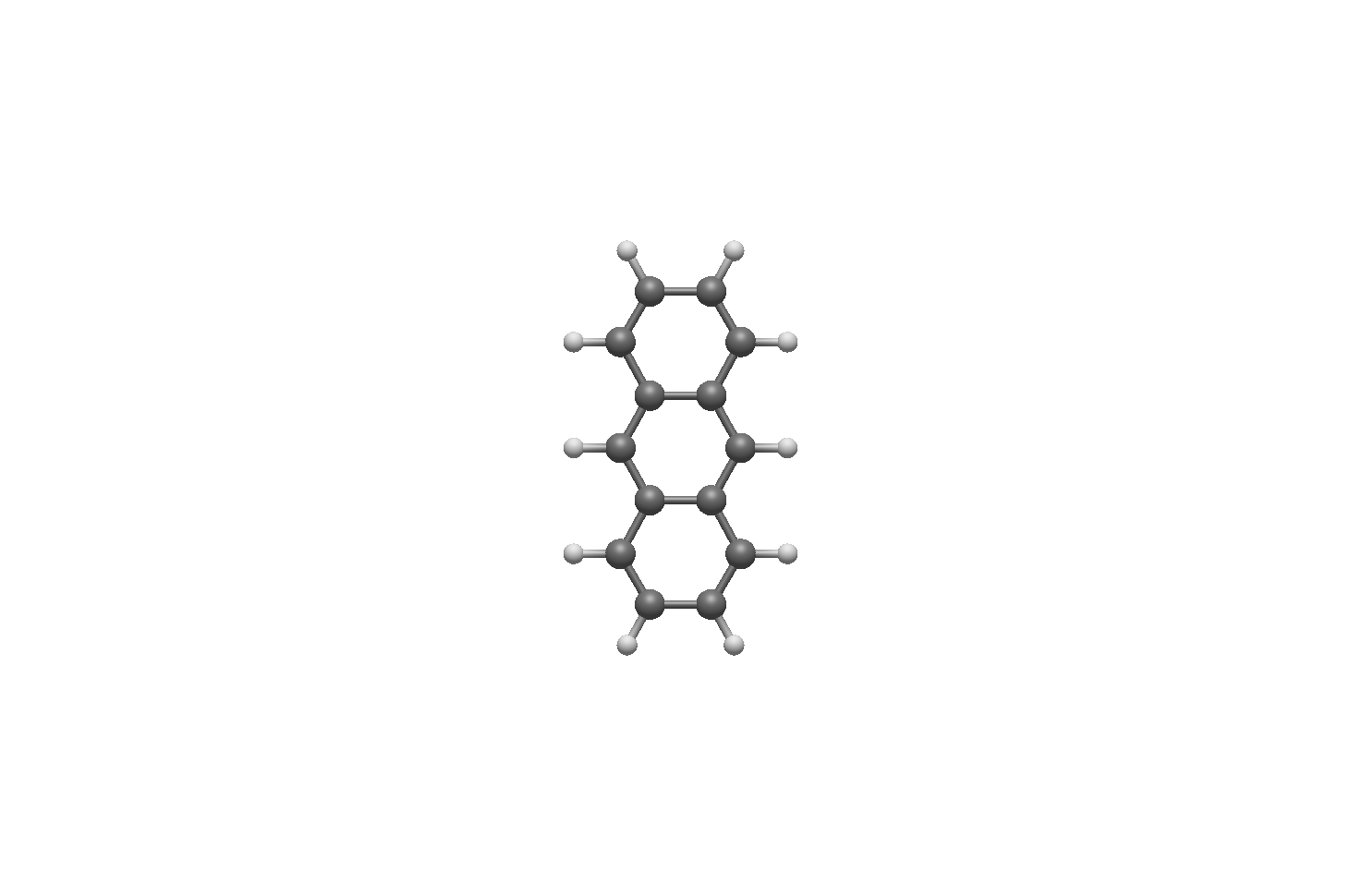}\par

        \medskip

        {\small Pentacene}\par
        \vspace{3pt}
        \adjincludegraphics[
            trim={0.4138\width} {0.1515\height}
                 {0.4145\width} {0.1515\height},
            clip,
            angle=90,
            height=\aceneH
        ]{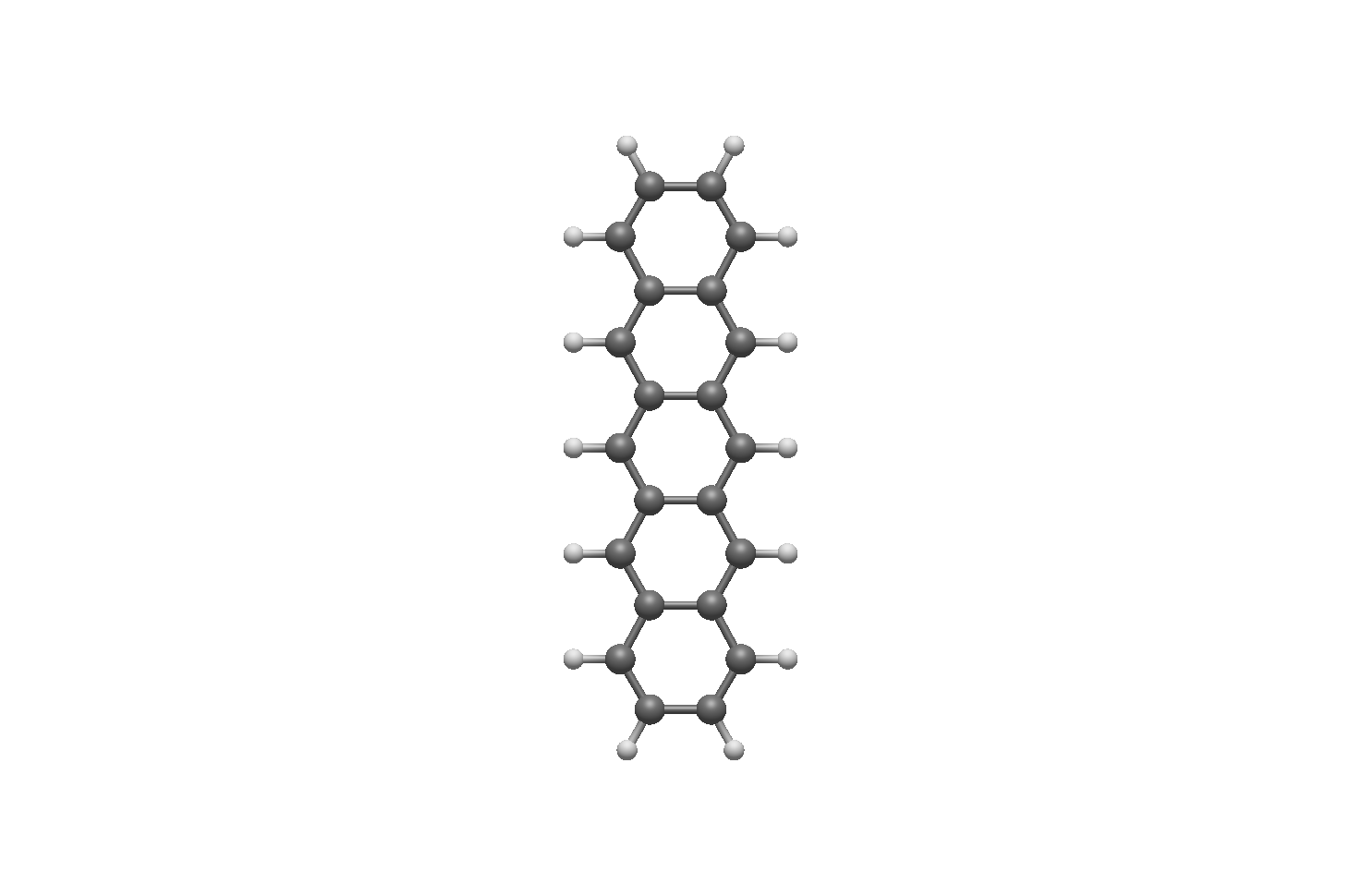}\par

        \medskip

        {\small Heptacene}\par
        \vspace{3pt}
        \usebox{\hepbox}
    \end{minipage}

    \caption{Molecular benchmarks and active spaces:
(a) TLD-1411 (chloride counterions not shown),
CAS(14e,14o), CAS(20e,20o), and CAS(26e,26o)
(28, 40, and 52 qubits, respectively);
(b) acenes, from top to bottom:
anthracene, CAS(14e,14o) (28 qubits);
pentacene, CAS(22e,22o) (44 qubits);
and heptacene, CAS(30e,30o) (60 qubits).
Atoms are coloured by element: carbon (dark grey), hydrogen (light grey), nitrogen (blue), sulfur (yellow), and ruthenium (teal).}
    \label{fig:molecular-benchmarks}
\end{figure}

Let us now assess the performance of VIME on molecular active spaces of the ruthenium complex TLD-1411 and linear acenes up to heptacene, spanning $28$ to $60$ qubits. Figure~\ref{fig:molecular-benchmarks} shows the molecular
structures of these compounds. 
At the quoted system sizes, exact statevector validation is no longer feasible, and we assess energy accuracy by comparing the MP-reported energy $E$ to the corresponding DMRG reference
energy $E_0$.

TLD-1411 is a ruthenium(II) polypyridyl complex and a close
structural analogue of the photosensitiser
TLD-1433~\cite{monro2019tld1433}. We use the active-space
Hamiltonians from Ref.~\cite{chakraborty2026adaptvmpe},
constructed in the def2-SVPD basis~\cite{rappoport2010property}.
We consider CAS(14e,14o), CAS(20e,20o), and CAS(26e,26o),
which include orbitals associated with both the ruthenium
centre and the extended thiophene
chain~\cite{tarocco2025aegiss,chakraborty2026adaptvmpe}
and correspond to $28$, $40$, and $52$ qubits, respectively.
The DMRG reference energies and ADAPT-VMPE comparison data are also taken from Ref.~\cite{chakraborty2026adaptvmpe}.

We first compare VIME with VMPE using single-layer
c-tUPS circuits. Both methods use blockwise
Majorana-length truncation with $w^\star=6$.
VIME has a substantially higher computational cost
per update, arising from its larger MP surrogate graph
and the evaluation of the Jacobian $J$
(see Appendix~\ref{app:cost-comparison} for details).
To make its computation feasible at $52$ qubits,
we therefore supplement the length cutoff with a
spatial-orbital support restriction $o^\star=4$
(Sec.~\ref{subsec: mp}).
This additional filter reduces the propagation graph
while retaining selected length-six contributions,
providing an intermediate level of truncation between
$w^\star=4$ and $w^\star=6$.
The less costly VMPE calculations require no additional
support filter.
We also explored sine-order truncation for VIME,
but it did not provide a satisfactory cost--accuracy
tradeoff in the tests performed.
In the remainder of this section VIME runs will terminate when $\lVert J^T v\rVert_2$ falls
below $10^{-3}$ of its initial value or the computed
update direction fails the energy-descent check
(Appendix~\ref{app:stopping}).

\begin{figure}[t]
    \centering
    \includegraphics[width=\columnwidth]
{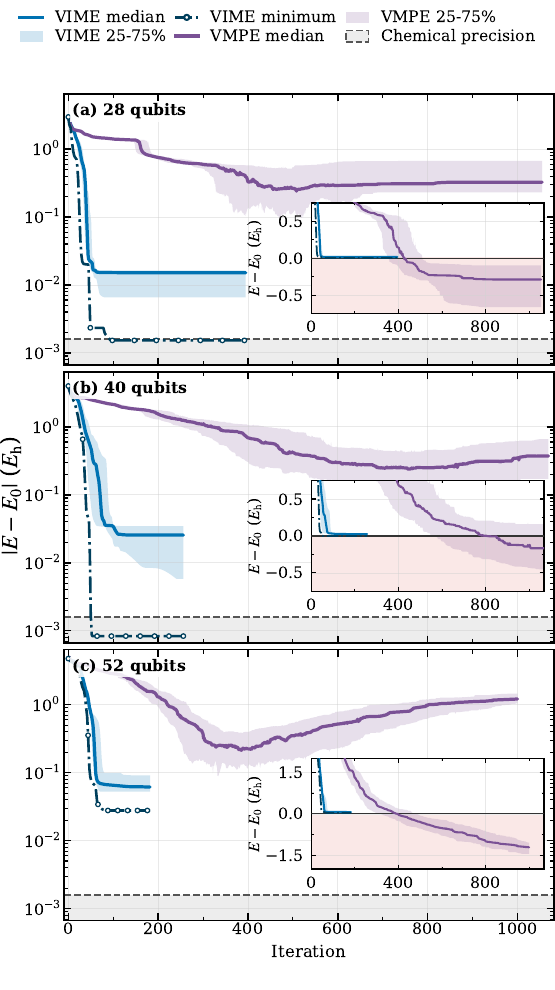}
\caption{\textbf{VIME and VMPE ensembles for TLD-1411.}
Single-layer c-tUPS calculations for (a) 28-, (b) 40-, and
(c) 52-qubit active spaces.
Solid curves and shaded bands show the median and 25--75\%
interval over 10 VIME runs (blue) and 100 VMPE runs (purple)
for each active space.
Main panels show absolute MP energy errors relative to the
DMRG reference $E_0$; insets show signed errors, with pink
shading indicating energies below $E_0$.
The dark-blue dash-dotted curve shows the pointwise minimum
VIME error across runs.
Gray shading bounded by dashed lines marks chemical precision.
Both methods use blockwise Majorana-length truncation
($w^\star=6$), retaining paired monomials.
For 52-qubit VIME, an additional spatial-orbital support
cutoff $o^\star=4$ is applied.}
\label{fig:vime-vs-vmpe-large}
\end{figure}

Figure~\ref{fig:vime-vs-vmpe-large} summarizes ensembles
of 10 VIME runs and 100 VMPE runs for each active-space
size, with each run initialized independently.
The smaller VIME ensemble reflects its  higher computational cost.
For both methods, we show the median and interquartile
interval to characterize the variation across runs.

For VIME, no MP energy falls below the DMRG reference at any point along the reported trajectories. Together with the statevector comparisons in
Sec.~\ref{subsec:surrogate-energy accuracy}, this behavior
suggests that VIME is more robust against the accumulation
of large negative MP-truncation errors observed during direct
energy minimization. This is an empirical finding for
the calculations considered here: it does not establish
a variational bound for $E_{\mathrm{MP}}$, and other VIME
runs could in principle cross below the reference.

For the observed VIME ensembles, however, remaining above the reference has a practical consequence: a lower MP
energy consistently corresponds to closer agreement with DMRG. We therefore select runs using only their final MP energies and report the lowest value for each
active space. Figure~\ref{fig:vime-vs-vmpe-large} also shows the pointwise minimum VIME error across runs
at each iteration.

The final median absolute VIME errors are $15.2$, $25.6$,
and $61.5\,\mathrm{m}E_{\mathrm h}$ for $28$, $40$, and
$52$ qubits, respectively. The corresponding minimum
errors are $1.5$, $0.834$, and $27.7\,\mathrm{m}E_{\mathrm h}$,
reaching chemical precision relative to DMRG for the two
smaller active spaces. For VMPE, the final signed errors are negative in the
majority of runs at all three sizes, including all
$100$ runs at $52$ qubits, where the median error
is approximately $-1200\,\mathrm{m}E_{\mathrm h}$.

Appendix~\ref{app: computational-details} presents
individual VIME trajectories and details of the
statistical presentation. The runs use different
initial parameters and values of the Tikhonov
regularization parameter $\lambda$
(introduced in Eq.~\eqref{eq: regularized OVQITE linear system}).
The variation in final energies across these runs
is consistent with the optimization difficulties reported
for tiled ans\"atze, where global basin-hopping and repeated
local optimizations were used to find low-energy
states~\cite{burton2024tups,ramoa2025ovpceo}.
Variational imaginary-time evolution can become trapped
in local minima and therefore does not guarantee
convergence to the lowest-energy state within a fixed
variational manifold~\cite{mcardle2019vqite}.

\begin{figure*}[t!]
    \centering
    \includegraphics[width=\textwidth]
{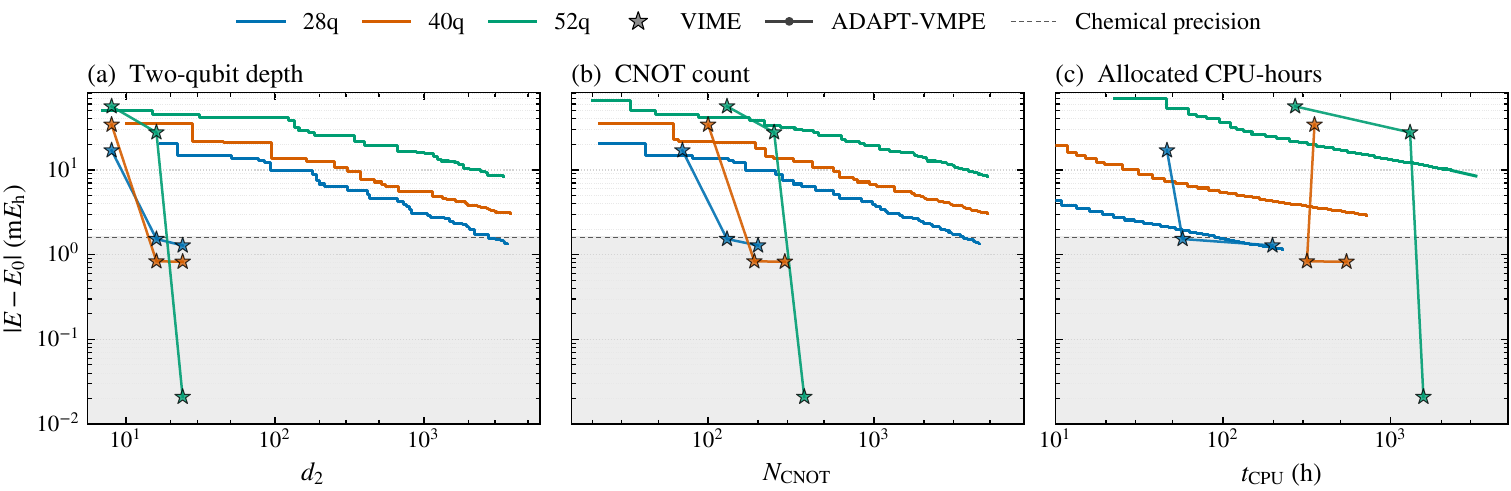}
    \caption{
Accuracy and resource costs for TLD-1411. Absolute MP energy errors relative to DMRG versus (a) two-qubit depth $d_2$, (b) CNOT count $N_{\mathrm{CNOT}}$, and (c) allocated CPU-hours $t_{\mathrm{CPU}}$(h). Stars show VIME endpoints at $L=0.5$, $1$, and $1.5$, connected as guides to the eye. Staircase curves show the resource--error Pareto frontiers of the ADAPT-VMPE runs from Ref.~\cite{chakraborty2026adaptvmpe}.
Circuit counts assume all-to-all connectivity and exclude
active orbital rotations.
CPU-hours include preceding runs for VIME continuations.
The dashed line and shading mark chemical precision. 
}
    \label{fig:vime-adapt-resource-efficiency}
\end{figure*}

\begin{table*}[t]
    \caption{
    Lowest reported absolute errors relative to DMRG and
    corresponding circuit resources.
    VIME minima occur at $L=1.5$, while ADAPT-VMPE minima
    occur at iteration $400$ of
    Ref.~\cite{chakraborty2026adaptvmpe}.
    Errors are given in $\mathrm{m}E_{\mathrm h}$, and $N_{\theta}$
    is the number of independent circuit parameters.
    All circuit quantities exclude active orbital rotations;
    $d_2$ and $N_{\mathrm{CNOT}}$ assume all-to-all connectivity.
    }
    \label{tab:vime-adapt-best-resources}
     \begin{ruledtabular}
        \begin{tabular}{c rrrr rrrr}
            & \multicolumn{4}{c}{VIME ($L=1.5$)}
            & \multicolumn{4}{c}{ADAPT-VMPE} \\
            Qubits
            & $|E-E_0|$ & $N_{\theta}$ & $d_2$ & $N_{\mathrm{CNOT}}$
            & $|E-E_0|$ & $N_{\theta}$ & $d_2$ & $N_{\mathrm{CNOT}}$ \\
            \colrule
            $28$ & $1.284$ & $60$  & $24$ & $200$
                 & $1.337$ & $400$ & $3651$ & $4300$ \\
            $40$ & $0.825$ & $87$  & $24$ & $290$
                 & $3.084$ & $400$ & $3810$ & $4854$ \\
            $52$ & $0.021$ & $114$ & $24$ & $380$
                 & $8.477$ & $400$ & $3436$ & $4803$ \\
        \end{tabular}
    \end{ruledtabular}
\end{table*}

Figure~\ref{fig:vime-adapt-resource-efficiency} compares VIME with c-tUPS circuits at $L=0.5$, $1$, and $1.5$ against ADAPT-VMPE in terms of MP-reported energy accuracy, quantum circuit resources, and classical computational cost. In this and the following comparisons, we report the VIME run with the lowest final MP energy among the runs performed for each active space and circuit depth, following the
selection criterion discussed above.

For ADAPT-VMPE, we use the reported runs from
Ref.~\cite{chakraborty2026adaptvmpe} for TLD-1411 and
the runs performed in this work for the acenes.
In both cases, the endpoints should not be interpreted
as limits on the attainable accuracy: further adaptive
growth could lower the MP-reported energies, but would
require additional classical computation and would
generally increase circuit resources.

The $L=0.5$ and $L=1$ circuits are optimized independently. For $L=1.5$, we start from the selected $L=1$ run, append a half-layer of c-tUPS tiles, and resume optimization of all parameters. This choice is motivated by the steep increase in VIME evaluation cost with circuit depth (Appendix~\ref{app: ctUPS VIME scaling}). Starting from an optimized one-layer circuit allows us to explore a larger variational space while limiting the number of expensive updates at the greater depth. The reported computational costs at $L=1.5$ include the preceding one-layer optimization.

Table~\ref{tab:vime-adapt-best-resources} summarizes
the errors and circuit resources of the selected $L=1.5$ VIME runs alongside the ADAPT-VMPE results at iteration $400$. These VIME runs reach chemical precision relative to DMRG for all three active-space sizes. Adding a half-layer yields only modest error reductions for $28$ and $40$ qubits, whose selected $L=1$ runs already achieve chemical precision.
For $52$ qubits, the improvement is substantial: the error drops from $27.7$ to $0.021\,\mathrm{m}E_{\mathrm h}$. VIME and ADAPT-VMPE achieve similar errors of approximately $1.3\,\mathrm{m}E_{\mathrm h}$ at $28$ qubits, whereas
the VIME errors at $40$ and $52$ qubits are smaller by
factors of approximately $3.7\times$ and $400\times$, respectively. The available ADAPT-VMPE runs do not reach chemical precision at the two larger active-space sizes.

\begin{figure*}[t]
    \centering
    \includegraphics[width=\textwidth]
{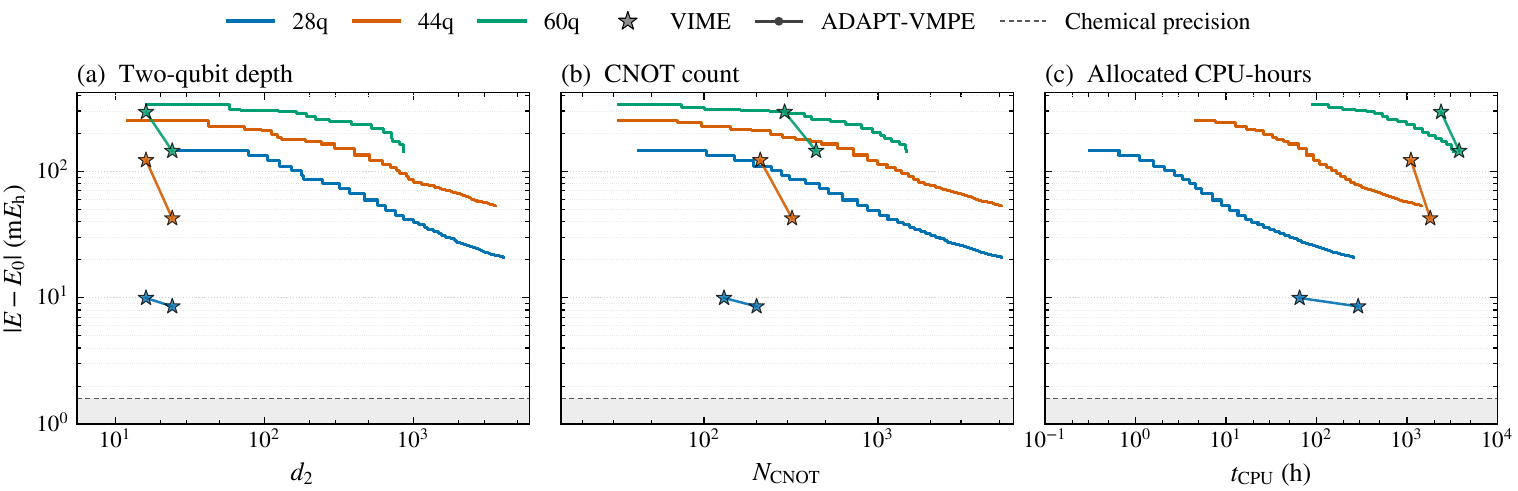}
    \caption{
    Accuracy and resource costs for anthracene ($28$ qubits),
    pentacene ($44$ qubits), and heptacene ($60$ qubits).
    Absolute MP energy errors relative to DMRG-SCF are shown
    versus (a) two-qubit depth, (b) CNOT count, and
    (c) allocated CPU-hours $t_{\mathrm{CPU}}$(h).
    Stars mark VIME endpoints at $L=1$ and $1.5$, connected
    as guides to the eye. Staircase curves show the resource--error Pareto frontiers of the ADAPT-VMPE calculations performed in this work. Curves end at the first iteration matching or improving on the $L=1.5$ VIME energy error, or at iteration $400$ if this target is not reached: iteration $400$ for anthracene and pentacene, and $134$ for heptacene.
    Circuit counts assume all-to-all connectivity and exclude active orbital rotations.
    VIME continuation costs include the preceding runs.
    The dashed line and shading mark chemical precision.
    }
    \label{fig:acene-resource-efficiency}
\end{figure*}

\begin{table*}[t]
    \caption{
    Absolute MP energy errors relative to DMRG-SCF and
    corresponding circuit resources for the selected
    $L=1.5$ VIME runs and ADAPT-VMPE comparison points
    in the acene series.
    ADAPT-VMPE results are taken at the first iteration
    reaching an error no larger than the corresponding
    VIME error, or at iteration $400$ if this target
    is not reached.
    The selected iterations are $400$, $400$, and $134$
    for anthracene ($28$ qubits), pentacene ($44$ qubits),
    and heptacene ($60$ qubits), respectively.
    Errors are given in $\mathrm{m}E_{\mathrm h}$, and
    $N_{\theta}$ counts independent circuit parameters.
    All circuit quantities exclude active orbital rotations;
    $d_2$ and $N_{\mathrm{CNOT}}$ assume all-to-all connectivity.
    }
    \label{tab:acene-best-resources}
    \begin{ruledtabular}
        \begin{tabular}{c rrrr rrrr}
            & \multicolumn{4}{c}{VIME ($L=1.5$)}
            & \multicolumn{4}{c}{ADAPT-VMPE} \\
            Qubits
            & $|E-E_0|$ & $N_{\theta}$ & $d_2$
            & $N_{\mathrm{CNOT}}$
            & $|E-E_0|$ & $N_{\theta}$ & $d_2$
            & $N_{\mathrm{CNOT}}$ \\
            \colrule
            $28$ & $8.533$   & $60$  & $24$ & $200$
                 & $20.920$  & $400$ & $4019$ & $5100$ \\
            $44$ & $42.458$  & $96$  & $24$ & $320$
                 & $53.329$  & $400$ & $3580$ & $5114$ \\
            $60$ & $144.864$ & $132$ & $24$ & $440$
                 & $144.434$ & $134$ & $856$ & $1460$ \\
        \end{tabular}
    \end{ruledtabular}
\end{table*}

Under all-to-all connectivity, each $\ctUPS$ tile requires
$10$ CNOTs and each half-layer contributes a two-qubit
depth of $8$.
Active orbital rotations are excluded from the counts
because they can be absorbed classically into the Hamiltonian
integrals and need not be executed on the quantum processor.
All three $L=1.5$ VIME circuits therefore have two-qubit
depth $24$, corresponding to a $143$--$159\times$
reduction relative to the final ADAPT-VMPE circuits
at iteration $400$, which yield the best reported
ADAPT-VMPE energies for these active spaces.
The corresponding CNOT reductions are approximately
$22\times$, $17\times$, and $13\times$ for the $28$-,
$40$-, and $52$-qubit active spaces, respectively.

We also compare the number of independent circuit parameters for the two methods, which serves
as a proxy for fault-tolerant circuit implementation cost~\cite{sayginel2024fault}.
Excluding active orbital rotations, the $L=1.5$ \ctUPS{} circuits use approximately $6.7\times$, $4.6\times$, and $3.5\times$ fewer parameters than ADAPT-VMPE at $28$, $40$, and $52$ qubits, respectively. 

Figure~\ref{fig:vime-adapt-resource-efficiency}(c) places
the quantum resource savings in the context of classical
pre-training cost.
For ADAPT-VMPE, we obtain the classical cost from TLD-1411 calculations performed using our own implementation of the algorithm.
We find that the selected $L=1.5$ VIME calculations require CPU-hours
of the same order of magnitude as these ADAPT-VMPE runs.

Next, we examine whether these circuit-resource advantages persist across the acene series, whose increasing multireference character complicates approximate ground-state preparation. Linear acenes are conjugated organic molecules relevant to
organic semiconductor applications, with an experimental preparation extending to heptacene through photogeneration in polymer and cryogenic inert-gas matrices \cite{Mondal2006,Mondal2009}. Along this experimentally accessible series, increasing molecular length provides a systematic test of growing active-space size and increasing multireference character in the $\pi$ manifold \cite{hachmann2007radical,Schriber-JCTC-2018}.
We consider the singlet ${}^{1}A_g$ ground states of
anthracene, pentacene, and heptacene in their full
$\pi/\pi^\ast$ active spaces, CAS(14e,14o), CAS(22e,22o), and CAS(30e,30o), corresponding to $28$, $44$, and $60$ qubits. For these systems, we generate the active-space Hamiltonians using DMRG-SCF-optimized orbitals in the cc-pVTZ basis. All the details of Hamiltonian construction and reference calculations are given in Appendix~\ref{app:molecular-benchmarks}. 

Figure~\ref{fig:acene-resource-efficiency} compares VIME
with $\ctUPS$ circuits at $L=1$ and $1.5$ against
ADAPT-VMPE, using the respective DMRG-SCF energies as
references $E_0$.
As in the preceding benchmarks, we report the VIME run
with the lowest final MP energy among the valid runs
performed for each system and circuit depth.
Unlike the TLD-1411 benchmarks, which use published
ADAPT-VMPE results, the acene calculations use our own
implementation of ADAPT-VMPE.
We compare the methods at the first ADAPT-VMPE iteration
that matches or improves on the $L=1.5$ VIME energy error,
with a maximum of $400$ iterations considered.
The VIME calculations use base Majorana-length cutoffs
of $w^\star=6$ for anthracene and pentacene and
$w^\star=4$ for heptacene.
Paired Majorana monomials are normally retained regardless
of their length, as described in Sec.~\ref{subsec: mp}.
For heptacene at $L=1.5$, we modify this truncation scheme
only within the added half-layer: truncation remains
blockwise with base cutoff $w^\star=4$, but newly generated
monomials exceeding this cutoff are discarded at block
boundaries, even if paired. These changes limit the classical propagation cost without altering the quantum circuit or its resource counts.

For all three acenes, increasing the c-tUPS circuit depth
from $L=1$ to $L=1.5$ brings the VIME energies closer
to the DMRG references.
Neither VIME nor ADAPT-VMPE, however, reaches chemical
precision for these systems, indicating a more demanding
benchmark than TLD-1411 under the computational settings
considered here.
Table~\ref{tab:acene-best-resources} summarizes the
resulting errors and circuit resources at the selected
comparison points. For anthracene and pentacene, ADAPT-VMPE does not reach the VIME accuracy within $400$ iterations; the VIME
errors are approximately $59\%$ and $20\%$ smaller,
respectively. For heptacene, ADAPT-VMPE first reaches the VIME accuracy at iteration $134$, with both methods yielding an
absolute MP energy error of approximately
$145\,\mathrm{m}E_{\mathrm h}$.

VIME achieves these energy accuracies with substantially
fewer quantum-circuit resources than ADAPT-VMPE.
Under the counting conventions defined above, VIME
with c-tUPS requires approximately $36$--$167\times$
lower two-qubit depth and $3.3$--$26\times$ fewer CNOTs.
For heptacene, nearly identical energy errors are obtained
with $36\times$ lower two-qubit depth and $3.3\times$
fewer CNOTs. Figure~\ref{fig:acene-resource-efficiency}(c) shows that
the allocated CPU-hours remain of the same order of
magnitude for both methods, including the preceding
one-layer calculations in the VIME totals at $L=1.5$.

The circuit-resource reductions reported above depend
in part on the operator pool used by ADAPT-VMPE.
Alternative pools, such as the coupled exchange operator (CEO) pool~\cite{ramoa2025ovpceo}, share with c-tUPS the aim of reducing the circuit cost of fermionic state preparation. In the linear H$_6$ benchmark of Ref.~\cite{ramoa2025ovpceo}, tUPS variants nevertheless reach chemical accuracy with fewer CNOTs than the enhanced CEO-ADAPT-VQE algorithm, although at substantially greater optimization and measurement costs. This comparison suggests that tiled ans\"atze can remain competitive in circuit resources even when more efficient adaptive pools are used. Whether such pools would narrow the resource differences observed here requires further benchmarks within ADAPT-VMPE for the larger active spaces considered.

A complementary question is whether VIME can improve the parameter optimization within adaptive ansatz construction. To examine this, we investigated ADAPT-VIME, which selects new gates according to the reduction they permit in the regularized operator-projected McLachlan objective and uses VIME to relax all parameters of the enlarged circuit. In the
adaptive calculations tested, its convergence closely tracked that of ADAPT-VMPE. Successive relaxations were dominated by changes in the newly added parameter, while previously optimized parameters changed only modestly. This behavior suggests limited scope for improved convergence from imaginary-time updates compared with direct energy minimization in the adaptive setting considered here.
Appendix~\ref{app:adapt vime} details the construction and
comparison of ADAPT-VIME and ADAPT-VMPE.

\section{Conclusions}
\label{sec:conclusions}

We have introduced the VIME algorithm, which extends OVQITE to molecular electronic-structure problems through MP. In combination with our newly developed c-tUPS circuit ansatz, VIME achieves higher ground-state overlaps than direct MP energy minimisation (VMPE) using the same circuits. Across both TLD-1411 and the acene series, the selected VIME calculations match or improve on the MP-reported energy accuracy of the ADAPT-VMPE runs,
while requiring substantially fewer quantum-circuit resources. To our knowledge, this is the first demonstration of classically trained variational quantum circuits that achieve high-fidelity ground-state preparation for challenging molecular electronic-structure problems up to $50-60$ qubits, while retaining circuit depths compatible with prospective near-term quantum hardware. 

A central methodological advancement is the use of a single-surrogate construction, which extracts all quantities required for the VIME update from a single compiled Hamiltonian-propagation graph. This eliminates the need to perform a separate propagation for each operator in the molecular pool, significantly reducing the associated computational overhead. The approach is further complemented by blockwise truncation, which preserves intermediate contributions within fermionic tiles that would otherwise be discarded.

Extending VIME beyond the active-space sizes and circuit depths explored in this work requires reducing its classical computational cost further, which is currently dominated by the evaluation of the Jacobian. Matrix-free iterative solvers offer one possible route by evaluating the action of $J$ and
$J^{T}$ on vectors without explicitly constructing the Jacobian, although their efficiency depends on the number of iterations required for convergence. Further, any improvements to derivative evaluation should be
combined with reductions in the propagation graph size to lower both computational cost and memory requirements. Selective pruning of paired monomials could provide such a reduction, provided that the accuracy of the surrogate graph and the resulting updates are maintained. 

Together with these computational improvements, future work could
explore combining MP with other approaches to variational
imaginary-time evolution, including conventional
VQITE~\cite{mcardle2019vqite,yuan2019variational},
DualQITE~\cite{gacon2024dual},
and the imaginary-time variant of bowtie
VarQTE~\cite{drudis2026bowtie}.
Comparing the accuracy and computational cost of these combinations
with those of VIME would help assess how the variational formulation
and the evaluation of its required quantities affect classical
pre-training.

An alternative approach is to retain OVQITE while replacing MP with a different classical representation. Tensor networks (TNs) \cite{orus2014practical} represent a natural alternative to MP for classically representing and evolving c-tUPS states. The low CNOT count and shallow circuit depth of c-tUPS motivate exploring this approach, since limited-depth circuits with local connectivity can admit efficient TN simulation~\cite{markov2008simulating}. However, replacing MP by TNs is non-trivial, as the required bond dimension can grow rapidly during the evolution, while the associated truncation errors may compromise the accuracy of the OVQITE dynamics.

Another possibility is Pauli propagation (PP), which
evaluates observables through truncated expansions
in Pauli strings \cite{begusic2025clifford,begusic2024fast,rudolph2023lowesa}.
With parameter-independent truncation and a terminal
basis spanning the operator pool, the single-surrogate
construction carries over to the Pauli basis, providing
the quantities required for OVQITE updates through
the same backward and forward passes. This would extend classical circuit pre-training to quantum spin models, returning to the setting in which OVQITE was originally demonstrated~\cite{anuar2024operator}. 

Along the same line, local lattice Hamiltonians offer potentially cheaper surrogate graph construction through their smaller number of terms, while
compatible lattice and circuit symmetries could further reduce
propagation costs through symmetry merging
~\cite{teng2025symmetry}. For fermionic lattice models, including
those relevant to condensed matter and high-energy physics,
MP remains a natural representation; whether c-tUPS efficiently
captures their correlated ground states is an interesting
direction for future work.

For molecular ground-state preparation, a further
priority is to improve the quality of the states
obtained with compact circuits. Starting from an optimized c-tUPS circuit, one could
adaptively append individual Majorana rotations or
additional c-tUPS tiles and then use VIME to optimize the enlarged circuit to capture correlations beyond the initial ansatz. Another direction is to initialize VIME from a correlated state rather than the Hartree--Fock reference, for example using separable-pair approximation (SPA) circuits~\cite{kottmann2022spa}. These strategies should be compared with continued VIME optimization using additional c-tUPS tiles while retaining the current reference state.

The practical value of VIME as a state-preparation step depends on the balance between preparation cost and the benefits of improved initial states for downstream quantum algorithms. The ground-state overlap is a key measure in this context, influencing, for example, the success probability of QPE. Although the small-system benchmarks demonstrate improved overlaps, establishing corresponding gains in larger active spaces and quantifying their impact on the total resource requirements of downstream algorithms remain necessary. The low CNOT count and shallow depth of c-tUPS circuits also motivate the investigation of how much of the predicted state quality is retained when executing VIME-pretrained circuits on current quantum hardware.

Beyond ground-state preparation, the classical simulation
methods discussed above could also
be combined with variational methods for real-time
evolution \cite{li2017efficient,yuan2019variational,barison2021efficient}.
The operator-projected formulation, OVQRTE~\cite{anuar2026ovqrte}, offers a particularly direct extension of the present framework: both
propagation methods could supply the observable
derivatives and commutator expectation values required
for its updates. Exploring these combinations would extend classical
circuit training to fermionic and spin dynamics,
building on related developments in Majorana-string
simulation~\cite{danna2025majoranastring}.

\section{Author contributions}
F.S. led the computational work, developed the
simulation code, performed the numerical simulations, and
wrote the initial version of the Manuscript. M.G.A. and F.Š. provided the initial Majorana-propagation code, and A.A.A. provided the initial OVQITE code, on which the
simulation implementation was built. A.M.W.
provided the molecular Hamiltonians and DMRG reference
data for the acene systems. P.S. proposed the
use of the tUPS ansatz. F.Š. provided
scientific direction and supervised the project.
All authors reviewed the manuscript and provided
feedback on its content and presentation.

\begin{acknowledgments}
We thank P. V. Sriluckshmy, Francesco Nappi and James Anglin for useful
discussions.
The authors used GPT-5.6 Sol and GPT-6 Astra to improve
the fluency and readability of the manuscript and to assist with code development and debugging. All assistance was directed by the authors, who reviewed the revised text, tested the resulting code, and take full responsibility for the published work.
\end{acknowledgments}

\bibliography{bibliography}

\clearpage
\onecolumngrid

\appendix
\section{The single-surrogate VIME update}
\label{app: single surrogate pass}

\noindent For a fixed operator pool
$S_\tau=\{\hat O_k\}_{k=1}^{|S_\tau|}$ and circuit parameters
$\theta\in\mathbb R^{N_\theta}$, an OVQITE update requires 

\begin{equation}
\begin{aligned}
  &  v(\hat O_k,\hat\rho_\theta) = -\langle\{\hat H,\hat O_k\}\rangle_{\hat\rho_\theta}
+ 2\langle\hat H\rangle_{\hat\rho_\theta}  \langle\hat O_k\rangle_{\hat\rho_\theta},\\
&J_{kj}(\theta)
=
\frac{\partial\langle\hat O_k\rangle_{\hat\rho_\theta}}
     {\partial\theta_j}
\end{aligned}
\end{equation}
where the quantities entering the update are defined as,
\begin{equation}
\begin{aligned}
E(\theta)
&=\operatorname{Tr}[\hat\rho_\theta\hat H],\\
o_k(\theta)
&=\operatorname{Tr}[\hat\rho_\theta\hat O_k],\\
C_k(\theta)
&=\operatorname{Tr}[\hat\rho_\theta\{\hat H,\hat O_k\}],\\
J_{kj}(\theta)
&=\partial_{\theta_j}o_k(\theta),
\end{aligned}
\label{eq: app update quantities}
\end{equation}
where
$\hat\rho_\theta=\hat U(\theta)\rho\hat U^\dagger(\theta)$ and
$\rho=\ket{\mathbf n}\bra{\mathbf n}$ is the fixed Fock reference.
The Hamiltonian and pool operators are independent of $\theta$.
We show how their VIME approximations are obtained from one
Hamiltonian coefficient back-propagation and a joint forward
traversal of the same compiled graph. Throughout this appendix,
the subscript $\mathrm{MP}$ distinguishes quantities obtained through Majorana Propagation from exact circuit expectation values.

\subsection{The compiled propagation graph}
\label{app: A graph}

\noindent Let $N_{\mathrm{rot}}$ denote the number of elementary
Majorana rotations in the circuit. We fix the ordering convention
\begin{equation}
\begin{aligned}
\hat U(\theta)
&=\hat U_{N_{\mathrm{rot}}}(\theta)\cdots\hat U_1(\theta),\\
\hat U_j(\theta)
&=\exp\!\left[
-\frac{i}{2}\varphi_j(\theta)\hat M_{\vec b_j}
\right],
\end{aligned}
\label{eq: app circuit convention}
\end{equation}
so that $\hat U_1$ acts first on the reference state.
The gate angles $\varphi_\ell(\theta)$ are differentiable functions of the independent variational parameters. This notation allows several rotations to share
one parameter, including any fixed multiplicative factors
in their angles. 

The compiled MP graph represents the Majorana expansions
retained during Heisenberg back-propagation. Its layers are
indexed by $j=0,\ldots,N_{\mathrm{rot}}$, with layer $j$
corresponding to the stage after processing
$\hat U_{N_{\mathrm{rot}}},\ldots,\hat U_{j+1}$.
At each layer, we fix an ordering of the distinct retained
monomials and write
\begin{equation}
\mathcal M_j
=\{\hat M_a^{(j)}\}_{a=1}^{D_j},
\qquad
\hat M_a^{(j)}=\hat M_{\vec b_{j,a}},
\qquad
D_j=|\mathcal M_j|.
\label{eq: app layer basis}
\end{equation}
Here, $\vec b_{j,a}\in\{0,1\}^{2N}$ specifies which
Majorana operators occur in the monomial:
\begin{equation}
\hat M_{\vec b_{j,a}}
=
i^{r_{\vec b_{j,a}}}
\hat m_1^{(\vec b_{j,a})_1}\cdots
\hat m_{2N}^{(\vec b_{j,a})_{2N}},
\end{equation}
as specified in Sec. \ref{subsec: mp}.
Each graph node is identified by a pair $(j,a)$ and represents
the monomial $\hat M_a^{(j)}$ at layer $j$.
Thus, the same monomial appearing at different layers
corresponds to distinct nodes.

The terminal layer $j=N_{\mathrm{rot}}$ precedes any
back-propagation. Its monomial set
$\mathcal M_{\mathrm{term}}\equiv\mathcal M_{N_{\mathrm{rot}}}$
is the union of the monomials appearing with nonzero
coefficients in the Majorana expansions of $\hat H$ and
the pool operators. In this common basis,
\begin{equation}
\begin{aligned}
\hat H
&=\sum_{a=1}^{D_{\mathrm{term}}}
\alpha_a\hat M_a^{(\mathrm{term})},\\
\hat O_k
&=\sum_{a=1}^{D_{\mathrm{term}}}
(A_k)_a\hat M_a^{(\mathrm{term})},
\end{aligned}
\label{eq: app terminal coefficients}
\end{equation}
where
$\hat M_a^{(\mathrm{term})}\equiv
\hat M_a^{(N_{\mathrm{rot}})}$,
$D_{\mathrm{term}}=D_{N_{\mathrm{rot}}}$, and
$\vec\alpha,\vec A_k\in\mathbb R^{D_{\mathrm{term}}}$
are the corresponding coefficient vectors.
For the Hamiltonian pool used here,
$\mathcal M_{\mathrm{term}}$ is simply the set of monomials
appearing in $\hat H$.

Starting from this terminal layer, we construct layer $j$
by conjugating each monomial at layer $j+1$ with
$\hat U_{j+1}$ according to
Eq.~\eqref{eq:majorana-propagation} and applying the
prescribed truncation.
Each retained contribution defines an edge from its parent
node at layer $j+1$ to the corresponding child node(s)
at layer $j$.
Contributions producing the same monomial at the same layer
share a single node, where their coefficients are added.
After all rotations have been processed, the root layer
$j=0$ contains the monomials whose expectation values
are evaluated on the reference state $\rho$.

The edges between layers $j+1$ and $j$ define a sparse
linear map that propagates the monomial coefficients,
\begin{equation}
\hat L_j(\theta)\in\mathbb R^{D_j\times D_{j+1}}.
\label{eq: app transition dimensions}
\end{equation}
Each column $p$ corresponds to a monomial in layer $j+1$
(the parent), and each row $c$ to a retained monomial in
layer $j$ (the child). Applying
Eq.~\eqref{eq:majorana-propagation} through $\hat U_{j+1}$
gives
\begin{equation}
(\hat L_j)_{cp}
=
\begin{cases}
1,
& \text{commuting, unchanged monomial},\\[2pt]
\cos\varphi_{j+1}(\theta),
& \text{anticommuting, unchanged monomial},\\[2pt]
\sigma_{j,cp}\sin\varphi_{j+1}(\theta),
& \text{anticommuting, branching},\\[2pt]
0,
& \text{no retained connection}.
\end{cases}
\label{eq: app transition entries}
\end{equation}
For an unchanged monomial, $\hat M_c^{(j)}=\hat M_p^{(j+1)}$. Here, $\sigma_{j,cp}=\pm1$ is the sign of the branch contribution in Eq.~\eqref{eq:majorana-propagation}.

\noindent The complete backward map is
\begin{equation}
\mathsf{\hat{P}}(\theta)
=\hat L_0(\theta)\cdots \hat L_{N_{\mathrm{rot}}-1}(\theta)
\in\mathbb R^{D_{\mathrm{root}}\times D_{\mathrm{term}}},
\qquad
D_{\mathrm{root}}=D_0,\quad
D_{\mathrm{term}}=D_{N_{\mathrm{rot}}}.
\label{eq: app propagation matrix}
\end{equation}
We apply this map through its sparse transitions without
forming the product explicitly. Coefficients travel from
terminal layer to root layer in the \emph{backward pass}; the
\emph{forward pass} traverses the same graph from root layer
to terminal layer using the transposed transitions $\hat L_i^T$.

\subsection{Expectation values of pool operators}
\label{app: A observables}

\noindent Using the terminal coefficient vectors defined
in Eq.~\eqref{eq: app terminal coefficients}, the  backward pass begins with
$\vec h_{\text{term}}=\vec h_{N_{\text{rot}}}\equiv\vec\alpha$ and evaluates
\begin{equation}
\vec h_i=L_i\vec h_{i+1},
\qquad
i=N_{\mathrm{rot}}-1,\ldots,0.
\label{eq: app backward recurrence}
\end{equation}
The output of this backward pass is the root coefficient vector
$\vec h_{\mathrm{root}}=\vec h_0\equiv\hat{\mathsf P}\vec\alpha$.
Define the reference expectation vector by
\begin{equation}
r_a=\operatorname{Tr}[\hat \rho \hat M_a^{(\text{root})}],
\qquad
\vec r\in\mathbb R^{D_{\mathrm{root}}}.
\label{eq: app reference vector}
\end{equation}
For a Fock reference, $r_a=0$ for unpaired monomials
and $r_a=\pm1$ for paired monomials. The MP-reported energy is
therefore
\begin{equation}
E_{\mathrm{MP}}
=\vec r^{\,T}\mathsf{\hat P}\vec\alpha.
\label{eq: app surrogate energy}
\end{equation}

\noindent To evaluate $o_{k,\mathrm{MP}}$ for every pool operator,
we first compute the MP expectation values of the terminal
basis monomials. Since the $a$th terminal monomial is mapped
to the root as
\begin{equation}
\hat M_a^{(\mathrm{term})}
\;\xrightarrow{\mathrm{MP}}\;
\sum_{b=1}^{D_{\mathrm{root}}}
\mathsf{\hat P}_{ba}\hat M_b^{(\mathrm{root})},
\end{equation}
its reference expectation value is
$\sum_b\hat{\mathsf P}_{ba}r_b$. Collecting these values for all
terminal monomials gives
\begin{equation}
\vec m_{\mathrm{MP}}^{\,\mathrm{term}}
=\mathsf{\hat P}^T\vec r= \vec u_{N_{\text{rot}}},
\label{eq: app terminal expectations}
\end{equation}
where
\begin{equation}
\label{eq: u vector}
\vec u_j
=
\left(\hat L_{j-1}^T\cdots\hat L_0^T\right)\vec r
\end{equation}
denotes the forward-propagated expectation vector at layer $j$. Accordingly,
$\vec m_{\mathrm{MP}}^{\,\mathrm{term}}$ collects the MP estimates of the
terminal monomial expectation values, with components
\begin{equation}
\left(\vec m_{\mathrm{MP}}^{\,\mathrm{term}}\right)_a
=
\left\langle \hat M_a^{(\mathrm{term})}\right\rangle_{\mathrm{MP}}.
\end{equation} Therefore, for each pool operator,
\begin{equation}
o_{k,\mathrm{MP}} =
\vec A_k^T\vec m_{\mathrm{MP}}^{\,\mathrm{term}}
=\vec A_k^T\vec u_{N_{\text{rot}}}.
\label{eq: forward pass operators}
\end{equation}
Thus, a single forward pass computes the MP expectation values of all terminal monomials. The expectation value of each pool operator is then obtained by taking their weighted sum with the operator's expansion coefficients $\vec A_k$.

\subsection{Expectation values of the anticommutators}
\label{app: A anticommutators}

\noindent Exact unitary conjugation preserves anticommutators:
\begin{equation}
\hat U^\dagger\{\hat H,\hat O_k\}\hat U
=
\{\hat U^\dagger\hat H\hat U,
  \hat U^\dagger\hat O_k\hat U\}.
\label{eq: app exact anticommutator}
\end{equation}
This identity allows the anticommutator expectation value
to be evaluated on the reference state using the separately
back-propagated operators $\hat U^\dagger\hat H\hat U$ and
$\hat U^\dagger\hat O_k\hat U$.
We approximate these two operators by applying the same
compiled MP map $\mathsf{\hat P}$ to their terminal
coefficient vectors:
\begin{equation}
\begin{aligned}
\hat U^\dagger\hat H\hat U
&\approx \widetilde H_{\mathrm{root}}
=\sum_a(\mathsf{\hat P}\vec\alpha)_a
\hat M_a^{(\mathrm{root})},\\
\hat U^\dagger\hat O_k\hat U
&\approx \widetilde O_{k,\mathrm{root}}
=\sum_a(\mathsf{\hat P}\vec A_k)_a
\hat M_a^{(\mathrm{root})}.
\end{aligned}
\label{eq: app propagated operators}
\end{equation}
Their coefficient vectors are
$\vec h_{\mathrm{root}}=\mathsf{\hat P}\vec\alpha$ and
$\vec o_{k,\mathrm{root}}\equiv \mathsf{\hat P}\vec A_k$,
respectively.
\noindent Introduce the real symmetric root-layer matrix
\begin{equation}
W_{ab}
=\operatorname{Tr}\!\left[
\hat \rho\{\hat M_a^{(\text{root})},\hat M_b^{(\text{root})}\}
\right],
\qquad
W\in\mathbb R^{D_{\mathrm{root}}\times D_{\mathrm{root}}}.
\label{eq: app root anticommutator matrix}
\end{equation}
The VIME anticommutator estimate is thus
\begin{equation}
\begin{aligned}
C_{k,\mathrm{MP}}
&\equiv
\operatorname{Tr}\!\left[
\hat \rho\{\widetilde H_{\mathrm{root}},
       \widetilde O_{k,\mathrm{root}}\}
\right]
\\
&=
\sum_{i,j}
(\vec h_{\mathrm{root}})_i
(\vec o_{k,\mathrm{root}})_j
\operatorname{Tr}\!\left[
\hat \rho\{\hat M_i^{\mathrm{root}},\hat M_j^{\mathrm{root}}\}
\right]
\\
&=
\vec o_{k,\mathrm{root}}^{\,T}
W\vec h_{\mathrm{root}}
\\
&=
A_k^T\mathsf {\hat P}^T W\mathsf {\hat P}\vec\alpha.
\end{aligned}
\label{eq: app surrogate anticommutator}
\end{equation}

\noindent Without truncation, this expression is exactly equal to $C_k$ in Eq.\eqref{eq: app update quantities}. Under truncation, it instead provides an estimate constructed from the separately propagated operators. Because truncated propagation does not, in general, preserve operator products, this estimate is not guaranteed to coincide with the result obtained by directly propagating the truncated operators ${\hat H,\hat O_k}$. Nevertheless, we empirically find that the two agree, as demonstrated in Sec.~\ref{subsec: VIME scaling}.

\noindent Equation~\eqref{eq: app surrogate anticommutator}
requires no separate backward pass for any pool
operator. Once $\vec h_{\mathrm{root}}$ is available,
we form the root seed
\begin{equation}
\vec B_0 \equiv W\vec h_{\mathrm{root}}
\label{eq: app anticommutator seed}
\end{equation}
and propagate it according to
\begin{equation}
\vec B_{i+1}=L_i^T\vec B_i.
\label{eq: app anticommutator recurrence}
\end{equation}
All anticommutator estimates then follow from
$C_{k,\mathrm{MP}}=A_k^T\vec B_{N_{\text{rot}}}$. A root monomial with $r_a=0$ does not contribute directly to an individual operator's expectation value. It can, however, contribute to an anticommutator expectation: the product of two root monomials can have a nonzero expectation value in the reference state even when the expectation value of each monomial vanishes. The root basis must therefore retain these monomials. Pruning them, or the propagation paths that contribute to their coefficients, solely because $r_a=0$ would remove potentially nonzero contributions to $C_{k,\mathrm{MP}}$.

\subsection{Structure of the root-layer anticommutator matrix}
\label{app: A W structure}

\noindent We now show how to compute the root seed
$\vec B_0=W\vec h_{\mathrm{root}}$ without constructing $W$.
The key observation is that each Majorana monomial maps
the Fock reference to a single Fock state, multiplied
by a phase.

For mode $\ell$, each of the Majorana operators
$\hat m_{2\ell-1}$ and $\hat m_{2\ell}$ changes the
occupation from $n_\ell$ to $1-n_\ell$.
A monomial therefore changes that occupation if it
contains exactly one of these two operators; if it
contains both or neither, the occupation is unchanged.
For root monomial $\hat M_a^{(\mathrm{root})}$, define the binary vector
$\vec p_a\in\{0,1\}^N$ by
\begin{equation}
(\vec p_a)_\ell
=
(\vec b_{0,a})_{2\ell-1}
\oplus
(\vec b_{0,a})_{2\ell},
\qquad \ell=1,\ldots,N,
\end{equation}
where $\vec b_{0,a}$ is its Majorana support vector
at the root layer and $\oplus$ denotes addition
modulo two. Thus, $(\vec p_a)_\ell=1$ precisely when
monomial $\hat M_a^{(\mathrm{root})}$ changes the occupation of mode $\ell$.
Its action on the reference state is
\begin{equation}
\hat M_a^{(\mathrm{root})}\ket{\mathbf n}
=
\zeta_a\ket{\mathbf n\oplus\vec p_a},
\qquad
\zeta_a\in\{1,-1,i,-i\}.
\label{eq: parity action}
\end{equation}
Here, $\mathbf n\oplus\vec p_a$ is obtained by flipping
the occupation of each mode $\ell$ for which
$(\vec p_a)_\ell=1$.
The phase $\zeta_a$ is fixed by the monomial, the reference
occupations, and the Majorana ordering and phase convention.

To evaluate $W_{ab}$, consider the overlap between the
states produced by monomials $a$ and $b$.
Since the monomials are Hermitian,
\begin{equation}
\begin{aligned}
\bra{\mathbf n}
\hat M_a^{(\mathrm{root})}\hat M_b^{(\mathrm{root})}
\ket{\mathbf n}
&=
\bigl(\hat M_a^{(\mathrm{root})}\ket{\mathbf n}\bigr)^\dagger
\bigl(\hat M_b^{(\mathrm{root})}\ket{\mathbf n}\bigr)\\
&=\zeta_a^*\zeta_b\,\delta_{\vec p_a,\vec p_b}.
\end{aligned}
\end{equation}
The factor $\delta_{\vec p_a,\vec p_b}$ is one when
the two occupation-change vectors are identical and
zero otherwise, because distinct Fock states are
orthogonal. Substituting these expectation values into
Eq.~\eqref{eq: app root anticommutator matrix} gives
\begin{equation}
\begin{aligned}
W_{ab}
&=(\zeta_a^*\zeta_b+\zeta_b^*\zeta_a)
  \delta_{\vec p_a,\vec p_b}\\
&=2\operatorname{Re}(\zeta_a^*\zeta_b)
  \delta_{\vec p_a,\vec p_b}.
\end{aligned}
\label{eq: W structure}
\end{equation}
If the two monomials change different occupations,
then $\vec p_a\neq\vec p_b$ and $W_{ab}=0$.
If they change exactly the same occupations,
the value depends on their phases:
equal phases give $W_{ab}=2$, opposite phases give
$W_{ab}=-2$, and a real phase paired with an imaginary
phase gives $W_{ab}=0$.

This structure allows us to evaluate
$\vec B_0=W\vec h_{\mathrm{root}}$ using one sum for
each distinct vector $\vec p$ represented in the root
basis. Define
\begin{equation}
z_{\vec p}
=
\sum_{b:\,\vec p_b=\vec p}
\zeta_b(\vec h_{\mathrm{root}})_b.
\end{equation}
The complex number $z_{\vec p}$ collects the
Hamiltonian coefficients of all root monomials that
produce $\ket{\mathbf n\oplus\vec p}$, including
their phases. Since $\vec h_{\mathrm{root}}$ is real,
Eq.~\eqref{eq: W structure} gives
\begin{equation}
\begin{aligned}
(\vec B_0)_a
&=\sum_b W_{ab}(\vec h_{\mathrm{root}})_b\\
&=2\operatorname{Re}\!\left[
\zeta_a^*z_{\vec p_a}
\right].
\end{aligned}
\label{eq: two pass}
\end{equation}

The computation therefore requires two passes over
the root monomials: the first accumulates the sums
$z_{\vec p}$, and the second evaluates each entry
of $\vec B_0$. The vectors $\vec p_a$, phases
$\zeta_a$, and assignment of monomials to groups
depend only on the root basis and the fixed reference
state. For a fixed root basis, they are computed once
and reused at each optimization step. The product
$\vec B_0=W\vec h_{\mathrm{root}}$ is then evaluated
directly from the group sums, without constructing
$W$, using $O(D_{\mathrm{root}})$ arithmetic operations
and $O(D_{\mathrm{root}})$ storage.

\subsection{The Jacobian of expectation values}
\label{app: A jacobian}

\noindent The MP-reported Jacobian is defined element-wise as
\begin{equation}
\begin{aligned}
(J_{\mathrm{MP}})_{ki}
&=\partial_{\theta_i}o_{k,\mathrm{MP}}\\
&=\vec A_k^T
\left(\partial_{\theta_i}\mathsf{\hat P}^T\right)\vec r\\
&= \vec A_k^T \partial_{\theta_i} \vec u_{N_{\text{rot}}}
\end{aligned}
\label{eq: app surrogate jacobian}
\end{equation}
with
$J_{\mathrm{MP}}\in\mathbb R^{|S_\tau|\times N_\theta}$.
The last line of Eq. (\ref{eq: app surrogate jacobian}) is obtained by recalling that $\vec u_{N_{\mathrm{rot}}}$ contains the MP-reported
expectation values of the terminal monomials:
\begin{equation}
(\vec u_{N_{\mathrm{rot}}})_a
=
\langle \hat M_a^{(\mathrm{term})}\rangle_{\mathrm{MP}}.
\end{equation}
\noindent For parameter $\theta_i$, define the set of
parameter-dependent transitions
\begin{equation}
\mathcal I_i
=
\left\{
j\in\{0,\ldots,N_{\mathrm{rot}}-1\}:
\partial_{\theta_i}\varphi_{j+1}
\not\equiv0
\right\}.
\label{eq: app parameter transitions}
\end{equation}
This definition includes every occurrence of a shared
parameter. Differentiating the transition entries
in Eq.~\eqref{eq: app transition entries} gives
\begin{equation}
\left(\partial_{\theta_i}\hat L_j\right)_{cp}
=
\partial_{\theta_i}\varphi_{j+1}
\begin{cases}
-\sin\varphi_{j+1},
&\text{anticommuting, unchanged monomial},\\[2pt]
\sigma_{j,cp}\cos\varphi_{j+1},
&\text{anticommuting, branching},\\[2pt]
0,
&\text{otherwise}.
\end{cases}
\label{eq: app transition derivatives}
\end{equation}
The factor $\partial_{\theta_i}\varphi_{j+1}$
accounts for the dependence of the elementary gate
angle on the variational parameter.

Due to the structure of Eq. (\ref{eq: app surrogate jacobian}), to obtain the Jacobian entries $(J_{\mathrm{MP}})_{ki}$ it is sufficient to propagate the derivatives $\vec u_j$ in the following manner. For each parameter $\theta_i$, define
\begin{equation}
\vec d_j^{\,(i)}
\equiv
\partial_{\theta_i}\vec u_j
\in\mathbb R^{D_j}.
\end{equation}
The vector $\vec d_j^{\,(i)}$ therefore records the derivative,
with respect to $\theta_i$, of every forward propagated value at layer $j$.
Because $\vec u_0=\vec r$ is parameter independent,
\begin{equation}
\vec d_0^{\,(i)}=0.
\end{equation}
Applying the  product rule to $\vec u_{j+1}=\hat L_j^T\vec u_j$ gives
\begin{equation}
\vec d_{j+1}^{\,(i)}
=
\hat L_j^T\vec d_j^{\,(i)}
+
\left(\partial_{\theta_i}\hat L_j^T\right)\vec u_j,
\qquad
j=0,\ldots,N_{\mathrm{rot}}-1.
\label{eq: app derivative recurrence}
\end{equation}
The first term propagates derivatives accumulated at preceding
transitions, whereas the second adds the derivative generated by
transition $j$. The latter vanishes for $j\notin\mathcal I_i$.
Consequently, occurrences of $\theta_i$ at different transitions
are all accumulated in the same vector. At the terminal layer,
\begin{equation}
(\vec d_{N_{\mathrm{rot}}}^{\,(i)})_a
=
\partial_{\theta_i}
\langle \hat M_a^{(\mathrm{term})}\rangle_{\mathrm{MP}}.
\end{equation}
Substitution into
Eq.~\eqref{eq: app surrogate jacobian} therefore yields
\begin{equation}
(J_{\mathrm{MP}})_{ki}
=
\vec A_k^T
\vec d_{N_{\mathrm{rot}}}^{\,(i)}.
\label{eq: app jacobian contraction}
\end{equation}

\subsection{VIME step summary}

The preceding subsections derive forward recurrences
for the pool expectation values, anticommutator
expectation values, and parameter derivatives.
All three use the same graph transitions, so the
vectors $\vec u_j$, $\vec B_j$, and
$\{\vec d_j^{\,(i)}\}_{i=1}^{N_\theta}$ can be
updated together during a single forward pass.

A VIME step therefore begins with one backward pass
to obtain $\vec h_{\mathrm{root}}$, from which we
compute $E_{\mathrm{MP}}$ and the initial vector
$\vec B_0$. The joint forward pass then propagates
$N_\theta+2$ vectors: one for the pool expectation
values, one for the anticommutator expectation values,
and one for each parameter derivative.
At the terminal layer, taking their weighted sums
with the coefficients $\vec A_k$ gives
\begin{equation}
\begin{aligned}
o_{k,\mathrm{MP}}
&=\vec A_k^T\vec u_{N_{\mathrm{rot}}},\\
C_{k,\mathrm{MP}}
&=\vec A_k^T\vec B_{N_{\mathrm{rot}}},\\
(J_{\mathrm{MP}})_{ki}
&=\vec A_k^T\vec d_{N_{\mathrm{rot}}}^{\,(i)}.
\end{aligned}
\label{eq: app joint outputs}
\end{equation}
The pool and anticommutator expectation values determine
the imaginary-time velocity components,
\begin{equation}
v_{k,\mathrm{MP}}
=-C_{k,\mathrm{MP}}
+2E_{\mathrm{MP}}o_{k,\mathrm{MP}}.
\end{equation}
Together with the Jacobian, these define
\begin{equation}
G_{\mathrm{MP}}=J_{\mathrm{MP}}^TJ_{\mathrm{MP}},
\qquad
b_{\mathrm{MP}}=J_{\mathrm{MP}}^Tv_{\mathrm{MP}},
\end{equation}
which enter the regularized linear system
\begin{equation}
(G_{\mathrm{MP}}+\lambda I)\dot\theta
=b_{\mathrm{MP}}.
\end{equation}
Let $K$ denote the total number of retained connections
in the compiled graph, including those whose weights
vanish at the current parameter point. If each elementary
rotation angle depends on at most one variational
parameter, the backward and joint forward passes require
$O((N_\theta+2)K)$ arithmetic operations, assuming
constant cost for evaluating each angle and its derivative.

\subsection{Adaptive imaginary-time integration}
\label{app:integration}

Solving the linear system above gives the parameter
velocity $\dot\theta$. We now need to  choose the step size used to update the circuit parameters.
Since $J_{\mathrm{MP}}\dot\theta$ need not reproduce
$v_{\mathrm{MP}}$ exactly, the energy-descent identity
in Eq.~\eqref{eq: variance descent} does not ensure
that this parameter update lowers the surrogate energy.
We therefore check the update direction and choose
the step size using Armijo backtracking
\cite{armijo1966minimization}.

For the computed direction $\dot\theta$, define the
directional energy derivative
\begin{equation}
s_E
=
\left.
\frac{dE_{\mathrm{MP}}(\theta+\eta\dot\theta)}
     {d\eta}
\right|_{\eta=0}
=
\sum_k c_k(J_{\mathrm{MP}}\dot\theta)_k.
\end{equation}
The last equality follows from
$\hat H=\sum_k c_k\hat O_k$.
A negative $s_E$ means that a sufficiently small step
along $\dot\theta$ lowers the surrogate energy.
We attempt an update only when $s_E<-\epsilon_E$,
where $\epsilon_E>0$ is the numerical tolerance;
otherwise, we reject the update.

For an accepted direction, we propose
\begin{equation}
\theta'=\theta+\eta\dot\theta,
\qquad 0<\eta\leq\eta_{\max},
\end{equation}
where $\eta$ is the imaginary-time step size and
$\eta_{\max}$ is its upper bound.
Keeping $\dot\theta$ fixed, we halve $\eta$ until
\begin{equation}
E_{\mathrm{MP}}(\theta')
\leq E_{\mathrm{MP}}(\theta)
+c_{\mathrm A}\eta s_E
\end{equation}
is satisfied. This condition requires an energy
decrease of at least a fraction $c_{\mathrm A}$
of the decrease predicted by the directional derivative.
We use $c_{\mathrm A}=10^{-4}$, which we
empirically found to work well in our calculations.
After accepting a step of size $\eta_{\mathrm{acc}}$,
we recompute the quantities entering the linear system
at the updated parameters. The next search starts with
\begin{equation}
\eta_{\mathrm{next}}
=
\min\!\left(2\eta_{\mathrm{acc}},\eta_{\max}\right),
\end{equation}
allowing the step size to increase again after
a successful update.

\subsection{Stopping criterion}
\label{app:stopping}

We monitor the norm of
$b_{\mathrm{MP}}=J_{\mathrm{MP}}^Tv_{\mathrm{MP}}$,
the right-hand side of the linear system used to
compute $\dot\theta$.
For a nonzero initial norm, the stopping criterion is
\begin{equation}
\label{eq: OVQITE convergence}
\frac{
\left\|b_{\mathrm{MP}}(\theta)\right\|_2
}{
\left\|b_{\mathrm{MP}}(\theta_0)\right\|_2
}
<10^{-3},
\end{equation}
where $\theta_0$ is the initial parameter vector.
Thus, iterations stop when this norm falls below
$0.1\%$ of its initial value.

Each column of $J_{\mathrm{MP}}$ describes how the pool
expectation values change when one circuit parameter
is varied. If $J_{\mathrm{MP}}^Tv_{\mathrm{MP}}=0$,
none of these changes has a component along the
IT velocity vector. The parameter-update system
then gives $\dot\theta=0$ for $\lambda>0$: no parameter
motion improves the fit to the target velocity within
this local linear approximation. This indicates
stationarity of the projected evolution, but does not
imply that the target velocity itself vanishes or that
the circuit prepares the ground state.

We also terminate when the computed direction fails
the energy-descent check. In our calculations, continuing
the evolution beyond this point consistently increased
the energy and led to a higher final energy.
Stopping at this point avoids further optimization steps
and preserves the lowest MP-reported energy reached
along the accepted trajectory.

\FloatBarrier
\section{Circuit synthesis of the two-orbital tiles}
\label{app: CNOT counts}

This appendix gives complete four-qubit circuits for one tUPS or c-tUPS tile on three hardware graphs. For each tile, we use the same search procedure to find circuits that implement its unitary using only CNOT
connections allowed by each hardware graph. Every circuit specifies all CNOTs and one-qubit gates as explicit functions of $(\theta_1,\theta_2,\theta_3)$, yielding the constructive
upper bounds in Table~\ref{tab: tile resources}.
The reported resources refer to individual tiles. Combining tiles may allow gate cancellations or additional parallel execution. The search is bounded and non exhaustive; the reported counts are not claims of global optimality.

\begin{table}[htbp]
\centering\small
\caption{CNOT count $N_{\mathrm{CNOT}}$ and two-qubit
depth $D_{2Q}$ for one tile, reported as
$N_{\mathrm{CNOT}}\,(D_{2Q})$. One-qubit gates are
not included in these resource counts.
All circuits are valid for arbitrary real parameters.
The c-tUPS row applies to both cp-tUPS and cm-tUPS.
For linear nearest-neighbor (LNN) connectivity, the
column headings specify the spin-mode order along
the qubit line, also used in the Jordan--Wigner mapping.
In the final column, the two spin modes of each spatial
orbital are adjacent, allowing tiles on successive
orbital pairs to act on consecutive qubits.
The counts in this column include two fermionic SWAP
gates (four additional CNOTs) used to convert the tile
circuit to this mode order.}
\label{tab: tile resources}
\setlength{\tabcolsep}{3pt}
\begin{tabular}{lcccc}
\toprule
Tile & All-to-all & Square grid & \multicolumn{2}{c}{LNN}\\
 & & &
 $\substack{p_\alpha-q_\alpha\\[-1pt]-p_\beta-q_\beta}$ &
 $\substack{p_\alpha-p_\beta\\[-1pt]-q_\alpha-q_\beta}$\\
\midrule
tUPS   & $14\,(8)$ & $15\,(9)$ & $16\,(10)$ & $20\,(14)$\\
c-tUPS & $10\,(8)$ & $11\,(8)$ & $12\,(9)$  & $16\,(13)$\\
\bottomrule
\end{tabular}
\end{table}

Section~\ref{app:common} defines the gates and fermionic operators used
throughout this appendix.
Section~\ref{app:hardware-synthesis} derives the Pauli rotations needed to
implement each tile and describes how circuits have been identified and verified.
Sections~\ref{app:ata}, \ref{app:square}, and~\ref{app:lnn} present the
resulting circuits for all-to-all connectivity, a square grid, and linear
nearest-neighbor (LNN) connectivity, respectively.
For notational simplicity, we omit hats on operators throughout this appendix.

\subsection{Conventions and generator algebra}
\label{app:common}

\subsubsection{Gate conventions and Clifford conjugation}

\label{app:conventions}

We quantify the resources required to execute each tile
on quantum hardware using the CNOT count
$N_{\mathrm{CNOT}}$ (total number of CNOTs) and
the two-qubit depth $D_{2Q}$ (number of CNOT layers,
allowing gates on disjoint qubit pairs to execute
in parallel). One-qubit gates are excluded from
these resource counts.
We write $\CX_{c\to t}$ for a CNOT with control
qubit $c$ and target qubit $t$.
We use the following Pauli rotation convention
and abbreviations:
\begin{equation}
R_\alpha(\phi)=e^{-\ii\phi\sigma_\alpha/2},\qquad
\Theta=\theta_1+\theta_3,\qquad
\Delta=\theta_1-\theta_3.
\label{eq:rotation-convention}
\end{equation}
where $\theta_1,\theta_2,\theta_3$ are the three
variational parameters of the original tile.
The combinations $\Theta$ and $\Delta$ simplify
the expressions for the one-qubit rotation angles
in the circuits below.

\noindent To construct the tile circuits, we use the identity
\begin{equation}
C\,e^{\theta K}C^{\dagger}=e^{\theta\,CKC^{\dagger}},
\label{eq:conjslide}
\end{equation}
valid for any unitary $C$ and following term by term from the power series. This allows us to transform a rotation by conjugating its generator. For a Clifford gate $C$, conjugation maps each Pauli operator to another Pauli operator, up to a sign. Clifford gates can be
constructed from $H$, $S$, and CNOT gates.
Equation~\eqref{eq:conjslide} is used below to transform generators into forms that admit
efficient circuit implementations. The required CNOT-conjugation rules are
\begin{equation}
X_c\mapsto X_cX_t,\quad Z_c\mapsto Z_c,\quad
X_t\mapsto X_t,\quad Z_t\mapsto Z_cZ_t
\label{eq:cxconj}
\end{equation}
for $\CX_{c\to t}$, together with $SXS^{\dagger}=Y$, $HXH=Z$ and $Y=\ii XZ$.

\subsubsection{Fermionic generators and physical mode order}
\label{app:generators}

Let the fermionic modes obey
$\{a_{p\sigma},a^\dagger_{q\tau}\}=\delta_{pq}\delta_{\sigma\tau}$ with all
other anticommutators vanishing, and for two spatial orbitals $p,q$ introduce
the spin-summed one-body excitation
$E_{pq}=a^\dagger_{p\alpha}a_{q\alpha}+a^\dagger_{p\beta}a_{q\beta}$ and the
anti-Hermitian generators
\begin{equation}
\kappa_1=E_{pq}-E_{qp},\qquad \kappa_2=E_{pq}^{\,2}-E_{qp}^{\,2}.
\label{eq:kappas}
\end{equation}
The square is not an independent definition: expanding $E_{pq}^2$, the
equal-spin terms vanish because $(a^\dagger_{p\sigma})^2=(a_{q\sigma})^2=0$ and
the two opposite-spin cross terms are equal after fermionic reordering, so that
\begin{equation}
E_{pq}^{\,2}=2P_{pq},\quad
P_{pq}=a^\dagger_{p\alpha}a^\dagger_{p\beta}a_{q\beta}a_{q\alpha},\quad
\kappa_2=2K_P,\quad K_P=P_{pq}-P_{pq}^\dagger .
\label{eq:pairtransfer}
\end{equation}
The \ctUPS{} family locks this pair transfer to the opposite-spin exchange,
\begin{equation}
F_{pq}=a^\dagger_{p\alpha}a^\dagger_{q\beta}a_{p\beta}a_{q\alpha},
\qquad K_F=F_{pq}-F_{pq}^\dagger,
\qquad \kappa_C^{(\pm)}=K_P\pm K_F ,
\label{eq:compressedfermion}
\end{equation}
giving the two tiles
\begin{align}
U_{\tUPS}&=e^{\theta_1\kappa_1}e^{\theta_2\kappa_2}e^{\theta_3\kappa_1},
\nonumber\\
U_{\ctUPS}^{(\pm)}&=e^{\theta_1\kappa_1}e^{\theta_2\kappa_C^{(\pm)}}
e^{\theta_3\kappa_1}.
\label{eq:tiles}
\end{align}
Since $\kappa_1$ appears twice, with angles $\theta_1$ and $\theta_3$, the
combinations $\Theta$ and $\Delta$ arise below whenever the middle factor acts
trivially and the two outer rotations merge.  Every monomial in $K_P$ and $K_F$
creates and annihilates one $\alpha$ (spin-up) and one $\beta$ (spin-down) electron, so
$[N,\kappa_C^{(\pm)}]=[S_z,\kappa_C^{(\pm)}]=0$, while the locking is not in
general a spin-scalar coupling and $[S^2,\kappa_C^{(\pm)}]\ne0$. 

\noindent We use the Jordan--Wigner convention
$a_j=(\prod_{k<j}Z_k)(X_j+\ii Y_j)/2$
with the spin-blocked local mode order
\begin{equation}
(p_\alpha,q_\alpha,p_\beta,q_\beta)
\equiv(A,B,\bar{A},\bar{B}).
\label{eq:wires}
\end{equation}
Here, $A,B,\bar A,\bar B$ label the four qubits in tensor-product
order; their physical connectivity is specified
separately for each hardware graph.

\subsection{Hardware-aware Clifford-frame synthesis}
\label{app:hardware-synthesis}

We first use identities of the fermionic generators to reduce the
number of Pauli rotations needed to implement each tile. We then
construct circuits for these rotations using a Clifford-frame search
restricted to the available two-qubit connectivity. The same algebraic
reductions and search procedure apply to all-to-all, square-grid,
and LNN connectivity. Finally, we verify that the resulting circuits
implement the tile unitary for arbitrary parameter values.

\subsubsection{Pauli rotations for the two-orbital tiles}
\label{app:reduced-targets}

We express each tile unitary as an ordered product of Pauli
rotations,
\begin{equation}
R_W(\phi)=\exp\!\left(-\frac{\ii\phi}{2}W\right).
\end{equation}
We specify the rotations in chronological order by
\begin{equation}
\mathcal R
=\bigl((W_1,\phi_1),\ldots,(W_m,\phi_m)\bigr),
\label{eq:synthesis-search-target-list}
\end{equation}
where $W_i$ is a four-qubit Pauli operator and $\phi_i$ is its
rotation angle.

To reduce the number of rotations, we identify components of
the two outer hopping generators that commute with the middle
generator. Their exponentials can then be moved through the
middle factor and combined. Below, we derive this reduction
for c-tUPS and tUPS.

\paragraph{Compressed tUPS.}
\label{app:compressed-reduction}
For the plus-sign compressed tile, we begin with
\begin{equation}
U_{\ctUPS}^{(+)}
=e^{\theta_1\kappa_1}
 e^{\theta_2\kappa_C^{(+)}}
 e^{\theta_3\kappa_1}.
\label{eq:compressed-synthesis-starting-product}
\end{equation}
Introduce the spin-resolved hopping generators
\begin{equation}
k_\alpha=\frac{\ii}{2}(XYII-YXII),\qquad
k_\beta=\frac{\ii}{2}(IIXY-IIYX),\qquad
\kappa_1=k_\alpha+k_\beta.
\label{eq:synthesis-spin-hopping-generators}
\end{equation}
For $\kappa_C^{(+)}$, direct Jordan--Wigner substitution
gives
\begin{equation}
\kappa_C^{(+)}
=\frac{\ii}{4}(XYXX+XYYY-YXXX-YXYY).
\label{eq:compressed-synthesis-generator}
\end{equation}

The four words in $\kappa_C^{(+)}$ commute. Each occurrence of
$e^{\theta_j\kappa_1}$ similarly decomposes into four commuting Pauli
rotations: two from $k_\alpha$ and two from $k_\beta$. A direct decomposition
of Eq.~\eqref{eq:compressed-synthesis-starting-product} therefore contains $4+4+4=12$
Pauli rotations. First, we note that 
\begin{equation}
\label{eq: commutator k_alpha, k_C}
[k_\alpha,\kappa_C^{(+)}]=0.
\end{equation}
For the beta-spin pair, define
\begin{equation}
b_\beta=-\ii\,IIYX,\qquad
c_\beta=\frac{\ii}{2}(IIXY+IIYX),\qquad
k_\beta=b_\beta+c_\beta.
\label{eq:synthesis-beta-split}
\end{equation}
These components satisfy
\begin{equation}
[b_\beta,c_\beta]=0,\qquad
[c_\beta,\kappa_C^{(+)}]=0.
\end{equation}
Splitting the two outer factors in
Eq.~\eqref{eq:compressed-synthesis-starting-product} and collecting these
commuting components gives
\begin{align}
U_{\ctUPS}^{(+)}
&=e^{\theta_1k_\alpha}
  e^{\theta_1b_\beta}
  e^{\theta_1c_\beta}
  e^{\theta_2\kappa_C^{(+)}}
  e^{\theta_3k_\alpha}
  e^{\theta_3b_\beta}
  e^{\theta_3c_\beta}
  \nonumber\\
&=e^{\Theta k_\alpha}
  e^{\theta_1b_\beta+\Theta c_\beta}
  e^{\theta_2\kappa_C^{(+)}}
  e^{\theta_3b_\beta},
\label{eq:compressed-synthesis-factorization}
\end{align}
Using
$R_W(\phi)=\exp(-\ii\phi W/2)$, the four factors in the second line of
Eq.~\eqref{eq:compressed-synthesis-factorization} decompose as
\begin{equation}
\begin{aligned}
e^{\Theta k_\alpha}
 &=R_{XYII}(-\Theta)R_{YXII}(\Theta),\\
e^{\theta_3b_\beta}
 &=R_{IIYX}(2\theta_3),\\
e^{\theta_2\kappa_C^{(+)}}
 &=R_{XYXX}(-\theta_2/2)
   R_{XYYY}(-\theta_2/2)
   R_{YXXX}(\theta_2/2)
   R_{YXYY}(\theta_2/2),\\
e^{\theta_1b_\beta+\Theta c_\beta}
 &=R_{IIXY}(-\Theta)R_{IIYX}(\Delta).
\end{aligned}
\label{eq:compressed-synthesis-rotation-factors}
\end{equation}

The two rotations generated by $k_\alpha$ commute with every other factor
and may therefore be placed first. Choosing an equivalent order within each
commuting group gives the chronological target list
\begin{align}
\mathcal R_C^{(+)}=\bigl(&
 (YXII,\Theta),
 (XYII,-\Theta),
 (IIYX,2\theta_3),
 \nonumber\\
&
 (XYXX,-\theta_2/2),
 (YXXX,\theta_2/2),
 (YXYY,\theta_2/2),
 (XYYY,-\theta_2/2),
 \nonumber\\
&
 (IIXY,-\Theta),
 (IIYX,\Delta)
 \bigr).
\label{eq:compressed-target}
\end{align}

The commuting $k_\alpha$ components reduce the four corresponding outer
rotations to two. Combining the two $c_\beta$ contributions reduces the four
beta-spin outer rotations to three, while the four middle rotations remain
unchanged. The search input therefore contains $2+3+4=9$ rotations instead of twelve.

\paragraph{Ordinary tUPS.}
\label{app:ordinary-reduction}
The ordinary tile starts from
\begin{equation}
U_{\tUPS}
=e^{\theta_1\kappa_1}
 e^{\theta_2\kappa_2}
 e^{\theta_3\kappa_1}.
\label{eq:ordinary-synthesis-starting-product}
\end{equation}
Its middle generator contains both compressed contributions,
\begin{equation}
\kappa_2=(K_P+K_F)+(K_P-K_F)=2K_P,
\end{equation}
and has the Pauli expansion
\begin{align}
\kappa_2=\frac{\ii}{4}\bigl(
 &XYXX+XXXY+YYXY+XYYY\nonumber\\
 &-XXYX-YXXX-YXYY-YYYX
 \bigr).
\label{eq:ordinary-synthesis-generator}
\end{align}
The eight Pauli words in $\kappa_2$ commute. Together with the four
rotations in each outer factor, a direct decomposition of
Eq.~\eqref{eq:ordinary-synthesis-starting-product} therefore contains $4+8+4=16$ Pauli rotations.

Unlike the compressed middle generator, $\kappa_2$ does not commute with all
of $k_\alpha$, so the complete $k_\alpha$ contributions cannot be combined.
Nevertheless, one component of each spin-resolved hopping generator commutes
through the middle factor. Define
\begin{align}
b_\alpha&=-\ii\,YXII,&
c_\alpha&=\frac{\ii}{2}(XYII+YXII),\nonumber\\
\mathcal B&=b_\alpha+b_\beta,&
\mathcal D&=c_\alpha+c_\beta,
\label{eq:ordinary-synthesis-split}
\end{align}
where $b_\beta$ and $c_\beta$ are defined in
Eq.~\eqref{eq:synthesis-beta-split}. These definitions give
\begin{equation}
\kappa_1=\mathcal B+\mathcal D,
\qquad
[\mathcal B,\mathcal D]=0.
\end{equation}

Each $c_\sigma$ acts only on the even-parity subspace of its corresponding
same-spin mode pair, whereas pair transfer acts only when both spin
sectors have odd parity. Their supports are therefore disjoint, implying
\begin{equation}
[c_\sigma,\kappa_2]=0
\quad(\sigma=\alpha,\beta),
\qquad
[\mathcal D,\kappa_2]=0.
\label{eq:ordinary-synthesis-commutation}
\end{equation}
The two $\mathcal D$ contributions can therefore be moved through the middle
factor and combined. Splitting the outer exponentials and collecting these
terms gives
\begin{align}
U_{\tUPS}
&=e^{\theta_1\mathcal B}
  e^{\theta_1\mathcal D}
  e^{\theta_2\kappa_2}
  e^{\theta_3\mathcal B}
  e^{\theta_3\mathcal D}
  \nonumber\\
&=e^{\theta_1\mathcal B+\Theta\mathcal D}
  e^{\theta_2\kappa_2}
  e^{\theta_3\mathcal B},
\label{eq:ordinary-synthesis-factorization}
\end{align}
Using
$R_W(\phi)=\exp(-\ii\phi W/2)$, the three factors in the second line of
Eq.~\eqref{eq:ordinary-synthesis-factorization} decompose as
\begin{equation}
\begin{aligned}
e^{\theta_3\mathcal B}
={}&R_{YXII}(2\theta_3)
   R_{IIYX}(2\theta_3),\\[2pt]
e^{\theta_2\kappa_2}
={}&R_{XYXX}(-\theta_2/2)
   R_{XXXY}(-\theta_2/2)
   R_{YYXY}(-\theta_2/2)
   R_{XYYY}(-\theta_2/2)\\
 &\times
   R_{XXYX}(\theta_2/2)
   R_{YXXX}(\theta_2/2)
   R_{YXYY}(\theta_2/2)
   R_{YYYX}(\theta_2/2),\\[2pt]
e^{\theta_1\mathcal B+\Theta\mathcal D}
={}&R_{YXII}(\Delta)
   R_{XYII}(-\Theta)
   R_{IIYX}(\Delta)
   R_{IIXY}(-\Theta).
\end{aligned}
\label{eq:ordinary-synthesis-rotation-factors}
\end{equation}
The rotations within each line commute, so their displayed internal order
may be chosen freely. Reading the three factors in chronological order gives
the target list
\begin{align}
\mathcal R_O=\bigl(&
 (YXII,2\theta_3),
 (IIYX,2\theta_3),
 \nonumber\\
&
 (XYXX,-\theta_2/2),
 (XXXY,-\theta_2/2),
 (YYXY,-\theta_2/2),
 (XYYY,-\theta_2/2),
 \nonumber\\
&
 (XXYX,\theta_2/2),
 (YXXX,\theta_2/2),
 (YXYY,\theta_2/2),
 (YYYX,\theta_2/2),
 \nonumber\\
&
 (YXII,\Delta),
 (XYII,-\Theta),
 (IIYX,\Delta),
 (IIXY,-\Theta)
 \bigr).
\label{eq:ordinary-target}
\end{align}

Combining the two $\mathcal D$ contributions reduces the eight outer
rotations to six: two from $e^{\theta_3\mathcal B}$ and four from
$e^{\theta_1\mathcal B+\Theta\mathcal D}$. Together with the unchanged
eight-word middle factor, the search input therefore contains $2+8+4=14$ rotations instead of sixteen.

\subsubsection{Hardware-aware Clifford-frame search}
\label{app:clifford-search}

The search takes the rotation sequence derived above,
$\mathcal R_C^{(+)}$ or $\mathcal R_O$, and a hardware graph
$G=(V,\mathsf E)$, where $V=\{A,B,\bar{A},\bar{B}\}$ and $\mathsf E$
specifies the allowed two-qubit connections. It preserves the
order of rotations whose generators do not commute; adjacent
commuting rotations may be exchanged.

We use Pauli-network synthesis~\cite{goubault2024faster} to
implement these rotations through a common sequence of Clifford
gates. Let $F$ be the product of the Clifford gates preceding
a given point in the circuit, with all parameterized rotations
omitted. The Clifford frame records $F^\dagger X_qF$ and
$F^\dagger Z_qF$ for each qubit $q$. A required rotation
$R_{W_i}(\phi_i)$ can be implemented at this point whenever
\begin{equation}
F^\dagger\sigma_qF=sW_i,
\qquad \sigma\in\{X,Y,Z\},\quad s\in\{+1,-1\}.
\label{eq:synthesis-localization}
\end{equation}
Here $s$ is the sign obtained by conjugation. The corresponding
one-qubit gate is $R_{\sigma_q}(s\phi_i)$, since
\begin{equation}
F^\dagger R_{\sigma_q}(s\phi_i)F=R_{W_i}(\phi_i).
\label{eq:synthesis-local-angle}
\end{equation}

For these four-qubit tiles, we replace the greedy construction
of Ref.~\cite{goubault2024faster} with a bounded bidirectional
search. Following the meet-in-the-middle strategy used in circuit
synthesis~\cite{amy2013meet}, we build partial circuits from both
ends of the rotation sequence. Each step adds a gate
\begin{equation}
T_{uv}(\sigma,\tau)
=\exp\!\left(-\frac{\ii\pi}{4}\sigma_u\tau_v\right),
\qquad \sigma,\tau\in\{X,Y,Z\},
\label{eq:synthesis-search-move}
\end{equation}
Each gate acts on an allowed edge $\{u,v\}\in\mathsf E$ and can
be implemented with one CNOT and one-qubit Clifford gates.
After adding a gate, we update the Clifford frame and check
which required rotations can now be implemented using
Eq.~\eqref{eq:synthesis-localization}.

The two searches construct the beginning and the end of the
circuit. We connect these parts when their Clifford frames
can be matched by one-qubit Clifford gates and together they
allow every required rotation to be implemented.

For c-tUPS, only the plus-sign tile requires a separate search.
Reversing the qubit order,
$\mathcal J:(A,B,\bar A,\bar B)
\mapsto(\bar B,\bar A,B,A)$, gives
\begin{equation}
U_{\ctUPS}^{(-)}(\theta_1,\theta_2,\theta_3)
=\mathcal J U_{\ctUPS}^{(+)}(-\theta_1,-\theta_2,-\theta_3)
\mathcal J^\dagger.
\label{eq:compressed-reflection}
\end{equation}
Thus, the minus-sign circuit is obtained by reflecting the gate
locations and negating the three parameters. This reflection
preserves all three hardware graphs, so no SWAP gates are needed
and the CNOT count and two-qubit depth are unchanged.

The explicit circuits establish the reported resource bounds.
The search is not exhaustive and therefore does not prove
optimality.

\FloatBarrier
\subsection{Circuit conventions and verification}
\label{app:circuit-verification}

For each connectivity, we present the c-tUPS circuit followed by the
ordinary tUPS circuit. Each circuit is specified by a diagram
showing all CNOTs and a table of one-qubit gates. We use
\begin{equation}
U=e^{\ii\gamma}\mathcal L_d\mathcal C_d\mathcal L_{d-1}
\cdots\mathcal C_1\mathcal L_0,
\qquad
\mathcal L_k=\bigotimes_{q\in\{A,B,\bar A,\bar B\}}\ell_{k,q}.
\label{eq:explicit-circuit-layers}
\end{equation}
The gates act in the order $\mathcal L_0,\mathcal C_1,\mathcal L_1,
\ldots,\mathcal C_d,\mathcal L_d$.
Here $\mathcal C_k$ is a CNOT layer, $\mathcal L_k$ is a layer
of one-qubit gates, and $\gamma$ is the global phase.
Gates on disjoint qubit pairs may act simultaneously, so
$d=D_{2Q}$. Within each table cell, one-qubit gates act from
left to right; $I$ indicates that no one-qubit gate is applied
in that layer. Unless stated otherwise,
the tensor-product order is $A,B,\bar A,\bar B$, as defined in
Eq.~\eqref{eq:wires}.

All circuits use four qubits without ancillas. We report
$N_{\mathrm{CNOT}}$, $D_{2Q}$, and $\gamma$ for each circuit.
The c-tUPS circuits shown below implement the plus-sign tile;
reflecting the circuit and negating the three parameters
gives the minus-sign tile with identical resources. We also report the one-qubit gate count $N_{1Q}$, counting each occurrence of $H$, $S$, $S^\dagger$, $X$, $Y$, $Z$, or $R_\alpha(\phi)$ in the tables
as one gate. Identity entries are excluded.
These counts refer to the circuits as written,
before cancellations or merging consecutive
one-qubit gates.

We verify analytically that each circuit implements
the intended tile for arbitrary parameter values.
As an additional check, numerical comparisons with
the full $16\times16$ Jordan--Wigner unitary at
1167 parameter triples give Frobenius errors below
$10^{-13}$, including the stated global phase.

\paragraph{Caveat: LNN circuits for overlapping tiles.}
The circuits above implement one tile, but an ansatz contains
tiles acting on successive orbital pairs, such as $(p,q)$ and
$(q,r)$. On a line, both tiles should act on consecutive qubits
without moving the modes between tiles. The original order
$(p_\alpha,q_\alpha,p_\beta,q_\beta)$ does not allow this for
every neighboring orbital pair. We therefore place the two
spin modes of each spatial orbital next to one another:
\begin{equation}
p_\alpha-p_\beta-q_\alpha-q_\beta-r_\alpha-r_\beta-\cdots.
\end{equation}
Each tile then acts on four consecutive qubits, locally ordered
$A-\bar A-B-\bar B$.

We use this mode order together with the Jordan--Wigner mapping.
Relative to the original four-mode order, the two middle modes
are exchanged. For fermions, this exchange introduces a minus
sign when both modes are occupied. It is described by the
fermionic SWAP (fSWAP) operation
\begin{equation}
\mathcal F\lvert n_1n_2\rangle
=(-1)^{n_1n_2}\lvert n_2n_1\rangle,
\qquad n_1,n_2\in\{0,1\}.
\label{eq:orbital-lnn-fswap}
\end{equation}
We obtain the tile circuit in the new order by applying this
operation to the two middle qubits before and after the
original circuit:
\begin{equation}
U_{\mathrm{new}}
=\mathcal F_{23}U_{\mathrm{original}}\mathcal F_{23}.
\label{eq:orbital-lnn-conjugation}
\end{equation}
Each fermionic SWAP uses two CNOTs, so this construction adds
four CNOTs per tile, included in the reported counts.
Each tile restores the mode assignment, allowing successive
tiles to overlap without additional SWAP gates.
The Hamiltonian and reference state must also be represented
using the new Jordan--Wigner order.

\FloatBarrier

\subsection{All-to-all circuits}

\label{app:ata}

\begin{figure}[H]
\centering
\begin{tikzpicture}[every node/.style={circle,draw,fill=white,minimum size=6mm,inner sep=1pt},thick]
\node (L) at (0,1.2) {$A$};\node (A) at (1.8,1.2) {$B$};\node (B) at (0,0) {$\bar A$};\node (R) at (1.8,0) {$\bar B$};
\draw (L)--(A)--(R)--(B)--(L);\draw (L)--(R);\draw (A)--(B);
\end{tikzpicture}
\caption{All-to-all connectivity.}
\label{fig:ata-connectivity}
\end{figure}
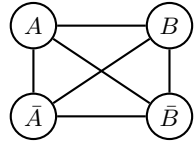

\FloatBarrier

\subsubsection{c-tUPS}

\label{app:compressed-ata}

\begin{equation}
N_{\mathrm{CNOT}}=10,\qquad D_{2Q}=8,\qquad N_{1Q}=61,\qquad \gamma=\pi/4.
\label{eq:compressed-ata-circuit}
\end{equation}

\begin{figure}[H]
\centering\scriptsize
\begin{adjustbox}{max width=\linewidth}
\begin{quantikz}[row sep={0.48cm,between origins},column sep=0.18cm]
\lstick{$A$} & \gate{\ell_{0,A}} & \ctrl{1} & \gate{\ell_{1,A}} & \qw      & \gate{\ell_{2,A}} & \qw      & \gate{\ell_{3,A}} & \ctrl{1} & \gate{\ell_{4,A}} & \qw      & \gate{\ell_{5,A}} & \qw      & \gate{\ell_{6,A}} & \ctrl{1} & \gate{\ell_{7,A}} & \qw      & \gate{\ell_{8,A}} & \qw \\
\lstick{$B$} & \gate{\ell_{0,B}} & \targ{}  & \gate{\ell_{1,B}} & \ctrl{2} & \gate{\ell_{2,B}} & \ctrl{1} & \gate{\ell_{3,B}} & \targ{}  & \gate{\ell_{4,B}} & \ctrl{1} & \gate{\ell_{5,B}} & \ctrl{2} & \gate{\ell_{6,B}} & \targ{}  & \gate{\ell_{7,B}} & \qw      & \gate{\ell_{8,B}} & \qw \\
\lstick{$\bar A$} & \gate{\ell_{0,\bar A}} & \ctrl{1} & \gate{\ell_{1,\bar A}} & \qw      & \gate{\ell_{2,\bar A}} & \targ{}  & \gate{\ell_{3,\bar A}} & \qw      & \gate{\ell_{4,\bar A}} & \targ{}  & \gate{\ell_{5,\bar A}} & \qw      & \gate{\ell_{6,\bar A}} & \ctrl{1} & \gate{\ell_{7,\bar A}} & \ctrl{1} & \gate{\ell_{8,\bar A}} & \qw \\
\lstick{$\bar B$} & \gate{\ell_{0,\bar B}} & \targ{}  & \gate{\ell_{1,\bar B}} & \targ{}  & \gate{\ell_{2,\bar B}} & \qw      & \gate{\ell_{3,\bar B}} & \qw      & \gate{\ell_{4,\bar B}} & \qw      & \gate{\ell_{5,\bar B}} & \targ{}  & \gate{\ell_{6,\bar B}} & \targ{}  & \gate{\ell_{7,\bar B}} & \targ{}  & \gate{\ell_{8,\bar B}} & \qw
\end{quantikz}
\end{adjustbox}
\caption{c-tUPS circuit on all-to-all connectivity. The boxes $\ell_{k,q}$ are specified in Table~\ref{tab:compressed-ata-locals}.}
\label{fig:compressed-ata}
\end{figure}
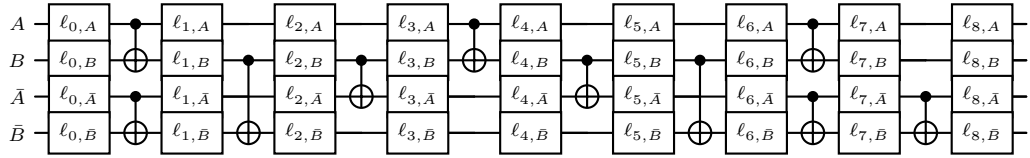

\begin{table}[H]
\centering\scriptsize
\caption{One-qubit gates for c-tUPS on all-to-all connectivity.}
\label{tab:compressed-ata-locals}
\begin{adjustbox}{max width=\linewidth}
\begin{tabular}{cllll}\toprule $k$ & $\ell_{k,A}$ & $\ell_{k,B}$ & $\ell_{k,\bar A}$ & $\ell_{k,\bar B}$\\
\midrule
 0 & $H$ & $H$ & $S^\dagger\,H$ & $H$\\
 1 & $H\,X$ & $H\,Z\,R_y(\Theta)$ & $H\,S^\dagger\,H$ & $H\,Z\,R_x(-2\theta_3)\,S^\dagger$\\
 2 & $I$ & $H\,X\,R_y(-\theta_2/2)\,H$ & $I$ & $I$\\
 3 & $I$ & $Z\,R_y(\theta_2/2)\,H$ & $X\,Z$ & $I$\\
 4 & $S\,H\,Z$ & $H\,S\,R_x(-\theta_2/2)$ & $I$ & $I$\\
 5 & $I$ & $H\,R_z(-\theta_2/2)\,H\,Z$ & $S\,H\,S^\dagger$ & $I$\\
 6 & $I$ & $R_x(-\Theta)\,H\,X$ & $I$ & $H\,S\,X\,Z$\\
 7 & $H\,S$ & $H$ & $H\,R_z(\Theta)\,H\,S\,X$ & $R_x(-\Delta)\,S\,H\,Z$\\
 8 & $I$ & $I$ & $H\,S$ & $S$\\
 
\bottomrule\end{tabular}
\end{adjustbox}\end{table}

\FloatBarrier

\subsubsection{tUPS}

\label{app:ordinary-ata}

\begin{equation}
N_{\mathrm{CNOT}}=14,\qquad D_{2Q}=8,\qquad N_{1Q}=84,\qquad \gamma=-3\pi/4.
\label{eq:ordinary-ata-circuit}
\end{equation}

\begin{figure}[H]
\centering\scriptsize
\begin{adjustbox}{max width=\linewidth}
\begin{quantikz}[row sep={0.48cm,between origins},column sep=0.18cm]
\lstick{$A$} & \gate{\ell_{0,A}} & \ctrl{1} & \gate{\ell_{1,A}} & \qw      & \gate{\ell_{2,A}} & \ctrl{2} & \qw      & \gate{\ell_{3,A}} & \ctrl{1} & \gate{\ell_{4,A}} & \ctrl{2} & \qw      & \gate{\ell_{5,A}} & \qw      & \gate{\ell_{6,A}} & \ctrl{1} & \gate{\ell_{7,A}} & \ctrl{1} & \gate{\ell_{8,A}} & \qw \\
\lstick{$B$} & \gate{\ell_{0,B}} & \targ{}  & \gate{\ell_{1,B}} & \ctrl{1} & \gate{\ell_{2,B}} & \qw      & \ctrl{2} & \gate{\ell_{3,B}} & \targ{}  & \gate{\ell_{4,B}} & \qw      & \ctrl{2} & \gate{\ell_{5,B}} & \ctrl{1} & \gate{\ell_{6,B}} & \targ{}  & \gate{\ell_{7,B}} & \targ{}  & \gate{\ell_{8,B}} & \qw \\
\lstick{$\bar A$} & \gate{\ell_{0,\bar A}} & \ctrl{1} & \gate{\ell_{1,\bar A}} & \targ{}  & \gate{\ell_{2,\bar A}} & \targ{}  & \qw      & \gate{\ell_{3,\bar A}} & \ctrl{1} & \gate{\ell_{4,\bar A}} & \targ{}  & \qw      & \gate{\ell_{5,\bar A}} & \targ{}  & \gate{\ell_{6,\bar A}} & \ctrl{1} & \gate{\ell_{7,\bar A}} & \ctrl{1} & \gate{\ell_{8,\bar A}} & \qw \\
\lstick{$\bar B$} & \gate{\ell_{0,\bar B}} & \targ{}  & \gate{\ell_{1,\bar B}} & \qw      & \gate{\ell_{2,\bar B}} & \qw      & \targ{}  & \gate{\ell_{3,\bar B}} & \targ{}  & \gate{\ell_{4,\bar B}} & \qw      & \targ{}  & \gate{\ell_{5,\bar B}} & \qw      & \gate{\ell_{6,\bar B}} & \targ{}  & \gate{\ell_{7,\bar B}} & \targ{}  & \gate{\ell_{8,\bar B}} & \qw
\end{quantikz}
\end{adjustbox}
\caption{tUPS circuit on all-to-all connectivity. The boxes $\ell_{k,q}$ are specified in Table~\ref{tab:ordinary-ata-locals}.}
\label{fig:ordinary-ata}
\end{figure}
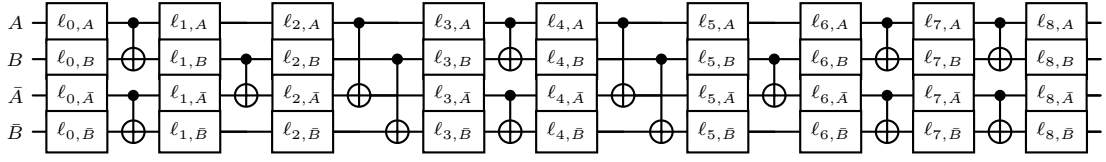

\begin{table}[H]
\centering\scriptsize
\caption{One-qubit gates for tUPS on all-to-all connectivity.}
\label{tab:ordinary-ata-locals}
\begin{adjustbox}{max width=\linewidth}
\begin{tabular}{cllll}\toprule $k$ & $\ell_{k,A}$ & $\ell_{k,B}$ & $\ell_{k,\bar A}$ & $\ell_{k,\bar B}$\\
\midrule
 0 & $S^\dagger\,H$ & $H$ & $I$ & $I$\\
 1 & $H\,S^\dagger$ & $H\,Z\,R_x(-2\theta_3)\,H$ & $Z\,R_y(-2\theta_3)\,S^\dagger$ & $H\,Z$\\
 2 & $I$ & $H\,X\,R_y(-\theta_2/2)\,H$ & $H\,S\,X\,R_x(-\theta_2/2)\,H$ & $I$\\
 3 & $S\,X$ & $H\,X\,R_y(\theta_2/2)$ & $H\,Z\,R_x(-\theta_2/2)$ & $X$\\
 4 & $X$ & $H\,S\,H\,R_z(\theta_2/2)\,H$ & $S\,H\,R_y(\theta_2/2)$ & $H\,S\,H$\\
 5 & $S\,H$ & $H\,R_z(\theta_2/2)\,H\,S\,X$ & $R_y(-\theta_2/2)\,X$ & $H\,S^\dagger\,H$\\
 6 & $I$ & $H\,Z$ & $R_x(\Delta)\,H\,S\,H$ & $I$\\
 7 & $H\,S\,R_z(\Theta)\,H\,S\,X$ & $R_x(-\Delta)\,S\,H\,Z$ & $H\,S\,R_z(\Theta)\,H\,Z$ & $H\,X\,Z$\\
 8 & $H\,S$ & $S$ & $I$ & $S$\\
 
\bottomrule\end{tabular}
\end{adjustbox}\end{table}

\FloatBarrier

\subsection{Square-grid circuits}

\label{app:square}

\begin{figure}[H]
\centering
\begin{tikzpicture}[every node/.style={circle,draw,fill=white,minimum size=6mm,inner sep=1pt},thick]
\node (L) at (0,1.2) {$A$};\node (A) at (1.8,1.2) {$B$};\node (B) at (0,0) {$\bar A$};\node (R) at (1.8,0) {$\bar B$};
\draw (L)--(A)--(R)--(B)--(L);
\end{tikzpicture}
\caption{Square-grid connectivity, with no diagonal connections.}
\label{fig:square-connectivity}
\end{figure}
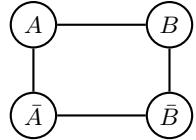

\FloatBarrier

\subsubsection{c-tUPS}

\label{app:compressed-square}

\begin{equation}
N_{\mathrm{CNOT}}=11,\qquad D_{2Q}=8,\qquad N_{1Q}=68,\qquad \gamma=\pi/4.
\label{eq:compressed-square-circuit}
\end{equation}

\begin{figure}[H]
\centering\scriptsize
\begin{adjustbox}{max width=\linewidth}
\begin{quantikz}[row sep={0.48cm,between origins},column sep=0.18cm]
\lstick{$A$}
& \gate{\ell_{0,A}}
& \ctrl{1}
& \gate{\ell_{1,A}}
& \targ{}
& \gate{\ell_{2,A}}
& \ctrl{1}
& \gate{\ell_{3,A}}
& \targ{}
& \gate{\ell_{4,A}}
& \qw
& \gate{\ell_{5,A}}
& \qw
& \gate{\ell_{6,A}}
& \qw
& \gate{\ell_{7,A}}
& \ctrl{1}
& \gate{\ell_{8,A}}
& \qw \\
\lstick{$B$}
& \gate{\ell_{0,B}}
& \targ{}
& \gate{\ell_{1,B}}
& \qw
& \gate{\ell_{2,B}}
& \targ{}
& \gate{\ell_{3,B}}
& \qw
& \gate{\ell_{4,B}}
& \qw
& \gate{\ell_{5,B}}
& \qw
& \gate{\ell_{6,B}}
& \ctrl{2}
& \gate{\ell_{7,B}}
& \targ{}
& \gate{\ell_{8,B}}
& \qw \\
\lstick{$\bar A$}
& \gate{\ell_{0,\bar A}}
& \targ{}
& \gate{\ell_{1,\bar A}}
& \ctrl{-2}
& \gate{\ell_{2,\bar A}}
& \targ{}
& \gate{\ell_{3,\bar A}}
& \ctrl{-2}
& \gate{\ell_{4,\bar A}}
& \targ{}
& \gate{\ell_{5,\bar A}}
& \targ{}
& \gate{\ell_{6,\bar A}}
& \qw
& \gate{\ell_{7,\bar A}}
& \targ{}
& \gate{\ell_{8,\bar A}}
& \qw \\
\lstick{$\bar B$}
& \gate{\ell_{0,\bar B}}
& \ctrl{-1}
& \gate{\ell_{1,\bar B}}
& \qw
& \gate{\ell_{2,\bar B}}
& \ctrl{-1}
& \gate{\ell_{3,\bar B}}
& \qw
& \gate{\ell_{4,\bar B}}
& \ctrl{-1}
& \gate{\ell_{5,\bar B}}
& \ctrl{-1}
& \gate{\ell_{6,\bar B}}
& \targ{}
& \gate{\ell_{7,\bar B}}
& \ctrl{-1}
& \gate{\ell_{8,\bar B}}
& \qw
\end{quantikz}
\end{adjustbox}
\caption{c-tUPS circuit on the square grid. The boxes $\ell_{k,q}$ are specified in Table~\ref{tab:compressed-square-locals}.}
\label{fig:compressed-square}
\end{figure}
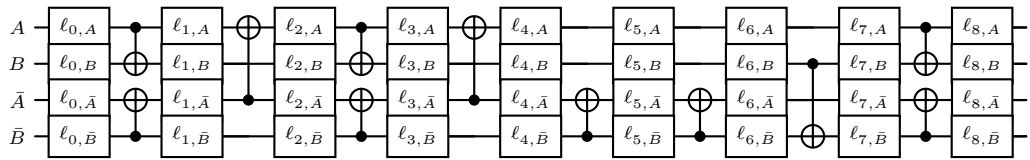

\begin{table}[H]
\centering\scriptsize
\caption{One-qubit gates for c-tUPS on the square grid.}
\label{tab:compressed-square-locals}
\begin{adjustbox}{max width=\linewidth}
\begin{tabular}{cllll} \toprule $k$ & $\ell_{k,A}$ & $\ell_{k,B}$ & $\ell_{k,\bar A}$ & $\ell_{k,\bar B}$\\
 \midrule
 0 & $S^\dagger\,H$ & $S^\dagger$ & $H$ & $H$\\
 1 & $S\,H\,Z\,R_z(\Theta)\,H$ & $H\,S\,X\,R_x(-\Theta)$ & $H\,Z\,R_y(-2\theta_3)\,S^\dagger\,H$ & $H\,X$\\
 2 & $X$ & $I$ & $S\,X\,R_x(\theta_2/2)\,H$ & $I$\\
 3 & $H\,S\,H$ & $H\,S^\dagger$ & $H\,S\,H\,R_y(-\theta_2/2)\,H$ & $S^\dagger$\\
 4 & $H\,S\,H$ & $I$ & $H\,R_y(\theta_2/2)\,H$ & $I$\\
 5 & $I$ & $I$ & $S\,R_x(\theta_2/2)\,S^\dagger\,H$ & $H\,S\,H$\\
 6 & $I$ & $I$ & $H\,R_z(\Theta)\,S^\dagger\,H\,S$ & $H\,S\,X$\\
 7 & $I$ & $X$ & $I$ & $R_x(-\Delta)\,S\,X$\\
 8 & $H\,S$ & $H$ & $S$ & $H\,S$\\
 
\bottomrule \end{tabular}
\end{adjustbox}
\end{table}

\FloatBarrier

\subsubsection{tUPS}

\label{app:ordinary-square}

\begin{equation}
N_{\mathrm{CNOT}}=15,\qquad D_{2Q}=9,\qquad N_{1Q}=90,\qquad \gamma=\pi/4.
\label{eq:ordinary-square-circuit}
\end{equation}

\begin{figure}[H]
\centering\scriptsize
\begin{adjustbox}{max width=\linewidth}
\begin{quantikz}[row sep={0.48cm,between origins},column sep=0.18cm]
\lstick{$A$}
& \gate{\ell_{0,A}}
& \ctrl{1}
& \gate{\ell_{1,A}}
& \qw
& \gate{\ell_{2,A}}
& \ctrl{1}
& \gate{\ell_{3,A}}
& \targ{} & \qw
& \gate{\ell_{4,A}}
& \ctrl{1}
& \gate{\ell_{5,A}}
& \targ{}
& \gate{\ell_{6,A}}
& \ctrl{1}
& \gate{\ell_{7,A}}
& \qw
& \gate{\ell_{8,A}}
& \ctrl{1}
& \gate{\ell_{9,A}}
& \qw \\
\lstick{$B$}
& \gate{\ell_{0,B}}
& \targ{}
& \gate{\ell_{1,B}}
& \ctrl{2}
& \gate{\ell_{2,B}}
& \targ{}
& \gate{\ell_{3,B}}
& \qw & \ctrl{2}
& \gate{\ell_{4,B}}
& \targ{}
& \gate{\ell_{5,B}}
& \qw
& \gate{\ell_{6,B}}
& \targ{}
& \gate{\ell_{7,B}}
& \ctrl{2}
& \gate{\ell_{8,B}}
& \targ{}
& \gate{\ell_{9,B}}
& \qw \\
\lstick{$\bar A$}
& \gate{\ell_{0,\bar A}}
& \targ{}
& \gate{\ell_{1,\bar A}}
& \qw
& \gate{\ell_{2,\bar A}}
& \targ{}
& \gate{\ell_{3,\bar A}}
& \ctrl{-2} & \qw
& \gate{\ell_{4,\bar A}}
& \targ{}
& \gate{\ell_{5,\bar A}}
& \ctrl{-2}
& \gate{\ell_{6,\bar A}}
& \targ{}
& \gate{\ell_{7,\bar A}}
& \qw
& \gate{\ell_{8,\bar A}}
& \targ{}
& \gate{\ell_{9,\bar A}}
& \qw \\
\lstick{$\bar B$}
& \gate{\ell_{0,\bar B}}
& \ctrl{-1}
& \gate{\ell_{1,\bar B}}
& \targ{}
& \gate{\ell_{2,\bar B}}
& \ctrl{-1}
& \gate{\ell_{3,\bar B}}
& \qw & \targ{}
& \gate{\ell_{4,\bar B}}
& \ctrl{-1}
& \gate{\ell_{5,\bar B}}
& \qw
& \gate{\ell_{6,\bar B}}
& \ctrl{-1}
& \gate{\ell_{7,\bar B}}
& \targ{}
& \gate{\ell_{8,\bar B}}
& \ctrl{-1}
& \gate{\ell_{9,\bar B}}
& \qw
\end{quantikz}
\end{adjustbox}
\caption{tUPS circuit on the square grid. The boxes $\ell_{k,q}$ are specified in Table~\ref{tab:ordinary-square-locals}.}
\label{fig:ordinary-square}
\end{figure}
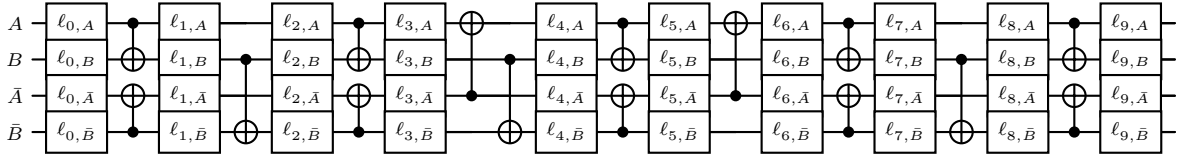

\begin{table}[H]
\centering\scriptsize
\caption{One-qubit gates for tUPS on the square grid.}
\label{tab:ordinary-square-locals}
\begin{adjustbox}{max width=\linewidth}
\begin{tabular}{cllll} \toprule $k$ & $\ell_{k,A}$ & $\ell_{k,B}$ & $\ell_{k,\bar A}$ & $\ell_{k,\bar B}$\\
 \midrule
 0 & $S^\dagger\,H$ & $H$ & $S^\dagger$ & $S^\dagger\,H$\\
 1 & $H\,S^\dagger$ & $H\,Z\,R_x(-2\theta_3)\,H$ & $H\,S\,X$ & $S\,X\,R_x(-2\theta_3)$\\
 2 & $I$ & $H\,X\,R_y(-\theta_2/2)\,H$ & $I$ & $X$\\
 3 & $H$ & $S^\dagger\,H\,R_z(-\theta_2/2)\,H$ & $H\,S\,R_y(\theta_2/2)$ & $I$\\
 4 & $H\,S^\dagger$ & $H\,X\,R_z(-\theta_2/2)\,H\,S^\dagger$ & $Z\,R_y(\theta_2/2)\,X\,Z$ & $H\,S$\\
 5 & $S\,H\,Z$ & $H\,R_y(\theta_2/2)\,H\,S^\dagger\,H$ & $S\,R_x(\theta_2/2)\,Z$ & $H\,Z$\\
 6 & $S\,H\,S^\dagger$ & $I$ & $R_x(-\theta_2/2)\,S\,H\,Z$ & $I$\\
 7 & $H\,R_z(\Theta)\,H\,S\,X$ & $H\,Z$ & $S\,R_y(-\Delta)\,X$ & $S^\dagger\,H\,S$\\
 8 & $I$ & $H\,R_x(-\Delta)\,S\,H\,Z$ & $I$ & $S\,R_y(\Theta)\,H\,Z$\\
 9 & $H\,S$ & $S$ & $I$ & $H$\\
 
\bottomrule \end{tabular}
\end{adjustbox}
\end{table}

\FloatBarrier

\subsection{Linear nearest-neighbor circuits}
\label{app:lnn}

\begin{figure}[H]
\centering
\begin{tikzpicture}[every node/.style={circle,draw,fill=white,minimum size=6mm,inner sep=1pt},thick]
\node (L) at (0,0) {$A$};\node (A) at (1.4,0) {$B$};\node (B) at (2.8,0) {$\bar A$};\node (R) at (4.2,0) {$\bar B$};
\draw (L)--(A)--(B)--(R);
\end{tikzpicture}
\caption{Linear nearest-neighbor connectivity in the order $A,B,\bar A,\bar B$.}
\label{fig:lnn-connectivity}
\end{figure}
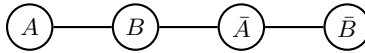

\FloatBarrier

\subsubsection{c-tUPS}

\label{app:compressed-lnn}

\begin{equation}
N_{\mathrm{CNOT}}=12,\qquad D_{2Q}=9,\qquad N_{1Q}=57,\qquad \gamma=-3\pi/4.
\label{eq:compressed-lnn-circuit}
\end{equation}

\begin{figure}[H]
\centering\scriptsize
\begin{adjustbox}{max width=\linewidth}
\begin{quantikz}[row sep={0.48cm,between origins},column sep=0.18cm]
\lstick{$A$} & \gate{\ell_{0,A}} & \ctrl{1} & \gate{\ell_{1,A}} & \qw      & \gate{\ell_{2,A}} & \ctrl{1} & \gate{\ell_{3,A}} & \qw      & \gate{\ell_{4,A}} & \qw      & \gate{\ell_{5,A}} & \qw      & \gate{\ell_{6,A}} & \ctrl{1} & \gate{\ell_{7,A}} & \qw      & \gate{\ell_{8,A}} & \ctrl{1} & \gate{\ell_{9,A}} & \qw \\
\lstick{$B$} & \gate{\ell_{0,B}} & \targ{}  & \gate{\ell_{1,B}} & \ctrl{1} & \gate{\ell_{2,B}} & \targ{}  & \gate{\ell_{3,B}} & \ctrl{1} & \gate{\ell_{4,B}} & \qw      & \gate{\ell_{5,B}} & \ctrl{1} & \gate{\ell_{6,B}} & \targ{}  & \gate{\ell_{7,B}} & \ctrl{1} & \gate{\ell_{8,B}} & \targ{}  & \gate{\ell_{9,B}} & \qw \\
\lstick{$\bar A$} & \gate{\ell_{0,\bar A}} & \ctrl{1} & \gate{\ell_{1,\bar A}} & \targ{}  & \gate{\ell_{2,\bar A}} & \qw      & \gate{\ell_{3,\bar A}} & \targ{}  & \gate{\ell_{4,\bar A}} & \ctrl{1} & \gate{\ell_{5,\bar A}} & \targ{}  & \gate{\ell_{6,\bar A}} & \ctrl{1} & \gate{\ell_{7,\bar A}} & \targ{}  & \gate{\ell_{8,\bar A}} & \ctrl{1} & \gate{\ell_{9,\bar A}} & \qw \\
\lstick{$\bar B$} & \gate{\ell_{0,\bar B}} & \targ{}  & \gate{\ell_{1,\bar B}} & \qw      & \gate{\ell_{2,\bar B}} & \qw      & \gate{\ell_{3,\bar B}} & \qw      & \gate{\ell_{4,\bar B}} & \targ{}  & \gate{\ell_{5,\bar B}} & \qw      & \gate{\ell_{6,\bar B}} & \targ{}  & \gate{\ell_{7,\bar B}} & \qw      & \gate{\ell_{8,\bar B}} & \targ{}  & \gate{\ell_{9,\bar B}} & \qw
\end{quantikz}
\end{adjustbox}
\caption{c-tUPS circuit on LNN connectivity. The boxes $\ell_{k,q}$ are specified in Table~\ref{tab:compressed-lnn-locals}.}
\label{fig:compressed-lnn}
\end{figure}
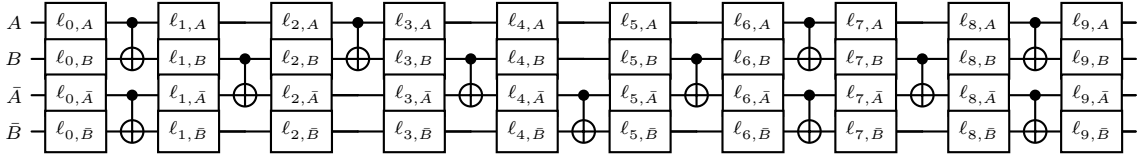

\begin{table}[H]
\centering\scriptsize
\caption{One-qubit gates for c-tUPS on LNN connectivity.}
\label{tab:compressed-lnn-locals}
\begin{adjustbox}{max width=\linewidth}
\begin{tabular}{cllll}\toprule $k$ & $\ell_{k,A}$ & $\ell_{k,B}$ & $\ell_{k,\bar A}$ & $\ell_{k,\bar B}$\\
\midrule
 0 & $S^\dagger\,H$ & $S^\dagger$ & $I$ & $I$\\
 1 & $S\,H\,Z\,R_z(\Theta)$ & $H\,S\,X\,R_x(-\Theta)\,H$ & $Z\,R_y(-2\theta_3)\,S^\dagger$ & $H\,Z$\\
 2 & $I$ & $Z$ & $H\,S\,X\,R_x(-\theta_2/2)\,H$ & $I$\\
 3 & $I$ & $S^\dagger$ & $I$ & $I$\\
 4 & $I$ & $I$ & $H\,Z\,R_x(-\theta_2/2)$ & $I$\\
 5 & $I$ & $I$ & $S\,H\,R_y(\theta_2/2)\,X$ & $S^\dagger$\\
 6 & $I$ & $S$ & $R_y(\theta_2/2)\,H\,Z$ & $I$\\
 7 & $H\,S\,X$ & $Z$ & $S\,H\,Z$ & $H\,R_y(\Theta)\,X$\\
 8 & $I$ & $S^\dagger\,H\,S$ & $S\,R_y(-\Delta)\,H\,Z$ & $I$\\
 9 & $H\,S$ & $S$ & $H$ & $I$\\
 
\bottomrule\end{tabular}
\end{adjustbox}\end{table}

\FloatBarrier

\subsubsection{tUPS}

\label{app:ordinary-lnn}

\begin{equation}
N_{\mathrm{CNOT}}=16,\qquad D_{2Q}=10,\qquad N_{1Q}=94,\qquad \gamma=3\pi/4.
\label{eq:ordinary-lnn-circuit}
\end{equation}

\begin{figure}[H]
\centering\scriptsize
\begin{adjustbox}{max width=\linewidth}
\begin{quantikz}[row sep={0.48cm,between origins},column sep=0.18cm]
\lstick{$A$} & \gate{\ell_{0,A}} & \ctrl{1} & \gate{\ell_{1,A}} & \qw      & \gate{\ell_{2,A}} & \ctrl{1} & \gate{\ell_{3,A}} & \qw      & \gate{\ell_{4,A}} & \ctrl{1} & \gate{\ell_{5,A}} & \qw      & \gate{\ell_{6,A}} & \ctrl{1} & \gate{\ell_{7,A}} & \ctrl{1} & \gate{\ell_{8,A}} & \qw      & \gate{\ell_{9,A}} & \ctrl{1} & \gate{\ell_{10,A}} & \qw \\
\lstick{$B$} & \gate{\ell_{0,B}} & \targ{}  & \gate{\ell_{1,B}} & \ctrl{1} & \gate{\ell_{2,B}} & \targ{}  & \gate{\ell_{3,B}} & \ctrl{1} & \gate{\ell_{4,B}} & \targ{}  & \gate{\ell_{5,B}} & \ctrl{1} & \gate{\ell_{6,B}} & \targ{}  & \gate{\ell_{7,B}} & \targ{}  & \gate{\ell_{8,B}} & \ctrl{1} & \gate{\ell_{9,B}} & \targ{}  & \gate{\ell_{10,B}} & \qw \\
\lstick{$\bar A$} & \gate{\ell_{0,\bar A}} & \ctrl{1} & \gate{\ell_{1,\bar A}} & \targ{}  & \gate{\ell_{2,\bar A}} & \ctrl{1} & \gate{\ell_{3,\bar A}} & \targ{}  & \gate{\ell_{4,\bar A}} & \ctrl{1} & \gate{\ell_{5,\bar A}} & \targ{}  & \gate{\ell_{6,\bar A}} & \ctrl{1} & \gate{\ell_{7,\bar A}} & \ctrl{1} & \gate{\ell_{8,\bar A}} & \targ{}  & \gate{\ell_{9,\bar A}} & \ctrl{1} & \gate{\ell_{10,\bar A}} & \qw \\
\lstick{$\bar B$} & \gate{\ell_{0,\bar B}} & \targ{}  & \gate{\ell_{1,\bar B}} & \qw      & \gate{\ell_{2,\bar B}} & \targ{}  & \gate{\ell_{3,\bar B}} & \qw      & \gate{\ell_{4,\bar B}} & \targ{}  & \gate{\ell_{5,\bar B}} & \qw      & \gate{\ell_{6,\bar B}} & \targ{}  & \gate{\ell_{7,\bar B}} & \targ{}  & \gate{\ell_{8,\bar B}} & \qw      & \gate{\ell_{9,\bar B}} & \targ{}  & \gate{\ell_{10,\bar B}} & \qw
\end{quantikz}
\end{adjustbox}
\caption{tUPS circuit on LNN connectivity. The boxes $\ell_{k,q}$ are specified in Table~\ref{tab:ordinary-lnn-locals}.}
\label{fig:ordinary-lnn}
\end{figure}
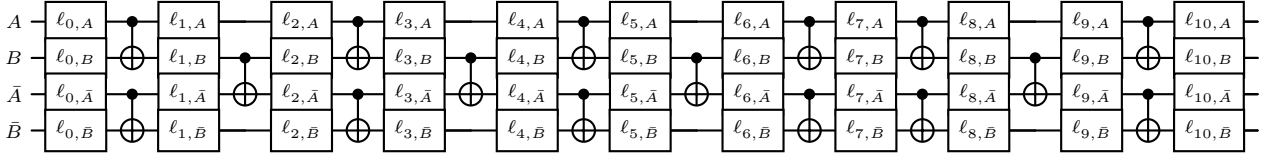

\begin{table}[H]
\centering\scriptsize
\caption{One-qubit gates for tUPS on LNN connectivity.}
\label{tab:ordinary-lnn-locals}
\begin{adjustbox}{max width=\linewidth}
\begin{tabular}{cllll}\toprule $k$ & $\ell_{k,A}$ & $\ell_{k,B}$ & $\ell_{k,\bar A}$ & $\ell_{k,\bar B}$\\
\midrule
 0 & $S^\dagger\,H$ & $H$ & $I$ & $I$\\
 1 & $H\,S^\dagger$ & $H\,Z\,R_x(-2\theta_3)\,H$ & $Z\,R_y(-2\theta_3)\,S^\dagger$ & $H\,Z$\\
 2 & $I$ & $H\,X\,R_y(-\theta_2/2)$ & $H\,S\,X\,R_x(-\theta_2/2)$ & $I$\\
 3 & $S\,X$ & $H\,S\,H\,R_z(\theta_2/2)\,H$ & $S\,H\,R_y(\theta_2/2)$ & $H\,S\,H$\\
 4 & $I$ & $H\,S^\dagger\,H$ & $X$ & $I$\\
 5 & $S$ & $H\,S^\dagger\,H$ & $H\,S\,H$ & $S\,X$\\
 6 & $I$ & $H\,R_z(\theta_2/2)\,H\,S$ & $Z$ & $I$\\
 7 & $S^\dagger$ & $H\,R_y(-\theta_2/2)\,R_x(\Delta)\,S\,H\,Z$ & $H\,R_x(\theta_2/2)\,S\,H$ & $H\,R_y(\theta_2/2)\,S\,H\,Z$\\
 8 & $H\,Z$ & $H\,S\,H$ & $X$ & $S\,R_x(-\Delta)\,S\,H\,Z$\\
 9 & $I$ & $H\,S\,R_y(\Theta)\,H\,X$ & $H\,R_z(\Theta)\,H\,S\,X$ & $I$\\
 10 & $H$ & $H$ & $H\,S$ & $S$\\
 
\bottomrule\end{tabular}
\end{adjustbox}\end{table}

\FloatBarrier
\subsection{Linear nearest-neighbor circuits for consecutive orbital pairs}
\label{app:lnn-orbital}

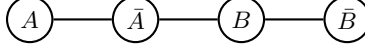
\begin{figure}[H]
\centering
\begin{tikzpicture}[every node/.style={circle,draw,fill=white,minimum size=6mm,inner sep=1pt},thick]
\node (L) at (0,0) {$A$};
\node (B) at (1.4,0) {$\bar A$};
\node (A) at (2.8,0) {$B$};
\node (R) at (4.2,0) {$\bar B$};
\draw (L)--(B)--(A)--(R);
\end{tikzpicture}
\caption{LNN connectivity with the two spin modes of each spatial
orbital adjacent: $(A,\bar A,B,\bar B)=(p_\alpha,p_\beta,q_\alpha,q_\beta)$.}
\label{fig:lnn-orbital-connectivity}
\end{figure}

\FloatBarrier
\subsubsection{c-tUPS}
\label{app:ctups-lnn-orbital}
\begin{equation}
N_{\mathrm{CNOT}}=16,\qquad D_{2Q}=13,\qquad N_{1Q}=77,\qquad\gamma=-5\pi/4.
\label{eq:ctups-lnn-orbital-circuit}
\end{equation}
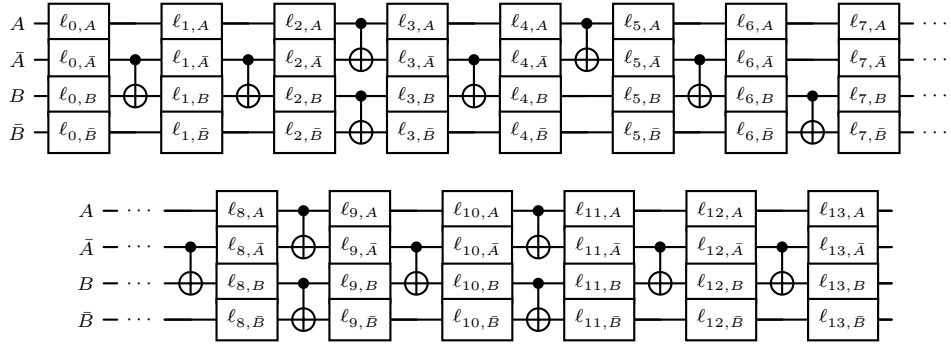
\begin{figure}[H]
\centering\scriptsize
\begin{adjustbox}{max width=\linewidth}
\begin{quantikz}[row sep={0.48cm,between origins},column sep=0.18cm]
\lstick{$A$} & \gate{\ell_{0,A}} & \qw & \gate{\ell_{1,A}} & \qw & \gate{\ell_{2,A}} & \ctrl{1} & \gate{\ell_{3,A}} & \qw & \gate{\ell_{4,A}} & \ctrl{1} & \gate{\ell_{5,A}} & \qw & \gate{\ell_{6,A}} & \qw & \gate{\ell_{7,A}} & \gate[style={draw=none,fill=none}]{\cdots} \\
\lstick{$\bar A$} & \gate{\ell_{0,\bar A}} & \ctrl{1} & \gate{\ell_{1,\bar A}} & \ctrl{1} & \gate{\ell_{2,\bar A}} & \targ{} & \gate{\ell_{3,\bar A}} & \ctrl{1} & \gate{\ell_{4,\bar A}} & \targ{} & \gate{\ell_{5,\bar A}} & \ctrl{1} & \gate{\ell_{6,\bar A}} & \qw & \gate{\ell_{7,\bar A}} & \gate[style={draw=none,fill=none}]{\cdots} \\
\lstick{$B$} & \gate{\ell_{0,B}} & \targ{} & \gate{\ell_{1,B}} & \targ{} & \gate{\ell_{2,B}} & \ctrl{1} & \gate{\ell_{3,B}} & \targ{} & \gate{\ell_{4,B}} & \qw & \gate{\ell_{5,B}} & \targ{} & \gate{\ell_{6,B}} & \ctrl{1} & \gate{\ell_{7,B}} & \gate[style={draw=none,fill=none}]{\cdots} \\
\lstick{$\bar B$} & \gate{\ell_{0,\bar B}} & \qw & \gate{\ell_{1,\bar B}} & \qw & \gate{\ell_{2,\bar B}} & \targ{} & \gate{\ell_{3,\bar B}} & \qw & \gate{\ell_{4,\bar B}} & \qw & \gate{\ell_{5,\bar B}} & \qw & \gate{\ell_{6,\bar B}} & \targ{} & \gate{\ell_{7,\bar B}} & \gate[style={draw=none,fill=none}]{\cdots}
\end{quantikz}
\end{adjustbox}
\par\vspace{8pt}
\begin{adjustbox}{max width=\linewidth}
\begin{quantikz}[row sep={0.48cm,between origins},column sep=0.18cm]
\lstick{$A$} & \gate[style={draw=none,fill=none}]{\cdots} & \qw & \gate{\ell_{8,A}} & \ctrl{1} & \gate{\ell_{9,A}} & \qw & \gate{\ell_{10,A}} & \ctrl{1} & \gate{\ell_{11,A}} & \qw & \gate{\ell_{12,A}} & \qw & \gate{\ell_{13,A}} & \qw \\
\lstick{$\bar A$} & \gate[style={draw=none,fill=none}]{\cdots} & \ctrl{1} & \gate{\ell_{8,\bar A}} & \targ{} & \gate{\ell_{9,\bar A}} & \ctrl{1} & \gate{\ell_{10,\bar A}} & \targ{} & \gate{\ell_{11,\bar A}} & \ctrl{1} & \gate{\ell_{12,\bar A}} & \ctrl{1} & \gate{\ell_{13,\bar A}} & \qw \\
\lstick{$B$} & \gate[style={draw=none,fill=none}]{\cdots} & \targ{} & \gate{\ell_{8,B}} & \ctrl{1} & \gate{\ell_{9,B}} & \targ{} & \gate{\ell_{10,B}} & \ctrl{1} & \gate{\ell_{11,B}} & \targ{} & \gate{\ell_{12,B}} & \targ{} & \gate{\ell_{13,B}} & \qw \\
\lstick{$\bar B$} & \gate[style={draw=none,fill=none}]{\cdots} & \qw & \gate{\ell_{8,\bar B}} & \targ{} & \gate{\ell_{9,\bar B}} & \qw & \gate{\ell_{10,\bar B}} & \targ{} & \gate{\ell_{11,\bar B}} & \qw & \gate{\ell_{12,\bar B}} & \qw & \gate{\ell_{13,\bar B}} & \qw
\end{quantikz}
\end{adjustbox}
\caption{c-tUPS circuit on the $A-\bar A-B-\bar B$ line in interleaved Jordan--Wigner order. The two panels are consecutive parts of one circuit. One-qubit gates are listed in Table~\ref{tab:ctups-lnn-orbital-locals}.}
\label{fig:ctups-lnn-orbital}
\end{figure}

\begin{table}[H]
\centering\scriptsize
\caption{One-qubit gates for c-tUPS on the $A-\bar A-B-\bar B$ line. Gates in each cell act from left to right.}
\label{tab:ctups-lnn-orbital-locals}
\begin{adjustbox}{max width=\linewidth}
\begin{tabular}{cllll}\toprule
$k$ & $\ell_{k,A}$ & $\ell_{k,\bar A}$ & $\ell_{k,B}$ & $\ell_{k,\bar B}$\\\midrule
0 & $S^\dagger\,H$ & $H\,S$ & $H\,S$ & $I$\\
1 & $I$ & $S^\dagger\,H$ & $H\,S$ & $I$\\
2 & $I$ & $H\,S\,S^\dagger$ & $I$ & $I$\\
3 & $S\,H\,Z\,R_z(\Theta)$ & $H\,S\,X\,R_x(-\Theta)\,H$ & $Z\,R_y(-2\theta_3)\,S^\dagger$ & $H\,Z$\\
4 & $I$ & $Z$ & $H\,S\,X\,R_x(-\theta_2/2)\,H$ & $I$\\
5 & $I$ & $S^\dagger$ & $I$ & $I$\\
6 & $I$ & $I$ & $H\,Z\,R_x(-\theta_2/2)$ & $I$\\
7 & $I$ & $I$ & $S\,H\,R_y(\theta_2/2)\,X$ & $S^\dagger$\\
8 & $I$ & $S$ & $R_y(\theta_2/2)\,H\,Z$ & $I$\\
9 & $H\,S\,X$ & $Z$ & $S\,H\,Z$ & $H\,R_y(\Theta)\,X$\\
10 & $I$ & $S^\dagger\,H\,S$ & $S\,R_y(-\Delta)\,H\,Z$ & $I$\\
11 & $H\,S$ & $S\,H\,S$ & $H\,H\,S$ & $I$\\
12 & $I$ & $S^\dagger\,H$ & $H\,S$ & $I$\\
13 & $I$ & $H\,S$ & $I$ & $I$\\
\bottomrule\end{tabular}
\end{adjustbox}
\end{table}
\FloatBarrier
\subsubsection{tUPS}
\label{app:tups-lnn-orbital}
\begin{equation}
N_{\mathrm{CNOT}}=20,\qquad D_{2Q}=14,\qquad N_{1Q}=114,\qquad\gamma=\pi/4.
\label{eq:tups-lnn-orbital-circuit}
\end{equation}
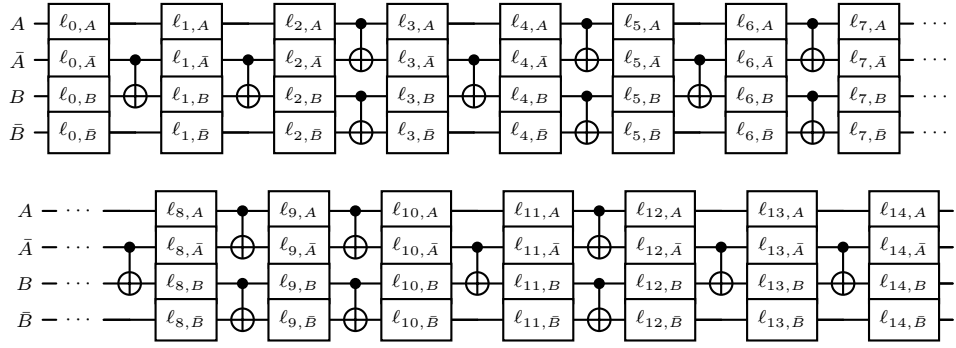
\begin{figure}[H]
\centering\scriptsize
\begin{adjustbox}{max width=\linewidth}
\begin{quantikz}[row sep={0.48cm,between origins},column sep=0.18cm]
\lstick{$A$} & \gate{\ell_{0,A}} & \qw & \gate{\ell_{1,A}} & \qw & \gate{\ell_{2,A}} & \ctrl{1} & \gate{\ell_{3,A}} & \qw & \gate{\ell_{4,A}} & \ctrl{1} & \gate{\ell_{5,A}} & \qw & \gate{\ell_{6,A}} & \ctrl{1} & \gate{\ell_{7,A}} & \gate[style={draw=none,fill=none}]{\cdots} \\
\lstick{$\bar A$} & \gate{\ell_{0,\bar A}} & \ctrl{1} & \gate{\ell_{1,\bar A}} & \ctrl{1} & \gate{\ell_{2,\bar A}} & \targ{} & \gate{\ell_{3,\bar A}} & \ctrl{1} & \gate{\ell_{4,\bar A}} & \targ{} & \gate{\ell_{5,\bar A}} & \ctrl{1} & \gate{\ell_{6,\bar A}} & \targ{} & \gate{\ell_{7,\bar A}} & \gate[style={draw=none,fill=none}]{\cdots} \\
\lstick{$B$} & \gate{\ell_{0,B}} & \targ{} & \gate{\ell_{1,B}} & \targ{} & \gate{\ell_{2,B}} & \ctrl{1} & \gate{\ell_{3,B}} & \targ{} & \gate{\ell_{4,B}} & \ctrl{1} & \gate{\ell_{5,B}} & \targ{} & \gate{\ell_{6,B}} & \ctrl{1} & \gate{\ell_{7,B}} & \gate[style={draw=none,fill=none}]{\cdots} \\
\lstick{$\bar B$} & \gate{\ell_{0,\bar B}} & \qw & \gate{\ell_{1,\bar B}} & \qw & \gate{\ell_{2,\bar B}} & \targ{} & \gate{\ell_{3,\bar B}} & \qw & \gate{\ell_{4,\bar B}} & \targ{} & \gate{\ell_{5,\bar B}} & \qw & \gate{\ell_{6,\bar B}} & \targ{} & \gate{\ell_{7,\bar B}} & \gate[style={draw=none,fill=none}]{\cdots}
\end{quantikz}
\end{adjustbox}
\par\vspace{8pt}
\begin{adjustbox}{max width=\linewidth}
\begin{quantikz}[row sep={0.48cm,between origins},column sep=0.18cm]
\lstick{$A$} & \gate[style={draw=none,fill=none}]{\cdots} & \qw & \gate{\ell_{8,A}} & \ctrl{1} & \gate{\ell_{9,A}} & \ctrl{1} & \gate{\ell_{10,A}} & \qw & \gate{\ell_{11,A}} & \ctrl{1} & \gate{\ell_{12,A}} & \qw & \gate{\ell_{13,A}} & \qw & \gate{\ell_{14,A}} & \qw \\
\lstick{$\bar A$} & \gate[style={draw=none,fill=none}]{\cdots} & \ctrl{1} & \gate{\ell_{8,\bar A}} & \targ{} & \gate{\ell_{9,\bar A}} & \targ{} & \gate{\ell_{10,\bar A}} & \ctrl{1} & \gate{\ell_{11,\bar A}} & \targ{} & \gate{\ell_{12,\bar A}} & \ctrl{1} & \gate{\ell_{13,\bar A}} & \ctrl{1} & \gate{\ell_{14,\bar A}} & \qw \\
\lstick{$B$} & \gate[style={draw=none,fill=none}]{\cdots} & \targ{} & \gate{\ell_{8,B}} & \ctrl{1} & \gate{\ell_{9,B}} & \ctrl{1} & \gate{\ell_{10,B}} & \targ{} & \gate{\ell_{11,B}} & \ctrl{1} & \gate{\ell_{12,B}} & \targ{} & \gate{\ell_{13,B}} & \targ{} & \gate{\ell_{14,B}} & \qw \\
\lstick{$\bar B$} & \gate[style={draw=none,fill=none}]{\cdots} & \qw & \gate{\ell_{8,\bar B}} & \targ{} & \gate{\ell_{9,\bar B}} & \targ{} & \gate{\ell_{10,\bar B}} & \qw & \gate{\ell_{11,\bar B}} & \targ{} & \gate{\ell_{12,\bar B}} & \qw & \gate{\ell_{13,\bar B}} & \qw & \gate{\ell_{14,\bar B}} & \qw
\end{quantikz}
\end{adjustbox}
\caption{tUPS circuit on the $A-\bar A-B-\bar B$ line in interleaved Jordan--Wigner order. The two panels are consecutive parts of one circuit. One-qubit gates are listed in Table~\ref{tab:tups-lnn-orbital-locals}.}
\label{fig:tups-lnn-orbital}
\end{figure}

\begin{table}[H]
\centering\scriptsize
\caption{One-qubit gates for tUPS on the $A-\bar A-B-\bar B$ line. Gates in each cell act from left to right.}
\label{tab:tups-lnn-orbital-locals}
\begin{adjustbox}{max width=\linewidth}
\begin{tabular}{cllll}\toprule
$k$ & $\ell_{k,A}$ & $\ell_{k,\bar A}$ & $\ell_{k,B}$ & $\ell_{k,\bar B}$\\\midrule
0 & $S^\dagger\,H$ & $H\,S$ & $H\,S$ & $I$\\
1 & $I$ & $S^\dagger\,H$ & $H\,S$ & $I$\\
2 & $I$ & $H\,S\,H$ & $I$ & $I$\\
3 & $H\,S^\dagger$ & $H\,Z\,R_x(-2\theta_3)\,H$ & $Z\,R_y(-2\theta_3)\,S^\dagger$ & $H\,Z$\\
4 & $I$ & $H\,X\,R_y(-\theta_2/2)$ & $H\,S\,X\,R_x(-\theta_2/2)$ & $I$\\
5 & $S\,X$ & $H\,S\,H\,R_z(\theta_2/2)\,H$ & $S\,H\,R_y(\theta_2/2)$ & $H\,S\,H$\\
6 & $I$ & $H\,S^\dagger\,H$ & $X$ & $I$\\
7 & $S$ & $H\,S^\dagger\,H$ & $H\,S\,H$ & $S\,X$\\
8 & $I$ & $H\,R_z(\theta_2/2)\,H\,S$ & $Z$ & $I$\\
9 & $S^\dagger$ & $H\,R_y(-\theta_2/2)\,R_x(\Delta)\,S\,H\,Z$ & $H\,R_x(\theta_2/2)\,S\,H$ & $H\,R_y(\theta_2/2)\,S\,H\,Z$\\
10 & $H\,Z$ & $H\,S\,H$ & $X$ & $S\,R_x(-\Delta)\,S\,H\,Z$\\
11 & $I$ & $H\,S\,R_y(\Theta)\,H\,X$ & $H\,R_z(\Theta)\,H\,S\,X$ & $I$\\
12 & $H$ & $H\,H\,S$ & $H\,S\,H\,S$ & $S$\\
13 & $I$ & $S^\dagger\,H$ & $H\,S$ & $I$\\
14 & $I$ & $H\,S$ & $I$ & $I$\\
\bottomrule\end{tabular}
\end{adjustbox}
\end{table}
\FloatBarrier

\FloatBarrier
\section{Ansatz variants and Majorana-propagation cost}
\label{app:  C majorana-block-analysis}

This appendix defines additional tUPS variants and
establishes the spin-exchange equivalence of cp-tUPS
and cm-tUPS for exact and blockwise MP energy evaluations.
We then compare the variants' MP energy accuracy during
VIME optimization and assess the compilation and evaluation
costs of tUPS and c-tUPS under blockwise and standard
truncation.

\subsection{Ansatz variants}
\label{app:masks-catalogue}

All variants use the tiling, reference state, and orbital
dressing of Sec. \ref{subsec: ctUPS} They differ only in
the two-orbital circuit applied at each tile. For spatial
orbitals $p$ and $q$, define
\begin{equation}
\begin{aligned}
\hat E_{pq}
&=\sum_{\sigma\in\{\alpha,\beta\}}
  \hat a^\dagger_{p\sigma}\hat a_{q\sigma},
&
\hat\kappa^{(1)}_{pq}
&=\hat E_{pq}-\hat E_{qp},
\\
\hat P
&=\hat a^\dagger_{p\alpha}\hat a^\dagger_{p\beta}
  \hat a_{q\beta}\hat a_{q\alpha},
&
\hat K_P
&=\hat P-\hat P^\dagger,
\\
\hat F
&=\hat a^\dagger_{p\alpha}\hat a^\dagger_{q\beta}
  \hat a_{p\beta}\hat a_{q\alpha},
&
\hat K_F
&=\hat F-\hat F^\dagger.
\end{aligned}
\end{equation}
Here $\hat K_P=\hat\kappa^{(2)}_{pq}/2$ generates pair
transfer and $\hat K_F$ generates opposite-spin exchange.
We write
\begin{equation}
\hat R_1(\theta)=e^{\theta\hat\kappa^{(1)}_{pq}},
\qquad
\hat R_P(\theta)
=e^{\theta\hat\kappa^{(2)}_{pq}}
=e^{2\theta\hat K_P}.
\end{equation}

Table~\ref{tab: C catalogue} lists the seven ansatz variants,
their two-orbital circuits, and the number of Majorana
rotations per tile. We use the convention of
Sec.~\ref{subsec: mp}: a Majorana rotation is
$e^{-i\theta\hat M_{\vec b}/2}$, where
$\hat M_{\vec b}$ is a Hermitian Majorana monomial
of length $w=\|\vec b\|_1$.

The tUPS circuit follows Ref.~\cite{burton2024tups}.
The reversed-Euler variant, rE-tUPS, interchanges the
roles of the one-body and pair-transfer rotations within
each tile:
\begin{equation}
\hat U_{pq}^{\mathrm{rE}}
=
\hat R_P(\theta_1)
\hat R_1(\theta_2)
\hat R_P(\theta_3).
\end{equation}
Thus, each tile contains a central one-body rotation
between two independently parameterized pair-transfer
rotations, while retaining the three parameters per
tile and brickwork connectivity of tUPS.

The compressed variants, cp-tUPS and cm-tUPS, use
$\hat K_P+\hat K_F$ and $\hat K_P-\hat K_F$, respectively;
we refer to them collectively as c-tUPS.
Both require four length-4 Majorana rotations per tile,
compared with eight for tUPS.

The independent-real variant, ir-tUPS, retains the tile
structure of c-tUPS, with two outer one-body rotations,
but assigns independent parameters to pair transfer and
opposite-spin exchange in the central factor:
\begin{equation}
\hat U_{pq}^{\mathrm{ir}}
=
\hat R_1(\theta_1)
e^{\theta_P\hat K_P+\theta_F\hat K_F}
\hat R_1(\theta_4).
\end{equation}
This gives four parameters per tile. Setting
$\theta_F=0$ recovers the tUPS family, with
$\theta_P=2\theta_2$ in the convention used for
$\hat R_P(\theta_2)$, while imposing
$\theta_F=\pm\theta_P$ recovers c-tUPS.

The remaining two variants, Kp-tUPS and Km-tUPS, use
the Kanamori-inspired rotations~\cite{kanamori1963electron} $\hat V_+$ and $\hat V_-$,
respectively, defined by
\begin{equation}
\label{eq: C kanamori}
\begin{aligned}
\hat H_P&=\hat P+\hat P^\dagger,
&
\hat H_F&=\hat F+\hat F^\dagger,
\\
\hat G_{K,\pm}
&=\hat H_P\pm(\hat H_{zz}-\hat H_F),
&
\hat V_\pm(\theta)&=e^{-i\theta\hat G_{K,\pm}},
\end{aligned}
\end{equation}
where
\begin{equation}
\begin{aligned}
\hat H_{zz}&=2\hat S_p^z\hat S_q^z,
\\
\hat S_u^z&=\tfrac12(\hat n_{u\alpha}-\hat n_{u\beta}),
\qquad
\hat n_{u\sigma}=\hat a^\dagger_{u\sigma}\hat a_{u\sigma}.
\end{aligned}
\end{equation}
Since $\hat H_{zz}-\hat H_F
=2\hat{\mathbf S}_p\cdot\hat{\mathbf S}_q$
and pair transfer is spin invariant, both Kanamori-inspired
variants conserve total spin, although they generally
produce complex amplitudes.

\begin{table*}[t]
\centering
\footnotesize
\caption{
Ansatz variants compared in
Fig.~\ref{fig: MP-Statevector error all variants}.
Each row gives the two-orbital circuit used at every tile.
All counts are per tile: $n_\theta$ is the number of
variational parameters, $n_{\mathrm{rot}}$ is the number
of Majorana rotations in the implementation, and $G_4$
counts those generated by length-4 Majorana monomials.
The columns ``real'' and $S^2$ indicate preservation of
real amplitudes and conservation of total spin squared.
The final column indicates whether a choice of parameters
recovers the tUPS circuit, with $=$ denoting equivalence.
All variants conserve the electron numbers
$\hat N_\alpha$ and $\hat N_\beta$ separately.
}
\label{tab: C catalogue}
\medskip
\begin{tabular}{llcccccc}
\hline\hline
variant & two-orbital circuit
& $n_\theta$ & $n_{\mathrm{rot}}$ & $G_4$
& real & $S^2$ & $\supset$ tUPS \\
\hline
tUPS
& $\hat R_1(\theta_1)\hat R_P(\theta_2)\hat R_1(\theta_3)$
& 3 & 16 & 8 & \checkmark & \checkmark & $=$ \\
rE-tUPS
& $\hat R_P(\theta_1)\hat R_1(\theta_2)\hat R_P(\theta_3)$
& 3 & 20 & 16 & \checkmark & \checkmark & $\times$ \\
cp-tUPS
& $\hat R_1(\theta_1)e^{\theta_2(\hat K_P+\hat K_F)}
   \hat R_1(\theta_3)$
& 3 & 12 & 4 & \checkmark & $\times$ & $\times$ \\
cm-tUPS
& $\hat R_1(\theta_1)e^{\theta_2(\hat K_P-\hat K_F)}
   \hat R_1(\theta_3)$
& 3 & 12 & 4 & \checkmark & $\times$ & $\times$ \\
ir-tUPS
& $\hat R_1(\theta_1)e^{\theta_P\hat K_P+\theta_F\hat K_F}
   \hat R_1(\theta_4)$
& 4 & 24 & 16 & \checkmark & $\times$ & \checkmark \\
Kp-tUPS
& $\hat R_1(\theta_1)\hat V_+(\theta_2)\hat R_1(\theta_3)$
& 3 & 16 & 8 & $\times$ & \checkmark & $\times$ \\
Km-tUPS
& $\hat R_1(\theta_1)\hat V_-(\theta_2)\hat R_1(\theta_3)$
& 3 & 16 & 8 & $\times$ & \checkmark & $\times$ \\
\hline\hline
\end{tabular}
\end{table*}

\subsection{Spin-exchange equivalence of cp-tUPS and cm-tUPS}
\label{app:ctups-spin-exchange}

We  now establish why cp-tUPS and cm-tUPS can be
represented by a single c-tUPS curve. Although their
central generators differ, global spin exchange relates
the two circuits and gives identical exact energy
landscapes for the Hamiltonians and references used here.
We show that blockwise MP truncation preserves this
equivalence, and identify the conditions under which
it also extends to VIME parameter updates.

Let $\mathcal S$ denote the fermionic unitary that exchanges
the two spin modes of every spatial orbital,
\begin{equation}
\begin{aligned}
\mathcal S\hat a_{p\alpha}\mathcal S^\dagger
&=\hat a_{p\beta},\\
\mathcal S\hat a_{p\beta}\mathcal S^\dagger
&=\hat a_{p\alpha}.
\end{aligned}
\label{eq:ctups-spin-exchange}
\end{equation}
Applying this transformation to the operators defined in
Eq.~\eqref{eq: KP KF}, with fermionic anticommutation
relations used to reorder the factors, gives
\begin{equation}
\mathcal S\hat P\mathcal S^\dagger=\hat P,
\qquad
\mathcal S\hat F\mathcal S^\dagger=\hat F^\dagger.
\end{equation}
Consequently,
\begin{equation}
\begin{aligned}
\mathcal S\hat K_P\mathcal S^\dagger&=\hat K_P,\\
\mathcal S\hat K_F\mathcal S^\dagger&=-\hat K_F,\\
\mathcal S\hat\kappa_{pq}^{(1)}\mathcal S^\dagger
&=\hat\kappa_{pq}^{(1)}.
\end{aligned}
\end{equation}
The compressed tiles therefore satisfy
\begin{equation}
\hat U_{pq}^{\mathrm{c},-\,(m)}
=
\mathcal S
\hat U_{pq}^{\mathrm{c},+\,(m)}
\mathcal S^\dagger
\label{eq:ctups-tile-spin-equivalence}
\end{equation}
at identical values of their three parameters.

The same global transformation applies to every tile,
and the spin-adapted active orbital rotations commute
with $\mathcal S$. Thus, for any number of layers,
the complete circuits obey
\begin{equation}
\hat U_-(\theta)
=
\mathcal S\hat U_+(\theta)\mathcal S^\dagger,
\label{eq:ctups-circuit-spin-equivalence}
\end{equation}
where $\hat U_+$ and $\hat U_-$ denote cp-tUPS and
cm-tUPS, respectively, including active orbital rotations.

We now show that the two circuits give the same energy
at the same parameters $\theta$. The Hamiltonians used
here are spin free, so exchanging $\alpha$ and $\beta$
spins leaves $\hat H$ unchanged. The reference state contains
only doubly occupied and empty spatial orbitals, so
the same exchange leaves its occupations unchanged
and can introduce only an overall phase. Its density
operator $\hat\rho=\ket{\Phi_0}\bra{\Phi_0}$ is therefore
also unchanged:
\begin{equation}
\mathcal S^\dagger\hat H\mathcal S=\hat H,
\qquad
\mathcal S^\dagger\hat\rho\mathcal S=\hat\rho.
\end{equation}
Since the two circuits are related by this spin exchange
[Eq.~\eqref{eq:ctups-circuit-spin-equivalence}], their
energy expectation values satisfy
\begin{equation}
\begin{aligned}
E_-(\theta)
&=\operatorname{Tr}\!\left[
\hat H\hat U_-(\theta)\hat\rho\hat U_-^\dagger(\theta)
\right]\\
&=\operatorname{Tr}\!\left[
\hat H\hat U_+(\theta)\hat\rho\hat U_+^\dagger(\theta)
\right]\\
&=E_+(\theta).
\end{aligned}
\label{eq:ctups-exact-energy-equivalence}
\end{equation}

The equality of exact energies does not automatically
carry over to a truncated approximation. Preserving it
requires both a truncation rule that respects spin exchange
and a choice of truncation points where the propagated
operators remain related by this symmetry.

The truncation rule in Eq.~\eqref{eq:paired-acceptance}
satisfies the first requirement because it depends only
on monomial length and pairedness. Spin exchange preserves
both: it relabels Majorana operators without changing
their number and maps the two operators of each mode
together to the opposite-spin mode. The rule therefore
retains or discards a monomial and its spin-exchanged
counterpart together.

Blockwise truncation satisfies the second requirement.
Equation~\eqref{eq:ctups-tile-spin-equivalence} relates
the exact propagation through complete tiles, but does
not by itself establish the same relation at intermediate
steps of their Majorana-rotation decompositions.
Applying truncation only after each complete tile
therefore preserves the spin-exchange relation throughout
the propagation, giving
\begin{equation}
E_{\mathrm{MP},-}^{(w^\star)}(\theta)
=
E_{\mathrm{MP},+}^{(w^\star)}(\theta).
\label{eq:ctups-mp-energy-equivalence}
\end{equation}
Since the exact energies also agree, the MP-truncation
errors coincide at the same parameters.

For VMPE, identical MP energies at every $\theta$ imply
identical energy gradients. The two variants therefore
follow the same parameter trajectory when initialized
and optimized identically, in exact arithmetic.

For VIME, we compare the parameter updates at the same
$\theta$. VIME chooses the parameter velocity $\dot\theta$
by minimizing the regularized McLachlan objective,

\begin{equation}
L_\lambda
=
\frac12\|J\dot\theta-v\|_2^2
+\frac{\lambda}{2}\|\dot\theta\|_2^2,
\end{equation}
where $v$ is the IT velocity and J is the Jacobian.

Our pool $S_H$ contains all operators in the Hamiltonian
expansion. Since the Hamiltonian is invariant under spin
exchange, each spin-exchanged operator also belongs to
the pool, up to a sign.

Spin exchange therefore only reorders the rows of $J$
and the entries of the IT velocity $v$, or changes the
sign of corresponding rows and entries together.
Neither operation changes $\|J\dot\theta-v\|_2^2$,
and the regularization term is also unchanged.
The two variants consequently give the same parameter
velocity at the same $\theta$ and regularization.
Starting from identical parameters and using the same
solver and integration settings, they follow identical
VIME trajectories in exact arithmetic.

\FloatBarrier
\subsection{MP energy accuracy across ansatz variants}

The ans\"atze in Table~\ref{tab: C catalogue} differ in both the generators used within each tile and the constraints on their variational parameters. These choices determine the states accessible to the circuit, but also affect the accuracy of truncated MP. We therefore compare their MP energies during VIME convergence under the same conditions, extending the tUPS--c-tUPS comparison of Fig.~\ref{fig: MP-Statevector error vs family} to the additional variants defined above.

\begin{figure}[t]
    \centering
    \includegraphics[width=\columnwidth]{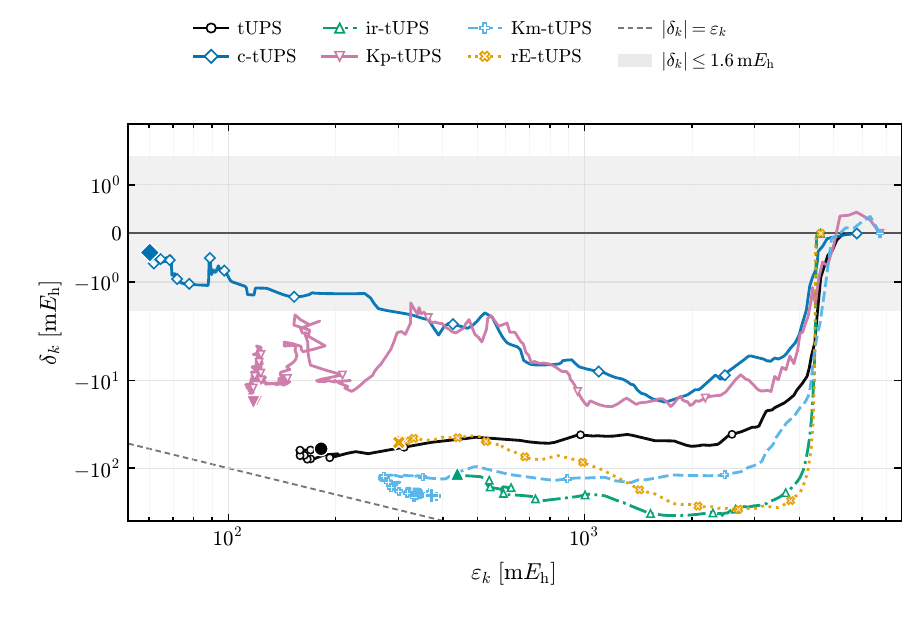}
    \caption{\textbf{MP energy accuracy for additional tUPS variants.}
VIME convergence curves for the two-orbital circuits listed in
Table~\ref{tab: C catalogue}, using one layer on H$_8$/STO-6G,
$w^\star=6$, and truncation at block boundaries.
The MP-truncation error
$\delta_k=E_{\mathrm{MP},k}-E_{\mathrm{SV},k}$
is plotted against the statevector energy error
$\varepsilon_k=E_{\mathrm{SV},k}-E_{\mathrm{ref}}$,
where $E_{\mathrm{ref}}$ is the exact active-space ground-state
energy.
Each point gives the median of each error over 20 runs at a
successive optimization step, progressing generally from right
to left; filled symbols mark the last step reached by all runs.
The cp-tUPS and cm-tUPS results are indistinguishable on this
scale and are represented by a single c-tUPS curve.
Spread bands are omitted for visual clarity.
The gray band marks
$|\delta_k|\leq1.6\,\mathrm{m}E_{\mathrm h}$.
The dashed line marks $\delta_k=-\varepsilon_k$, corresponding
to $E_{\mathrm{MP},k}=E_{\mathrm{ref}}$ for an individual run.}
    \label{fig: MP-Statevector error all variants}
\end{figure}

Figure~\ref{fig: MP-Statevector error all variants} shows
this comparison for one-layer circuits on H$_8$/STO-6G.
For each variant, the median MP-truncation error is initially
close to zero, grows in magnitude during the first few updates,
and is predominantly negative thereafter.
Despite the additional choices of generators and parameters,
none of the other variants improves on c-tUPS in either
the magnitude of the final median MP-truncation error or
the final median statevector energy error.
The differences between the curves also allow us to examine
how MP energy accuracy depends on the number and choice
of length-four rotations within each tile.

The data suggests a broad association between
MP-truncation error and $G_4$, the number of rotations
generated by length-four Majorana monomials in each
two-orbital block.
At comparable statevector energy errors, c-tUPS, with
$G_4=4$, generally gives smaller MP-truncation errors than
variants containing more such rotations.
As discussed in Sec.~\ref{subsec: mp}, length-two rotations
preserve monomial length, whereas length-four rotations can
increase it and generate terms beyond the truncation cutoff.
A larger $G_4$ therefore provides more opportunities to
generate contributions that MP may subsequently discard.

However, $G_4$ alone does not determine the truncation error.
The original tUPS circuit and the Kanamori-inspired variants
Kp-tUPS and Km-tUPS all have $G_4=8$, yet their
MP-truncation errors differ substantially.
The number of length-four rotations is thus a useful
consideration when choosing circuits for MP, but the
particular rotations and their parameter values also matter.

MP energy accuracy must also be distinguished from the
expressivity of an ansatz.
The independent-real variant, ir-tUPS, recovers tUPS when
$\theta_F=0$ and both compressed variants when
$\theta_F=\pm\theta_P$, with the corresponding rescaling
of the pair-transfer parameter.
Its variational manifold therefore contains those of tUPS and c-tUPS at the same number of layers.
Nevertheless, the magnitude of its median MP-truncation error
remains above $10^2\,\mathrm{m}E_{\mathrm h}$ during the later
stages of optimization.
Its curve ends at a median statevector energy error
approximately $2.4$ times that reached by tUPS and $7.3$
times that reached by c-tUPS.
The larger variational manifold therefore does not translate
into better convergence under truncated MP in this benchmark.
The substantial discrepancy between MP and statevector
energies suggests that inaccurate MP estimates hinder VIME
from reaching the lower-energy states available within the
ir-tUPS manifold.

Beyond these comparisons of MP accuracy and convergence,
the compressed two-body structure of c-tUPS also motivates
its use for quantum execution. We do not expect the other
variants to offer lower CNOT counts or two-qubit depth
at the same number of layers under comparable connectivity
and compilation conditions, although establishing this
ordering would require dedicated synthesis of each variant.

\begin{figure*}[!tb]
\centering
\includegraphics[width=\textwidth]{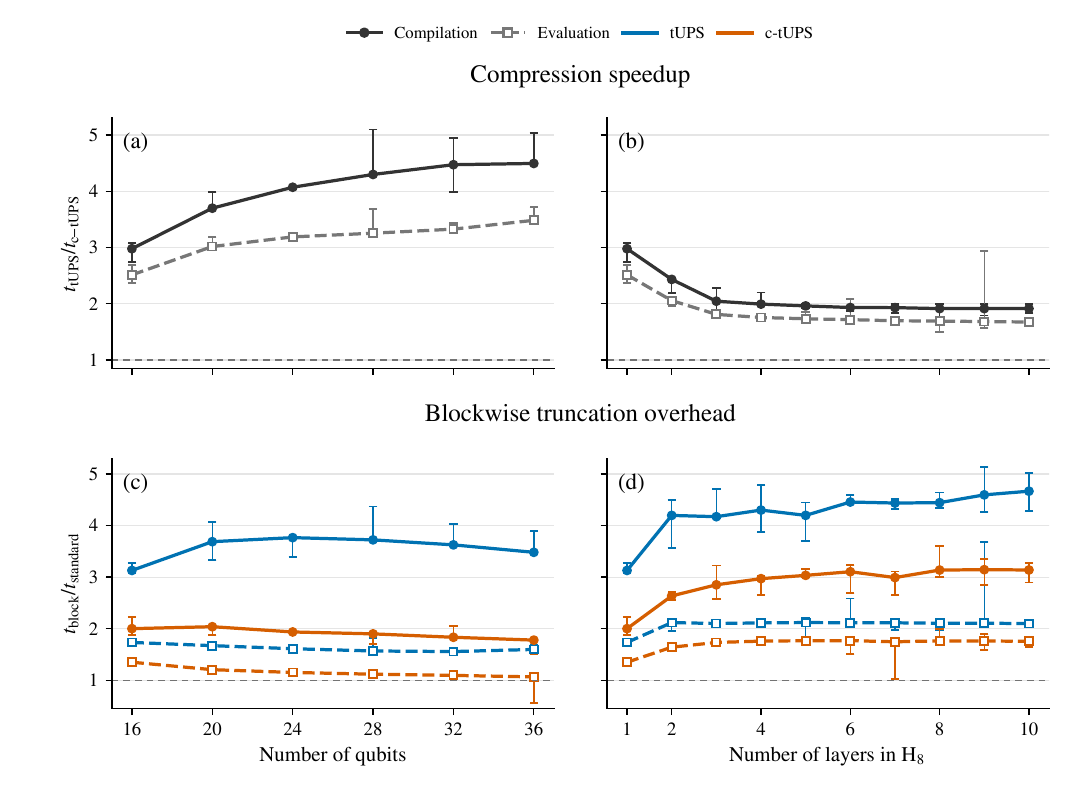}
\caption{
\textbf{Computational benefit of compression and cost of
blockwise truncation.}
Timings for tUPS and c-tUPS on hydrogen chains in STO-6G
at $w^\star=6$.
Left column: system-size dependence for one layer.
Right column: layer dependence for H$_8$ (16 qubits).
(a,b) Compression speedup
$t_{\mathrm{tUPS}}/t_{\mathrm{c\text{-}tUPS}}$
under blockwise truncation.
(c,d) Overhead of blockwise relative to standard truncation,
$t_{\mathrm{block}}/t_{\mathrm{standard}}$.
Solid circles denote compilation; dashed squares denote
evaluation of the IT velocity and full parameter Jacobian.
Markers show the median of five timing ratios, each computed
from a pair of corresponding runs; error bars span the
minimum and maximum ratios.
}
\label{fig:tups-compression-block-cost}
\end{figure*}

\subsection{Computational cost of compression and blockwise truncation}
\label{app:block-cost}

Blockwise truncation applies the length cutoff after
all rotations in a tile have been propagated.
A monomial can therefore grow beyond the cutoff and
return to a length at or below it before truncation
is applied. Standard truncation instead applies the cutoff after each
Majorana rotation.
Figure~\ref{fig:tups-compression-block-cost} compares
the computational savings from compression and the
additional cost of blockwise truncation.

Under blockwise truncation, c-tUPS reduces both compilation
and evaluation times throughout the tested ranges, as can be seen in panels (a,b). For one-layer circuits, increasing the
system size from 16 to 36 qubits raises the compilation
speedup from approximately $3.0$ to $4.5$ and the
evaluation speedup from $2.5$ to $3.5$.
For H$_8$, increasing the depth from one to ten layers
instead reduces these speedups to approximately
$1.9$ and $1.7$, respectively.

As shown in panels (c,d), blockwise truncation incurs less overhead for
c-tUPS than for tUPS across both the qubit-count and layer-count sweeps.
For one-layer circuits, the compilation overhead is approximately
$3.1$--$3.8$ for tUPS, compared with $1.8$--$2.0$ for c-tUPS.
The evaluation overhead generally declines as the qubit count increases,
reaching approximately $1.6$ and $1.07$, respectively, at 36 qubits.
For H$_8$ at ten layers, the compilation overhead reaches approximately
$4.7$ for tUPS and $3.2$ for c-tUPS, while the evaluation overhead is
approximately $2.1$ and $1.8$, respectively. The lower evaluation
overhead of c-tUPS is especially relevant to repeated parameter updates,
which reuse the compiled graph.

\FloatBarrier
\section{Computational cost and numerical behavior of VIME}
\label{app:cost-comparison}

This appendix examines the computational requirements
and numerical behavior of VIME.
We first compare its computational cost with direct VMPE,
identifying the additional work associated with the
propagation graph and parameter derivatives.
We then investigate how compilation time, evaluation time,
and memory requirements depend on system size and c-tUPS
circuit depth.
Finally, we examine how different combinations of initial
parameters and Tikhonov regularization affect the VIME
trajectories for the considered systems.

\subsection{Cost comparison with direct VMPE}
\label{app:vime-vmpe-comparison}

\begin{table}[tb]
\caption{
Wall-clock times and compiled graph sizes for the
single c-tUPS layer TLD-1411 calculations.
VIME times are medians over complete integration steps;
VMPE times are medians over energy-and-gradient evaluations.
Graph sizes count stored nodes in the initially compiled
MP graphs.
}
\label{tab:vime-vmpe-cost-large}
\centering
\small
\setlength{\tabcolsep}{5pt}
\begin{tabular}{ccccc}
\toprule
$N_q$ &
\multicolumn{2}{c}{Wall-clock time (s)} &
\multicolumn{2}{c}{Graph nodes} \\
\cmidrule(lr){2-3}
\cmidrule(lr){4-5}
& VIME & VMPE & VIME & VMPE \\
\midrule
28 & $61.5$ & $0.861$
   & $1.93{\times}10^{8}$ & $2.84{\times}10^{7}$ \\
40 & $608$ & $2.26$
   & $1.37{\times}10^{9}$ & $1.89{\times}10^{8}$ \\
52 & $1.40{\times}10^{3}$ & $6.66$
   & $2.03{\times}10^{9}$ & $8.23{\times}10^{8}$ \\
\bottomrule
\end{tabular}
\end{table}

VMPE minimizes the MP reported energy $E_{\mathrm{MP}}$
and requires its value and parameter gradient. VIME
additionally requires the pool expectation values
$o_{k,\mathrm{MP}}$, the anticommutator estimates
$C_{k,\mathrm{MP}}$, and the surrogate Jacobian
$J_{\mathrm{MP}}$. These quantities are defined in
Eqs.~\eqref{eq: app surrogate energy},
\eqref{eq: forward pass operators},
\eqref{eq: app surrogate anticommutator}, and
\eqref{eq: app surrogate jacobian}, respectively.
Their evaluation increases the computational cost in
two ways: VIME retains a larger propagation graph and
propagates one derivative vector per variational parameter.

The difference in graph size follows from the
anticommutator construction. For VMPE, a root monomial
with $r_a=0$, where $\vec r$ is the reference expectation
vector of Eq.~\eqref{eq: app reference vector}, contributes
neither to the surrogate energy nor to its gradient.
Such monomials can therefore be removed, together with
graph nodes and connections that contribute only to them.
For VIME, however, vanishing individual expectation values
do not imply a vanishing anticommutator expectation value.
The root-layer anticommutator matrix of
Eq.~\eqref{eq: app root anticommutator matrix},
\begin{equation}
W_{ab}
=
\operatorname{Tr}\!\left[
\hat\rho
\{\hat M_a^{(\mathrm{root})},
  \hat M_b^{(\mathrm{root})}\}
\right],
\end{equation}
can have nonzero entries even when $r_a=r_b=0$.
Here, $\hat M_a^{(\mathrm{root})}$ denotes a monomial
in the root-layer basis defined in
Eq.~\eqref{eq: app layer basis}.
VIME must retain these monomials and their contributing
paths to construct the root seed
$\vec B_0=W\vec h_{\mathrm{root}}$ of
Eq.~\eqref{eq: app anticommutator seed}.

The additional propagation cost comes mainly from
evaluating the full surrogate Jacobian. One backward
pass, given by Eq.~\eqref{eq: app backward recurrence},
provides the root coefficient vector
$\vec h_{\mathrm{root}}$ used to evaluate
$E_{\mathrm{MP}}$ and construct $\vec B_0$.
The subsequent joint forward pass propagates
$N_\theta+2$ vectors: the forward-propagated expectation
vector $\vec u_j$ of Eq.~\eqref{eq: u vector}, the
anticommutator vector $\vec B_j$ propagated according to
Eq.~\eqref{eq: app anticommutator recurrence}, and one
derivative vector
$\vec d_j^{\,(i)}=\partial_{\theta_i}\vec u_j$
per independent variational parameter, following
Eq.~\eqref{eq: app derivative recurrence}.
Their terminal contractions give the pool expectation
values, anticommutator estimates, and Jacobian entries
in Eq.~\eqref{eq: app joint outputs}.
For the $28$-, $40$-, and $52$-qubit circuits considered
here, $N_\theta=130$, $247$, and $400$, respectively.

Table~\ref{tab:vime-vmpe-cost-large} reports the measured
wall-clock times and compiled graph sizes for the
single-layer TLD-1411 calculations in
Fig.~\ref{fig:vime-vs-vmpe-large}.
The VIME time is the median wall-clock interval between
successive points recorded along the VIME trajectory;
each interval includes a complete integration step.
The VMPE time is the median duration of an
energy-and-gradient evaluation, pooled over the $100$
independently initialized runs. Thus, the timings compare
one complete VIME step with one VMPE energy-and-gradient
evaluation.

At $28$, $40$, and $52$ qubits, the VIME graphs contain
approximately $6.8$, $7.3$, and $2.5$ times as many
stored nodes as the corresponding VMPE graphs.
The smaller ratio at $52$ qubits reflects the additional
spatial-orbital support restriction $o^\star=4$ applied
only to VIME to make the calculation feasible; VMPE
uses no corresponding support filter.

The measured VIME step times are approximately $71$,
$269$, and $210$ times the corresponding VMPE
energy-and-gradient evaluation times.
The larger graphs and the propagation of
$N_\theta$ derivative vectors account for the main
additional computational work.
These timing ratios also include the remaining operations
within a complete VIME step, including the adaptive
imaginary-time integration described in
Appendix~\ref{app:integration}. VIME typically needs fewer
integration steps than VMPE needs energy-and-gradient
evaluations to reach a comparable energy, so the per-operation
ratios do not represent total time-to-solution ratios.
The higher cost per step explains why fewer independent
VIME trajectories were performed.

\subsection{System-size and circuit-depth dependence}
\label{app: ctUPS VIME scaling}

\begin{figure*}[t]
    \centering
    \includegraphics[width=\textwidth]
    {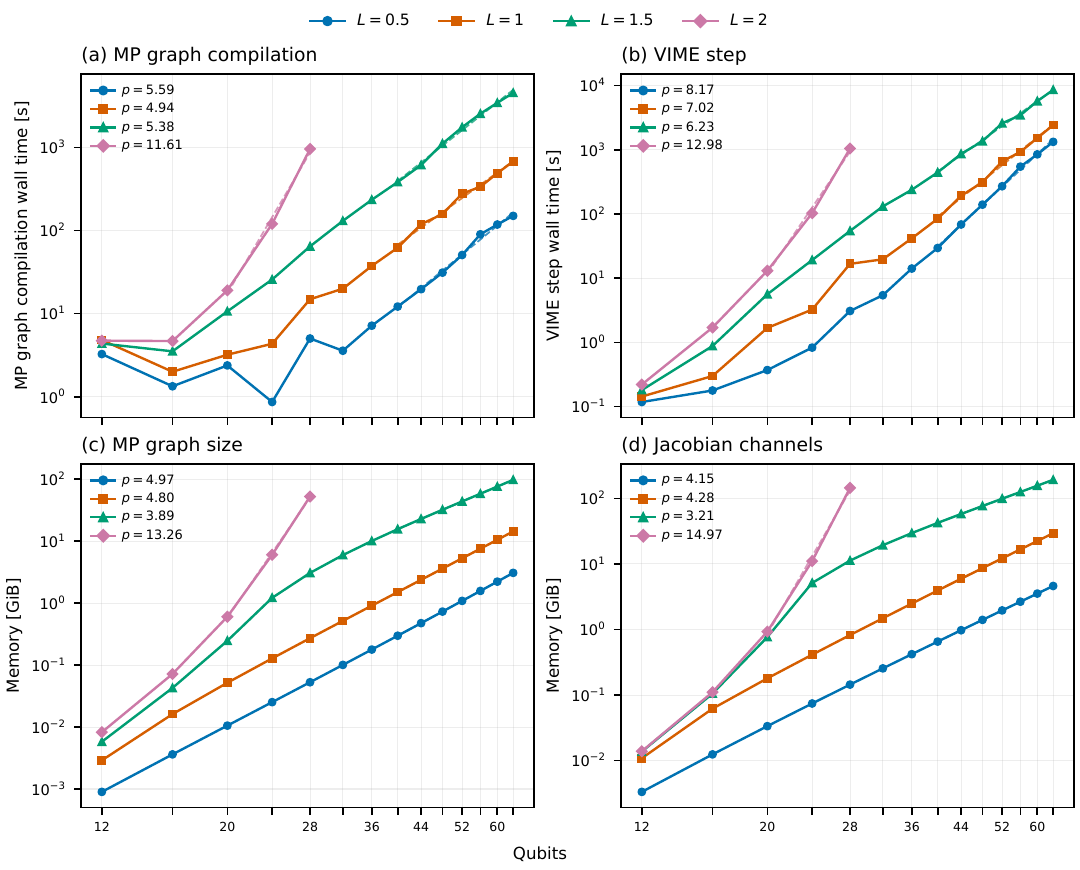}
\caption{
\textbf{System-size and circuit-depth dependence of
VIME computational resources.}
Hydrogen-chain benchmarks in the STO-6G basis for c-tUPS
circuits, where $N$ is the number of qubits and $L$ is
the number of circuit layers.
Panels show (a) MP graph compilation time,
(b) evaluation time for the ITE quantities and full pool
Jacobian, excluding the solution of the linear system
for the parameter update,
(c) compiled graph storage, and
(d) memory allocated for Jacobian evaluation with a fixed
batch size of $16$.
Panels (c,d) do not represent total peak process memory.
Markers show median values, connected by solid lines;
error bars indicate 10th--90th percentile ranges.
Dashed curves are power-law fits $aN^p$ over
$40\leq N\leq64$ for $L\leq1.5$;
legends give the fitted exponents.
Two-layer fits use only $N=20,24,28$ and serve
as descriptive guides.
Larger two-layer calculations exceeded the available
$1\,\mathrm{TiB}$ memory.
}
    \label{fig:ctups-resource-scaling}
\end{figure*}

The practical reach of VIME depends both on the memory required
to compile and evaluate the MP graph and on the cost of repeatedly
evaluating the quantities needed along an imaginary-time trajectory.
Figure~\ref{fig:ctups-resource-scaling} examines these requirements
for hydrogen chains in the STO-6G basis using c-tUPS circuits with
$L=0.5$, $1$, $1.5$, and $2$.
We distinguish the one-time graph compilation cost from the
repeated evaluation of the ITE quantities and full pool Jacobian.
We also report the compiled graph storage and the memory allocated
for Jacobian evaluation, with derivatives evaluated in batches of $16$.
These memory measures account for only part of the implementation's
total peak memory usage.

Both system size and circuit depth strongly increase the
computational cost.
For $L\leq1.5$, power-law fits over
$40\leq N\leq64$ yield effective exponents of
$4.94$--$5.59$ for compilation time and $6.23$--$8.17$
for repeated evaluation.
The steeper size dependence of evaluation is consistent with
the additional work required to propagate parameter derivatives
through the growing graph.
It is important to clarify that a smaller fitted exponent does not imply a lower cost:
the deeper circuits remain substantially more expensive over the entire range studied.

The absolute evaluation times make this limitation particularly
clear. At $64$ qubits, evaluating the ITE quantities and full
pool Jacobian takes approximately $40$ minutes for one layer
and $2.4$ hours for one-and-a-half layers, excluding the time
needed to solve the linear system for the parameter update.
Although compilation is performed only once for a fixed circuit
and truncation scheme, this evaluation cost is incurred at every
VIME step. At the measured one-and-a-half-layer rate, even
$100$ steps would require approximately ten days of evaluation
time alone.
Thus, a calculation can fit in memory while a long
imaginary-time trajectory remains computationally impractical.

Memory imposes a further restriction on accessible circuit depths.
Over the same fitting window, the effective exponents are
$3.89$--$4.97$ for compiled graph storage and
$3.21$--$4.28$ for the memory allocated for Jacobian evaluation.
For two layers, calculations beyond $28$ qubits exceeded the
available $1\,\mathrm{TiB}$ memory.
The corresponding fits therefore use only
$N=20,24,28$ and serve as descriptive guides; they cannot
be directly compared with fits over $40$--$64$ qubits or
used to establish asymptotic scaling.
Together, the timing and memory results identify substantial
limits on both system size and circuit depth in the present
implementation.

These costs also motivate the layerwise continuation strategy
used in this work.
Evolving from the initial reference to a state with a small
energy error can require many VIME steps.
Starting directly with one-and-a-half layers would incur the
higher evaluation cost throughout this initial evolution.
Instead, we first optimize a one-layer circuit, append a
half-layer, and resume the evolution from the previously
optimized parameters, with the new parameters initialized
at or near zero.
This places the initial relaxation in the less expensive
variational manifold and reserves the deeper circuit for
further refinement.
Continuation does not reduce the cost of an individual
one-and-a-half-layer step; its computational motivation is
to reduce the number of such expensive steps needed to
reach a low-error state.

\FloatBarrier
\subsection{Dependence on initialization and regularization}
\label{app: computational-details}

\begin{figure*}[t]
    \centering
    \includegraphics[width=\textwidth]
    {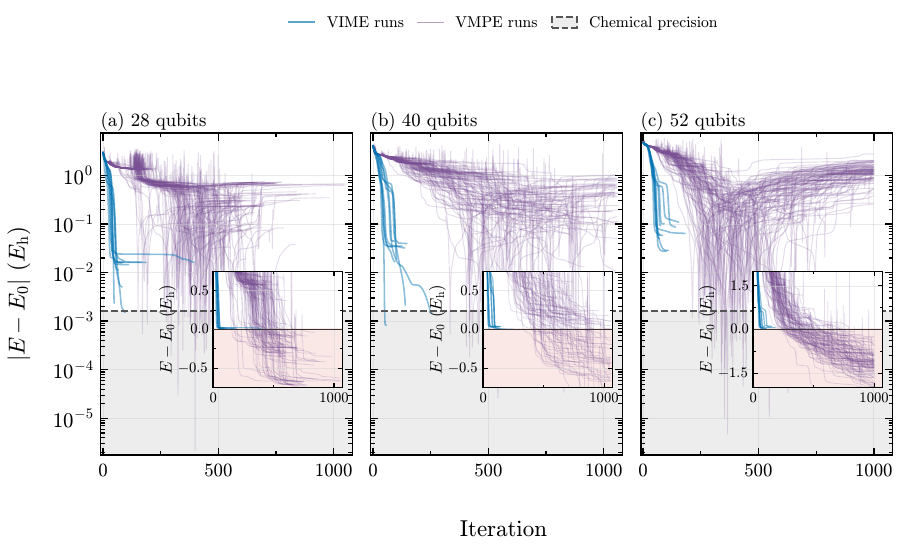}
    \caption{
    \textbf{Individual VIME and VMPE trajectories for TLD-1411.}
    Single-layer c-tUPS calculations for
    (a) 28-, (b) 40-, and (c) 52-qubit active spaces.
    Blue and purple curves show the 10 VIME and 100 VMPE
    runs, respectively, underlying
    Fig.~\ref{fig:vime-vs-vmpe-large}.
   Both methods are initialized with random parameters,
and the VIME runs additionally differ in the Tikhonov
regularization parameter $\lambda$.
    Main panels show absolute MP energy errors relative
    to the DMRG reference $E_0$; insets show signed errors,
    with pink shading marking energies below $E_0$.
    Gray shading marks chemical precision.
    Each curve ends at its last recorded iteration.
    Truncation settings are the same as in
    Fig.~\ref{fig:vime-vs-vmpe-large}.
    }
    \label{fig:tld-individual-vime}
\end{figure*}

Figure~\ref{fig:tld-individual-vime} shows the individual
trajectories underlying Fig.~\ref{fig:vime-vs-vmpe-large}:
10 VIME runs and 100 VMPE runs for each active space.
Both methods show variation in their trajectories and
final recorded energies across random initial parameters.
For VIME, the Tikhonov regularization parameter $\lambda$
also varies between runs, so the effects of initialization
and regularization cannot be separated.

For VIME, initialization determines the starting state
and the local Jacobian, while regularization changes
the parameter-update direction. To see the latter,
write the singular-value decomposition as
$J_{\mathrm{MP}}=U\Sigma V^T$. The regularized update
is then
\begin{equation}
\dot\theta
=
\sum_{k:\sigma_k>0}
\frac{\sigma_k}{\sigma_k^2+\lambda}
\left(u_k^T v_{\mathrm{MP}}\right)V_k ,
\end{equation}
where $\sigma_k$, $u_k$, and $V_k$ are the singular values
and corresponding left and right singular vectors.
Increasing $\lambda$ suppresses contributions associated
with small singular values, reducing the amplification
of errors but also slowing motion along the corresponding
parameter directions. It can therefore change both the
trajectory and the convergence rate, rather than merely
rescaling the imaginary-time step.
These effects, together with the optimization difficulties
reported for tiled ans\"atze~\cite{burton2024tups,ramoa2025ovpceo},
offer possible explanations for the variation among VIME
runs. The present results do not determine whether
individual trajectories are limited by local minima,
Jacobian conditioning, or MP truncation.

Despite the variation within both ensembles, their behavior
relative to the DMRG reference differs clearly.
All recorded VIME energies remain above the reference,
whereas the final VMPE energies lie below it in the
majority of runs for each active space, including all
100 runs at 52 qubits.
The signed-error insets show that the sharp dips in
absolute VMPE error occur as the energy crosses below
the DMRG reference and continues to decrease.
These dips therefore reflect a transient agreement with
the reference rather than convergence.
Without an independently known reference energy, such
crossings could not be identified from the VMPE energy
trajectory alone and would provide no criterion for
selecting an accurate state or stopping the optimization.
The absence of such crossings in the recorded VIME runs
supports its greater robustness against large negative
MP truncation errors in these calculations, despite its
sensitivity to initialization and regularization.
However, the higher computational cost of VIME limits
its ensemble to rouhgly these 10 runs per active space, compared with
100 for VMPE. As discussed in the main text, the absence
of below-reference VIME energies in this finite sample
does not establish a variational bound for
$E_{\mathrm{MP}}$.

In Fig.~\ref{fig:vime-vs-vmpe-large}, medians and quartiles
are calculated at each iteration, separately for absolute
and signed errors. For shorter runs, the last recorded
energy is retained when calculating subsequent statistics.
The pointwise minimum uses the same convention and can
come from different runs at different iterations.
In Fig.~\ref{fig:tld-individual-vime}, individual VIME
and VMPE curves end at their last recorded iteration.

\FloatBarrier
\section{Linear acenes: benchmark systems and computational details}
\label{app:molecular-benchmarks}

\begin{table}[tbh]
    \centering
    \caption{
        Full $\pi/\pi^\ast$ active spaces and total
        DMRG-SCF reference energies for the linear acenes,
        using cc-pVTZ and maximum bond dimension
        $\chi=2000$. $n$ denotes the number of fused rings.
    }
    \label{tab:acene-reference-energies}
    \begin{ruledtabular}
        \begin{tabular}{lcccr}
            Molecule & $n$ & Active space & Qubits
            & $E_{\mathrm{ref}}$ ($E_{\mathrm h}$) \\
            \colrule
            Anthracene & 3 & CAS(14e,14o) & 28
                & $-536.2770791$ \\
            Pentacene & 5 & CAS(22e,22o) & 44
                & $-841.7116496$ \\
            Heptacene & 7 & CAS(30e,30o) & 60
                & $-1147.126562$ \\
        \end{tabular}
    \end{ruledtabular}
\end{table}

Linear acenes are polycyclic aromatic hydrocarbons composed
of $n$ linearly fused benzene rings, with general formula
$\mathrm{C}_{4n+2}\mathrm{H}_{2n+4}$.
Their electronic properties make them relevant to organic
semiconductor applications, including organic field-effect
transistors, light-emitting diodes, and photovoltaic
devices~\cite{Bendikov-ChemRev-2004,Anthony-ChemRev-2006,
Anthony-AngewChemIntEd-2008}.
Here we consider anthracene
($n=3$, $\mathrm{C}_{14}\mathrm{H}_{10}$),
pentacene ($n=5$, $\mathrm{C}_{22}\mathrm{H}_{14}$),
and heptacene ($n=7$, $\mathrm{C}_{30}\mathrm{H}_{18}$),
using the cc-pVTZ basis for all three systems.
Their structures are shown in
Fig.~\ref{fig:molecular-benchmarks}.

The acene series provides a systematic benchmark for
electronic-structure methods as both the size of the
$\pi$ system and its multireference character increase.
The smaller members have predominantly closed-shell
singlet ground states, while increasing chain length
is associated with growing open-shell singlet and
radical character
\cite{hachmann2007radical,Yang-PNAS-2016,Schriber-JCTC-2018}.
The singlet--triplet energy gap also decreases along
the series, and a balanced treatment of static and
dynamical correlations is important for describing
the ground-state character and excitation
energies~\cite{Schriber-JCTC-2018,Luo-JCTC-2018}.
These features, together with the growing active spaces
required to describe the full $\pi/\pi^\ast$ manifold,
make the longer acenes challenging tests of
electronic-correlation methods.
The present benchmarks focus on the preparation of the
singlet ${}^{1}A_g$ ground state within this active space.

The molecular geometries belong to the $D_{2h}$ point
group. Initial structures were taken from
Ref.~\cite{Schriber-JCTC-2018} and optimized for the
singlet ground state using density functional theory
in NWChem~7.2.0~\cite{nwchem-2020}.
The calculations employed the PBE0 hybrid
functional~\cite{pbe0}, the D3 dispersion correction
with Becke--Johnson damping
\cite{dft-d3-2010,dft-d3bj-2011}, and the
cc-pVTZ basis set~\cite{cc-pvtz-1989},
with tight convergence criteria.

Restricted Hartree--Fock calculations on the optimized
geometries provided the starting orbitals.
The active spaces were selected using the atomic valence
active space (AVAS) procedure~\cite{avas-2017}, targeting
carbon $2p_z$ atomic orbitals to span the full
$\pi/\pi^\ast$ manifold.
Table~\ref{tab:acene-reference-energies} lists the resulting
active spaces and qubit counts.

The orbitals were then optimized using DMRG-SCF,
with DMRG as the active-space solver, targeting the
${}^{1}A_g$ ground state.
These calculations used
PySCF~2.14.0~\cite{pyscf-2020,pyscf-2018}
and Block2~2.0~\cite{zhai2023block2}, with a maximum
DMRG bond dimension of $\chi=2000$.
The optimized orbitals were used to generate the
one- and two-electron integrals defining the active-space
Hamiltonians employed in the VIME benchmarks.

The total DMRG-SCF energies in
Table~\ref{tab:acene-reference-energies} serve as
finite-bond-dimension numerical references.
The reported VIME energy errors therefore measure the
agreement with these references within the selected
$\pi/\pi^\ast$ active spaces; they do not quantify
the recovery of correlation outside these spaces.

\FloatBarrier
\section{ADAPT-VIME}
\label{app:adapt vime}

ADAPT-VIME extends the fixed-circuit VIME method of
Sec.~\ref{subsec: vime} by adding Majorana rotations one at a time to better match the target IT velocity, following the incremental ansatz construction of ADAPT-VQE and ADAPT-VMPE \cite{grimsley2019adaptvqe,chakraborty2026adaptvmpe}.

For the Hamiltonian operator pool $S_\tau=S_H$, the single
compiled MP graph of Appendix~\ref{app: single surrogate pass}
supplies the energy $E_{\mathrm{MP}}$, the expectation values
$o_{k,\mathrm{MP}}$ and $C_{k,\mathrm{MP}}$ of $\hat O_k$ and
$\{\hat H,\hat O_k\}$, and their required parameter derivatives.
Throughout this appendix, we write $J=J_{\mathrm{MP}}$ and
$v=v_{\mathrm{MP}}$ to simplify the notation. Thus,
\begin{equation}
J_{kj}=\frac{\partial o_{k,\mathrm{MP}}}{\partial\theta_j},
\qquad
v_k=2E_{\mathrm{MP}}o_{k,\mathrm{MP}}-C_{k,\mathrm{MP}}.
\end{equation}
The circuit produces the rates $J\dot\theta$, and VIME determines
the parameter velocity from solving the linear system
\begin{equation}
(G+\lambda I)\dot\theta_*=b,
\end{equation}
where $G=J^TJ$, $b=J^Tv$ and $\lambda\geq0$ is the Tikhonov regularisation parameter.
The \emph{residual}
\begin{equation}
\label{eq:adapt-vime-residual}
\boldsymbol{\vec \epsilon}=v-J\dot\theta_*
\end{equation}
is the difference between the target and circuit-produced
IT velocities for the pool expectation values.

ADAPT-VIME selects the Majorana rotation that most reduces
$\tfrac12\|\boldsymbol{\vec \epsilon}\|_2^2$. When
$\lambda>0$, this comparison also includes the penalty
$\tfrac{\lambda}{2}\|\dot\theta\|_2^2$ on all parameter velocities, consistent with the regularized VIME update.

These Majorana rotations are chosen from a generator pool
$\mathcal P$ of Majorana monomials of length two and four, following Ref. ~\cite{chakraborty2026adaptvmpe}. For each generator $\hat M_{\vec b_a}\in\mathcal P$, we consider
\begin{equation}
\hat R_a(\phi)=\exp(-i\phi\hat M_{\vec b_a}/2),
\qquad
\hat U^{(a)}(\theta,\phi)=\hat U(\theta)\hat R_a(\phi).
\label{eq:adapt-vime-trial}
\end{equation}
With the ordering of Eq.~\eqref{eq: app circuit convention},
the candidate rotation acts directly on the reference state.
Each candidate is evaluated at $\phi=0$ with $\theta$ fixed.
After selection, VIME evolves both $\theta$ and the new angle $\phi$, which is initialized at zero.

\subsection{Computing the candidate column}

At $\phi=0$, the candidate rotation is the identity.
Thus $E_{\mathrm{MP}}$, $o_{k,\mathrm{MP}}$, $C_{k,\mathrm{MP}}$, $v$, and the existing Jacobian columns remain unchanged. The new parameter $\phi$ adds instead one column $j_a$ to the Jacobian,
\begin{equation}
\label{eq:adapt-vime-column}
J_a=[J\;j_a],
\qquad
(j_a)_k=
\left.
\frac{\partial o_{k,\mathrm{MP}}^{(a)}(\theta,\phi)}
     {\partial\phi}
\right|_{\phi=0},
\end{equation}
where $o_{k,\mathrm{MP}}^{(a)}$ is the expectation value of
$\hat O_k$ with the candidate rotation included.
Each candidate is therefore represented by such a column $j_a$.

We compute $j_a$ using the derivative propagation of
Appendix~\ref{app: A jacobian}. The candidate rotation acts
first on the reference state and is therefore encountered
last during Heisenberg back-propagation. Propagating the
current root monomials through this rotation adds a new
root layer.

Let $\hat L_a^{\mathrm{ins}}(\phi)$ be the matrix that
propagates monomial coefficients from the old root to the
new root. In the extended graph, the new root is layer $0$,
the old root is layer $1$, and the terminal layer is
$N_{\mathrm{rot}}+1$, where $N_{\mathrm{rot}}$ counts the
rotations before candidate insertion. As in Eq.~\eqref{eq: u vector},
let $\vec u_j^{\,(a)}$ denote the forward-propagated
expectation vector, and define
\begin{equation}
\vec d_j^{\,(a)}
=
\left.
\partial_\phi\vec u_j^{\,(a)}(\theta,\phi)
\right|_{\phi=0}.
\end{equation}

The reference expectation vector $\vec r^{\,(a)}$ at the
new root is independent of $\phi$, so
$\vec d_0^{\,(a)}=0$. Applying
Eq.~\eqref{eq: app derivative recurrence} through the
candidate rotation gives
\begin{equation}
\vec d_1^{\,(a)}
=
\left.
\frac{\partial
      [\hat L_a^{\mathrm{ins}}(\phi)]^T}{\partial\phi}
\right|_{\phi=0}
\vec r^{\,(a)}.
\end{equation}
Since $\partial_\phi\cos\phi|_{\phi=0}=0$ and
$\partial_\phi\sin\phi|_{\phi=0}=1$, only retained
anticommuting branches contribute, with the signs specified
by Eq.~\eqref{eq:majorana-propagation}.

The remaining propagation matrices do not depend on $\phi$.
The derivative vector therefore follows
\begin{equation}
\vec d_{j+1}^{\,(a)}
=\hat L_{j-1}^T\vec d_j^{\,(a)},
\qquad j=1,\ldots,N_{\mathrm{rot}},
\end{equation}
where $\hat L_{j-1}$ belongs to the original graph.
At the terminal layer,
Eq.~\eqref{eq: app jacobian contraction} gives
\begin{equation}
\label{eq:adapt-vime-candidate-pass}
(j_a)_k
=\vec A_k^T\vec d_{N_{\mathrm{rot}}+1}^{\,(a)},
\end{equation}
where $\vec A_k$ contains the expansion coefficients of
$\hat O_k$ in the terminal Majorana basis.
Computing $j_a$ therefore requires one forward pass of
the derivative vector through the existing graph.

\subsection{Accurate candidate scoring}

Once $j_a$ is known, we ask how much better the enlarged Jacobian
$[J\;j_a]$ can reproduce the same target $v$. 
For the unregularized loss $L$ of
Eq.~\eqref{eq: OVQITE McLachlan distance}, the improvement is
\begin{equation}
\label{eq:adapt-vime-selection}
\Delta L_a=\frac12\|\boldsymbol{\vec \epsilon}\|_2^2
-\frac12\min_{\dot\theta,\dot\phi}
\|J\dot\theta+j_a\dot\phi-v\|_2^2,
\end{equation}
where $\boldsymbol{\vec \epsilon}$ is evaluated with $\lambda=0$. Both the existing parameter
velocities $\dot\theta$ and the candidate velocity $\dot\phi$
are varied in the minimisation. Only the component of $j_a$ outside
the existing column space adds a new accessible direction. Defining
$\Pi_J=JJ^+$, with $J^+$ the Moore--Penrose pseudoinverse, gives
\begin{equation}
\label{eq:adapt-vime-rank-one}
\begin{aligned}
z_a&=(I-\Pi_J)j_a,\\
\Delta L_a&=\frac{(j_a^T\boldsymbol{\vec \epsilon})^2}
{2\|z_a\|_2^2},\qquad \|z_a\|_2>0.
\end{aligned}
\end{equation}
A candidate with $\|z_a\|_2=0$ adds no independent observable variation and
has zero unregularized improvement. For a nonredundant candidate,
the numerator measures alignment with the residual, while the
denominator normalises the added direction. For consistency with Eq.~\eqref{eq: regularized OVQITE linear system},
the regularized score is defined using
\begin{equation}
\label{eq:adapt-vime-regularized-loss}
L_\lambda(\dot\theta,\dot\phi)
=\frac12\|J\dot\theta+j_a\dot\phi-v\|_2^2
+\frac{\lambda}{2}
\left(\|\dot\theta\|_2^2+\dot\phi^2\right).
\end{equation}
Set $G_\lambda=G+\lambda I$, $q_a=J^Tj_a$, and evaluate
$\boldsymbol{\vec \epsilon}$ using Eq.~\eqref{eq:adapt-vime-residual}.
After adding candidate $a$ and reoptimising all parameter velocities,
the minimum of the regularized least-squares objective decreases by
\begin{equation}
\label{eq:adapt-vime-schur}
\begin{aligned}
\Delta L_a^{(\lambda)}
&=\frac{(j_a^T\boldsymbol{\vec \epsilon})^2}{2\chi_a},\\
&\chi_a=j_a^Tj_a+\lambda-q_a^TG_\lambda^{-1}q_a.
\end{aligned}
\end{equation}
For $\lambda>0$, $\chi_a$ is positive. At zero regularisation the pseudoinverse form recovers Eq.~\eqref{eq:adapt-vime-rank-one}.
These identities assume the corresponding linear system is solved accurately.

The selected Majorana rotation is the one with the largest
positive $\Delta L_a^{(\lambda)}$, which we call the accurate selection score.

\subsection{Cheap screening of the generator pool}

Evaluating the accurate score for every generator requires
$|\mathcal P|$ candidate columns, each obtained by a forward
derivative pass through the graph. For $N$ fermionic modes,
a pool of length-two and length-four Majorana generators
contains at most $O(N^4)$ elements. To reduce this cost,
we adopt the two-stage selection strategy of
Ref.~\cite{chakraborty2026adaptvmpe}: rank the full pool
using a relatively cheaper screening score, then evaluate the retained subset $\widetilde{\mathcal P}$ with the accurate score.

For ADAPT-VIME, we use the cheap screening score
\begin{equation}
s_a=|j_a^T\boldsymbol{\vec \epsilon}|,
\end{equation}
whose square is the numerator of
Eq.~\eqref{eq:adapt-vime-schur}. Unlike the accurate score,
$s_a$ does not account for the magnitude of the candidate
column or how much of it can already be reproduced by
the existing Jacobian columns.

The cheap score can be obtained without constructing
$j_a$ explicitly. We define
\begin{equation}
\label{eq:adapt-vime-residual-observable}
\hat R_{\epsilon}
=\sum_{\hat O_k\in S_H}\epsilon_k\hat O_k,
\end{equation}
where $\epsilon_k$ is the $k$th component of the residual
vector $\boldsymbol{\vec \epsilon}$. The screening score for candidate $a$ can then be written as the magnitude of the derivative of this observable's expectation value with respect to the candidate angle:
\begin{equation}
\label{eq:adapt-vime-probe}
\begin{aligned}
s_a
&=
\left|
\left.
\partial_\phi
\langle\hat R_{\epsilon}\rangle_{\mathrm{MP}}^{(a)}
\right|_{\phi=0}
\right|
=
\left|\sum_k\epsilon_k(j_a)_k\right|
=
|j_a^T\boldsymbol{\vec \epsilon}|.
\end{aligned}
\end{equation}
A large score indicates that the candidate produces an observable
variation strongly aligned with the current residual. 
To evaluate this quantity efficiently, we adapt the fast
Heisenberg-picture gradient probe of Appendix C of
Ref.~\cite{chakraborty2026adaptvmpe}, replacing the Hamiltonian
by $\hat R_{\epsilon}$. We first back-propagate the Majorana coefficients of
$\hat R_\epsilon$ through the current circuit:
\begin{equation}
\label{eq:adapt-vime-residual-backward}
\vec h_{\epsilon,\mathrm{root}}
=\mathsf{\hat P}\sum_k\epsilon_k\vec A_k.
\end{equation}
These coefficients define the back-propagated operator
\begin{equation}
\hat R_{\epsilon,\mathrm{root}}
=\sum_b(\vec h_{\epsilon,\mathrm{root}})_b
\hat M_b^{(\mathrm{root})}.
\end{equation}
Here $\vec A_k$ contains the terminal coefficients of
$\hat O_k$, and $\mathsf{\hat P}$ is the propagation matrix
of Eq.~\eqref{eq: app propagation matrix}.
This calculation is performed once for all candidates.

The candidate rotation acts directly on the reference state,
so only its action on $\hat R_{\epsilon,\mathrm{root}}$
remains to be evaluated. Differentiating with respect to
its angle at $\phi=0$ gives
\begin{equation}
\label{eq:adapt-vime-root-probe}
\begin{aligned}
s_a
&=
\left|
\left.
\partial_\phi
\operatorname{Tr}\!\left[
\hat\rho\hat R_a^\dagger(\phi)
\hat R_{\epsilon,\mathrm{root}}
\hat R_a(\phi)
\right]\right|_{\phi=0}
\right|\\
&=
\frac12\left|
\operatorname{Tr}\!\left[
\hat\rho[\hat M_{\vec b_a},
\hat R_{\epsilon,\mathrm{root}}]
\right]\right|.
\end{aligned}
\end{equation}
Thus, each screening score is obtained from the commutator
of the candidate generator $\hat M_{\vec b_a}$ with the same back-propagated
operator $\hat R_{\epsilon,\mathrm{root}}$. We form $\widetilde{\mathcal P}$ by retaining a chosen number
or fraction of the candidates with the largest $s_a$.
For these candidates, we compute the Jacobian columns
[Eq.~\eqref{eq:adapt-vime-candidate-pass}] and select the
Majorana rotation with the largest positive
$\Delta L_a^{(\lambda)}$ [Eq.~\eqref{eq:adapt-vime-schur}]. Following Ref.~\cite{chakraborty2026adaptvmpe}, we repeat
the full-pool screening every $k$ growth steps.

\subsection{Adaptive loop and relation to ADAPT-VMPE}

The algorithm starts from a Fock reference followed by active
orbital rotations, as in Ref.~\cite{chakraborty2026adaptvmpe}. At first, VIME evolves all parameters until the convergence criterion
of Eq.~\eqref{eq: OVQITE convergence}, based on
$\|J^Tv\|_2$, is satisfied.

If the residual $\|\boldsymbol{\vec \epsilon}\|_2$ remains above
a chosen tolerance, we select an additional Majorana rotation
using the screening and accurate scoring procedures described
in the preceding subsections. The selected rotation is inserted
at $\phi=0$, the MP graph is extended as described above,
and its generator is removed from the available pool.
VIME then resumes from the current parameters, evolving both
the new angle and all existing parameters. These steps are repeated until the residual after VIME convergence is below tolerance, no candidate has a positive accurate score, or the circuit budget is reached.

\begin{figure*}[t]
\centering
\includegraphics[width=\textwidth]{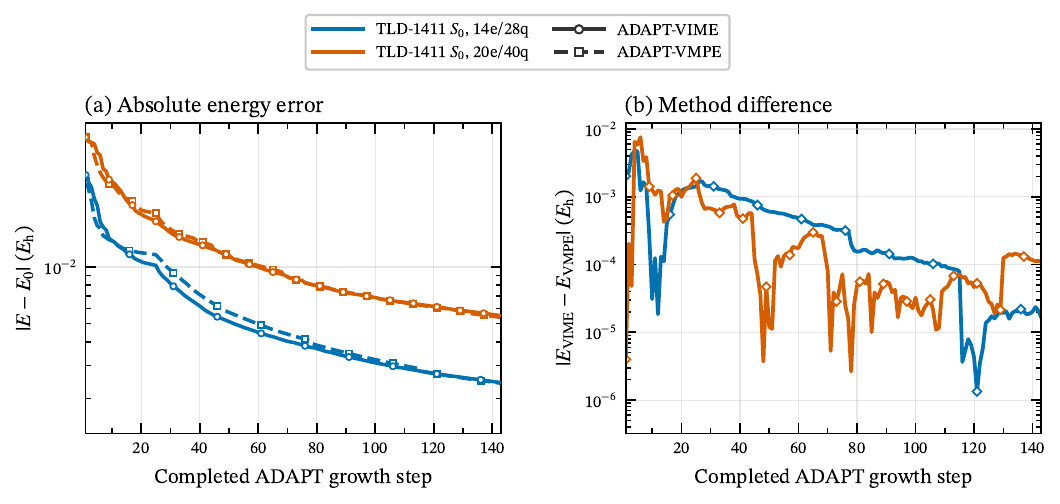}
\caption{
Energy convergence of ADAPT-VIME and ADAPT-VMPE for the
28- and 40-qubit active spaces of TLD1411 in the $S_0$ state.
(a) Absolute MP energy error relative to the reference energy
$E_0$.
(b) Absolute difference between the MP energies obtained by the
two methods.
Energies are compared after relaxation at the same number of
added gates, over the first 143 completed growth steps.
The methods exhibit similar energy convergence despite selecting
different operator sequences.
}
\label{fig:adapt-vime-vmpe}
\end{figure*}

Figure~\ref{fig:adapt-vime-vmpe} compares ADAPT-VIME and ADAPT-VMPE
for the 28- and 40-qubit active spaces of TLD1411 in the $S_0$
state. The energy curves closely follow each other despite
the different selection criteria, which lead to different
Majorana generators at almost every step. Over 143 growth
steps, the choices coincide at only one step for the 28-qubit
system and four steps for the 40-qubit system. Even disregarding
selection order, only 86 and 47 generators, respectively, are
common to both methods. The close energy agreement therefore
persists despite differences in both the generators selected
and their order.

To understand this phenomenon, we compare
the new angle $\phi$ with the changes in the previously existing parameters $\theta$.
At growth step $n$, let $\phi_n$ be the final angle of the
added Majorana rotation, measured from zero, and let
$\Delta\theta^{(n)}$ be the change in the previously existing parameters
between convergence of the preceding and enlarged circuits. We define
\begin{equation}
\label{eq:adapt-parameter-ratios}
R_{\mathrm{mean}}^{(n)}
=
\frac{|\phi_n|}
{\operatorname{mean}_i|\Delta\theta_i^{(n)}|},
\qquad
R_2^{(n)}
=
\frac{|\phi_n|}{\|\Delta\theta^{(n)}\|_2}.
\end{equation}
These ratios compare the new angle with the mean absolute
change and the total Euclidean norm of the changes in the
existing parameters, respectively.

\begin{table}[t]
\centering
\caption{
Medians of the ratios in
Eq.~\eqref{eq:adapt-parameter-ratios} over the first 143
completed growth steps. Existing parameters include active
orbital rotations. The new angle is measured from zero,
including the greedy gradient-free (GGF) initialization
in ADAPT-VMPE.
}
\label{tab:adapt-parameter-updates}
\begin{tabular}{rlrr}
\hline\hline
Qubits & Method & $R_{\mathrm{mean}}$ & $R_2$ \\
\hline
28 & ADAPT-VIME & 27.6 & 0.62 \\
28 & ADAPT-VMPE & 113  & 4.23 \\
40 & ADAPT-VIME & 33.3 & 0.44 \\
40 & ADAPT-VMPE & 164  & 4.19 \\
\hline\hline
\end{tabular}
\end{table}

Table~\ref{tab:adapt-parameter-updates} indicates that the
ADAPT-VMPE updates are dominated by the new angle, whose
magnitude exceeds the combined changes in the existing
parameters by a median factor of approximately four.
This suggests that the optimization after adding a rotation
is largely a one-dimensional problem.

ADAPT-VMPE addresses this problem through greedy
gradient-free (GGF) selection. With the existing parameters
$\theta$ fixed, each candidate has energy
\begin{equation}
\label{eq:adapt-vime-single-angle}
E_{\mathrm{MP}}(\theta,\phi)
=c_0+c_1\cos\phi+c_2\sin\phi.
\end{equation}
Three energy evaluations determine the optimal angle
analytically~\cite{chakraborty2026adaptvmpe}. The candidate
giving the lowest energy is selected and initialized
at this angle before all parameters are optimized together.
In these runs, the new angle changes negligibly during
joint optimization, indicating that its adjustment is
essentially completed during GGF selection.
This provides a plausible explanation for the similar
energy convergence: direct energy minimisation is already
effective in this regime, leaving limited scope for
ADAPT-VIME's imaginary-time updates to improve the energy
accuracy per added rotation.

\end{document}